\documentclass[11pt,draftclsnofoot,onecolumn]{IEEEtran}
\usepackage{cite}

\ifCLASSINFOpdf
  \usepackage[pdftex]{graphicx}
\else
\fi
\usepackage{amsmath}
\usepackage{amssymb}

\usepackage[pdftex]{hyperref}

\newtheorem{theorem}{Theorem}
\newtheorem{lemma}{Lemma}
\newtheorem{corollary}{Corollary}

\newcommand{\lt}{\left}
\newcommand{\rt}{\right}
\newcommand{\cn}{\,|\,}
\newcommand{\1}[1]{\boldsymbol{1}_{\lt\{#1\rt\}}}
\newcommand{\Pe}[1]{P_{e,#1}^{(n)}}
\newcommand{\Ex}[1]{\mathbb{E}\!\left[\,#1\,\right]}
\newcommand{\dz}[2]{Z_{(#1)}^{(#2)}}
\newcommand{\dzs}[1]{Z^{(#1)}}
\newcommand{\dv}[2]{V_{(#1)}^{(#2)}}
\newcommand{\dvs}[1]{V^{(#1)}}
\newcommand{\dvl}[1]{V_{(#1)}}
\newcommand{\lv}[2]{v_{(#1)}^{(#2)}}
\newcommand{\lvs}[1]{v^{(#1)}}

\newcommand{\C}{\overline{C}}
\newcommand{\uC}{\underline{C}}

\newcommand{\Ps}{\mathcal{P}}
\newcommand{\St}{\mathcal{S}}
\newcommand{\U}{\mathcal{U}}
\newcommand{\V}{\mathcal{V}}
\newcommand{\X}{\mathcal{X}}
\newcommand{\Y}{\mathcal{Y}}
\newcommand{\Z}{\mathcal{Z}}

\newcommand{\sAe}{A_{\epsilon}^{*(m)}}
\newcommand{\Ae}{A_{\epsilon}^{(m)}}
\newcommand{\infg}{\inf_{\gamma>0}}
\newcommand{\bH}{\bar{H}}
\newcommand{\bP}{\bar{P}}
\newcommand{\bS}{\bar{S}}
\newcommand{\bX}{\bar{X}}
\newcommand{\bU}{\bar{U}}
\newcommand{\bV}{\bar{V}}
\newcommand{\bs}{\bar{s}}
\newcommand{\bx}{\bar{x}}
\newcommand{\bu}{\bar{u}}
\newcommand{\bv}{\bar{v}}
\newcommand{\bxi}{\bar{\xi}}
\newcommand{\bta}{\bar{\eta}}

\newcommand{\cta}{\breve{\eta}}

\newcommand{\hG}{\hat{G}}
\newcommand{\hX}{\hat{X}}

\newcommand{\tC}{\tilde{C}}
\newcommand{\tP}{\tilde{P}}
\newcommand{\tp}{\tilde{p}}
\newcommand{\tq}{\tilde{q}}
\newcommand{\ts}{\tilde{s}}
\newcommand{\tX}{\tilde{X}}
\newcommand{\tV}{\tilde{V}}
\newcommand{\tpi}{\tilde{\pi}}
\newcommand{\tphi}{\tilde{\phi}}

\begin{document}
%
\title{Models and information-theoretic bounds for nanopore sequencing}
%
%
%

\author{Wei Mao, Suhas Diggavi, and Sreeram Kannan
\thanks{Wei Mao was with the Department of Electrical Engineering, University of California, Los Angeles, CA 90095 USA. He is now with Intel Corporation, Santa Clara, CA 95054 USA (email: wmao.tsinghua@gmail.com).}%
\thanks{Suhas Diggavi is with the Department of Electrical Engineering, University of California, Los Angeles, CA 90095 USA (e-mail: suhasdiggavi@ucla.edu).}
\thanks{Sreeram Kannan is with the Department of Electrical Engineering, University of Washington, Seattle, WA 98195 USA (e-mail: ksreeram@uw.edu).}
\thanks{This paper was presented in part at ISIT 2017. The work of Wei Mao and Suhas Diggavi was supported in part by NSF grants 1423271 and 1705077. The work of Sreeram Kannan was supported by NSF Career award 1651236 and NSF grant 1703403. We would also like to acknowledge the technical discussions and data from Jens Gundlach's UW Nanopore group, which enabled some of the mathematical models as well as numerical evaluations in this paper.}}

\maketitle

\begin{abstract}
Nanopore sequencing is an emerging new technology for sequencing DNA, which can read long fragments of DNA ($\sim$50,000 bases) in contrast to most current short-read sequencing technologies which can only read hundreds of bases. While nanopore sequencers can acquire long reads, the high error rates ($20\%$--$30\%$) pose a technical challenge. In a nanopore sequencer, a DNA is migrated through a nanopore and current variations are measured. The DNA sequence is inferred from this observed current pattern using an algorithm called a {\em base-caller}. In this paper, we propose a mathematical model for the ``channel'' from the input DNA sequence to the observed current, and calculate bounds on the information extraction capacity of the nanopore sequencer. This model incorporates impairments like (non-linear) inter-symbol interference, deletions, as well as random response. These information bounds have two-fold application: (1) The decoding rate with a uniform input distribution can be used to calculate the average size of the plausible list of DNA sequences given an observed current trace. This bound can be used to benchmark existing base-calling algorithms, as well as serving a performance objective to design better nanopores. (2) When the nanopore sequencer is used as a reader in a DNA storage system, the storage capacity is quantified by our bounds.
\end{abstract}


%
\IEEEpeerreviewmaketitle

\section{Introduction}
\label{sec:intro}

DNA sequencing technology has undergone a major revolution with the
cost of sequencing plummeting nearly six orders of magnitude. Much of
this improvement was made possible by second generation sequencers,
utilizing massive parallelization, but these machines can only read
short fragments of DNA, typically a few hundred bases long. These
short reads are then stitched together with algorithms exploiting the
overlap between reads to assemble them into long DNA sequences. This
assembly is unreliable because of repeated regions which commonly
occur in genomic DNA. These repeated regions play an important role in
evolution, development and in the genetics of many diseases
\cite{manolio2009finding,renton2011hexanucleotide,walsh2008rare}.

Nanopore sequencing promises to address this problem, by increasing
the read lengths by orders of magnitude ($\approx$ 10--100K bases) \cite{deamer2016three}.  The
technology is based on DNA transmigrated through a nanopore, and the
ion current variations through the pore are measured 
\cite{kasianowicz1996characterization,branton2008potential,GundlachNature14,deamer2016three}. The sequence of DNA bases is inferred from the observed current traces using an algorithm termed as base-caller \cite{david2016nanocall,bovza2016deepnano}. Nanopore-sequencing technology is also beginning to be commercialized by Oxford Nanopore Technologies \cite{eisenstein2012oxford}. Nanopore sequencers have enabled new applications such as rapid virology \cite{eisenstein2012oxford,cao2016streaming,greninger2015rapid} and
forensics \cite{zaaijer2016democratizing}. 

However, despite recent progress, there is an important bottleneck;
nanopore sequencers built to date, have high error rate for \emph{de
  novo} sequencing\footnote{\emph{De novo} sequencing refers to
  decoding the DNA sequence without the help of any reference DNA. For
  example, state-of-the-art nanopore sequencers have nearly $20$--$30\%$
  single-read sequencing error-rate \cite{sovic2016fast}. Error-rates of $10$--$15\%$ are
  reported using double-stranded sequencing, but are still high for
  many applications \cite{jain2015improved,attar2015techniques}.
}. It is unclear whether the present error-rate is fundamental to the nanopore or due to the limitations of present base-calling algorithms. Thus one goal of the present direction of work is to understand the information-theoretic limits of the nanopore sequencer, and help benchmark base-calling algorithms. To achieve this goal, standardized signal-level models are needed to analyze such sequencers, which are currently unavailable. Another important benefit of such information theoretic understanding is that it provides a way to optimize the nanopore sequencer (for maximum information decoding capacity).


Motivated by this, our first contribution is in developing a mathematical model for the nanopore sequencer. We use the physics of the nanopore sequencers and experimental data provided by our collaborators \cite{GundlachNature14} to develop a signal-level model for the \emph{nanopore channel}, which captures (non-linear) inter-symbol interference, signal-level synchronization errors, and random response of the nanopore to the same inputs.

In de-novo DNA sequencing, the DNA sequence is arbitrary and only the decoder is under our control. In such a case, a performance metric of interest is the number of plausible input sequences given that the output is observed. When the mapping between the DNA sequence and observed current is noiseless and fully invertible, the number of plausible sequences is exactly one, which is the true sequence. In a noisy and imperfect DNA sequencer, as in the model described above, many sequences will be plausible and thus the size of the uncertainty list is a measure of the decoding capability of a sequencer: the smaller the list size is, the better the sequencer is at extracting the information. For an i.i.d. uniform input distribution, each input sequence is (weakly) typical. The reliable decoding rate calculated with an i.i.d. uniform input distribution, gives a count of the number of sequences that can be reliably distinguished from all possible sequences, and therefore the remaining list size is the difference between the total number of sequences and this reliable rate. This list size is therefore a quantifiable performance metric for the sequencer.

Our second contribution is the reliable information rate analysis for the nanopore channel. We first generalize Dobrushin's multi-letter capacity formula \cite{Dobrushin-channel-sync-error} for channels with synchronization errors to the case with inter-symbol interference, which includes the nanopore channel as a special case. Then we develop single-letter computable lower bounds on the nanopore channel capacity using techniques for deletion channels. As argued before, these lower bounds are used for estimating the size of uncertainty list of DNA sequences after observing the output current trace, giving a metric for the performance of the nanopore sequencer. We can also use these ideas to assess information storage capability in DNA using nanopore sequencers as decoders, by optimizing the input distribution, which can be done if we can design DNA information storage codes.

As another contribution, we also develop novel \emph{computable} upper bounds for the decoding capacity of nanopore channel, using a combination of upper-bounding techniques for deletion channels and finite-state channels.

To apply these ideas to assess the performance of the nanopore sequencer, we numerically evaluate these bounds for both synthetic data using nanopore models as well as measured responses from nanopore data, which could give insights into the physical design of nanopore having good decoding performance. Using channel models obtained from measured data, we show that the reliable decoding rate of the nanopore is $1.6$--$1.9$ bits for channel parameters obtained in the data\footnote{The parameters, like  deletion probability, noise, etc., used in this statement are in the range of the data in \cite{GundlachNature14}, i.e., deletion probability 0.01--0.1. See Fig.~\ref{fig:nanopore-channel} for the block model and Fig.~\ref{fig:forward-pore-linear} and~\ref{fig:forward-pore-log} for more complete numerics.}. Note that the exponent of the average list size is $\log|\mathcal{X}|-R=2-R$, where $R$ is the reliable decoding rate and $\mathcal{X}=\{A,C,G,T\}$. This means, from our numerics stated above (and explored in Sections \ref{sec:main-results},\ref{sec:Numerics}), a list size exponent of between $0.1$--$0.4$ is potentially achievable. In addition, through a model for the nanopore response we exhibit a trade-off for the achievable rates of nanopore sequencers when the nanopore length $K$ increases: larger $K$ provides more protection against deletion errors, but also introduces more noise in the $K$-mer map responses, therefore providing a sweet spot for $K$ with respect to achievable rates. Therefore the reliable decoding rate, and the list size, give a quantifiable performance metric, which can enable analysis of such design trade-offs.

The major technical challenge for the analysis of the nanopore channel lies in the fact that it belongs to the category of channels with synchronization errors. The study of such channels dates back to Gallager \cite{gallager1961sequential} and Dobrushin \cite{dobrushin1967shannon}, and interest in the problem was revived due to new bounds in \cite{Diggavi-Grossglauser-finite-buffer-channel} (see also \cite{DG01}). See \cite{Mercier-DMC-sync-sub-error} for a survey of results. However, even the simplest i.i.d. deletion channel capacity is unresolved. In this paper we develop novel lower and upper bounds for a channel model inspired by nanopore sequencing. We believe that several of the techniques and results could also be of independent interest, beyond its application to nanopore sequencing. In \cite{duda2016fundamental}, a nanopore sequencer is modeled at a {\em hard-decision} level by a simplified insertion-deletion channel, where no run of DNA bases is deleted (similar to \cite{DMP07}) or inserted, to understand how to combine multiple reads. Our channel model, however, is aimed at designing base-callers (algorithms for decoding DNA from current trace) and therefore operates at fine-grained signal level.

The paper is organized as follows. Section \ref{sec:model} develops the signal-level model for the nanopore channel. We outline the main technical results of this paper in Section~\ref{sec:main-results}. We develop the proof of the multi-letter capacity formula in Section~\ref{sec:multi-letter-capacity}, the achievable rates for the nanopore channel in Section~\ref{sec:ach-rate}, and the upper bounds in Section~\ref{sec:upper-bounds}. Numerical evaluation of these bounds, including channel model derived from real measurements in nanopore sequencers provided by UW nanopore lab \cite{GundlachNature14} are given in Sections~\ref{sec:main-results} and~\ref{sec:Numerics}.

\section{Model, Notation and Problem formulation}
\label{sec:model}

\subsection{Nanopore sequencer}

\begin{figure}[!t]
\centering
\includegraphics[width=2in]{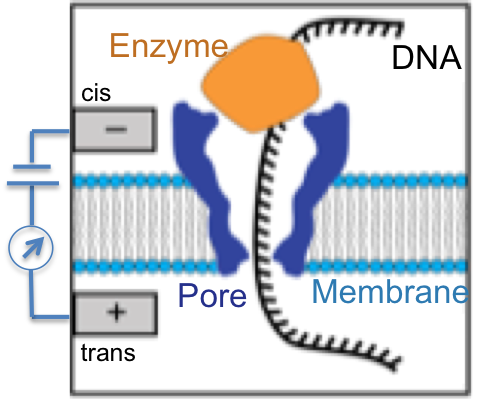}
\caption{Nanopore Sequencer schematic, adapted from \cite{GundlachNature14}}
\label{fig:schematic}
\end{figure}

We will use the nanopore sequencer shown in the simplified schematic in Fig.~\ref{fig:schematic}, with details in
\cite{GundlachNature14,GundlachNature12}.  A salt solution is divided into two wells: {\em cis} well and {\em trans} well by a lipid bilayer membrane. A nanopore is inserted into the bilayer, and a voltage applied between the cis and trans wells results in an ionic current. The (single-strand) DNA sequence to be measured is prefixed and affixed with short known DNA pieces (called adaptors) and enters the nanopore. As the DNA migrates through the pore, modulations of the ionic current caused by different nucleotides partially blocking the pore are measured. An enzyme controls the DNA motion through the nanopore, slowing it down so that the current variations can be measured accurately.\footnote{In the experiments of \cite{GundlachNature14,GundlachNature12}, our collaborators use a biological nanopore called Mycobacterium smegmatis porin A (MspA) and a Phi29 DNA polymerase as the enzyme.}

\begin{figure}[!t]
\centering
\includegraphics[width=0.8\textwidth]{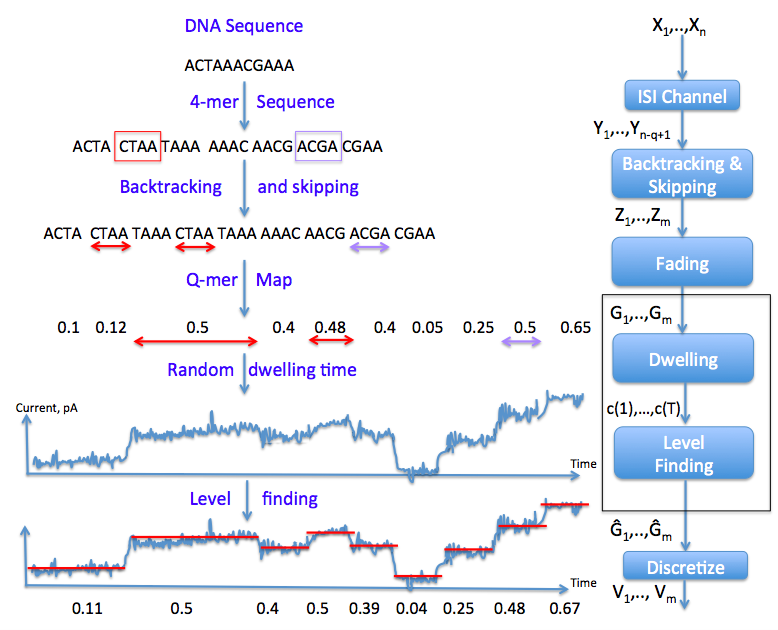}
\caption{Model of the Nanopore Sequencer for a toy DNA sequence. See Section~\ref{subsec:model-notation} for a description of the level-finding and discretizing process.}
\label{fig:model}
\end{figure}

\begin{figure}[!t]
\centering
\includegraphics[width=0.8\textwidth]{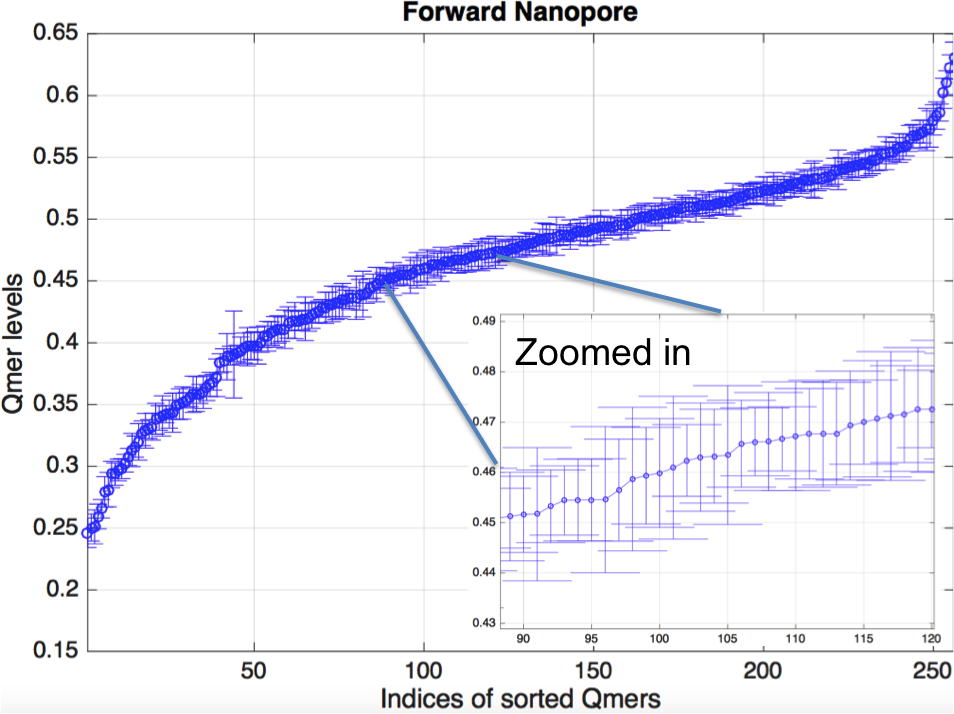}
\caption{Mean and standard deviation of 256 Q-mer maps obtained
  from our nanopore. Notice the significant overlap of Q-mer
  maps.}
\label{fig:q-mer}
\end{figure}

An ideal nanopore sequencing system would ensure that DNA migrates through the pore at a constant rate, that only one base of DNA affects the current at a given time and from the observed current, the nucleotide of DNA can be decoded unambiguously with high probability. However, the following phenomena occur due to the physics of the nanopore and the enzyme (see Fig.~\ref{fig:model}). 
 (i) \emph{Inter-symbol interference:} Each observed current value is influenced by multiple bases or ``$K$-mer'' ($K$ bases), because the constriction of the nanopore is thicker than a single nucleotide. In our experiments $K \approx 4$, so we also call them quadromers or ``Q-mers'' for short. As in DNA sequences there are 4 different type of nucleotides, there are $4^{4} = 256$ different Q-mers all together (see Fig.~\ref{fig:model}).
(ii) \emph{Random dwelling time:} Each Q-mer of the DNA sequence may spend a varying amount of time in the nanopore. This means that the measured current segment corresponding to a Q-mer transitions to the next Q-mer segment at random times (see Fig.~\ref{fig:model}). In fact the speed at which the DNA sequence migrates through the enzyme is a stochastic process.
 (iii) \emph{Backtracking and skippings:} The nucleotides are not necessarily read in order by the nanopore - there is some backtracking and significant mis-stepping in the nanopore induced by the enzyme that draws the DNA through the nanopore. This results in segments that are repeated as well as segments that pass through without registering a current reading, resulting in backstepping as well as skipping (deletion) of segments (see Fig.~\ref{fig:model}).
 (iv) \emph{Q-mer map fading:} For each segment $i$, the measured current level $G_{i}$ is a function of the Q-mer $z_{i}$ dwelling in the nanopore. However, this function is not deterministic: each time the same dwelling Q-mer produces a current level with some variation, which resembles the \emph{fading} effect in communication channels. In Fig.~\ref{fig:q-mer}, we plot the mean value of the 256 random Q-mer map, together with their variances as error-bar.
 (v) \emph{Noisy samples:} On top of the Q-mer map fading, within the same segment $i$, each sample of the same current level is subject to a random noise. Hence $G_{i}$ is estimated\footnote{The estimation is usually accurate, since the sampling rate is usually high enough ($\sim$ 50kHz) to produce a large number of points for each segment (duration $>$ 40 ms on average).} by $\hG_{i}$, the average of all samples within segment $i$ (see Fig.~\ref{fig:model}). Usually we model the noise as an AWGN process with some variance $\sigma^{2}_{i}$, which may vary from segment to segment. Hence if at each sample time $t$ the measured current value is $c(t)$, then $G_{i} \approx \sum_{T_{i-1}<t\leq T_{i}}c(t)/(T_{i}-T_{i-1})$, where $T_{i-1}$ and $T_{i}$ denote the boundary of the samples within segment $i$.

\subsection{Channel model and Notation}\label{subsec:model-notation}

We can model the nanopore experiment as a communication process, with the input being the DNA sequence to be measured and the output being the current samples produced.
However, the sample-resolution channel is too complicated to admit any useful analysis and algorithm development---since the random dwelling time demands an infinite memory for the channel. In practice\cite{GundlachNature14}, a \emph{separation-based} approach is often taken to solve this issue. In this scheme the data-processing of the measured current consists of two stages: 1) finding the transitions of the dwelling Q-mers using a change-point detection algorithm, or similar algorithms, to determine the boundaries of each step; 2) estimating the mean value of the current samples during each step, which gives the current level corresponding to the dwelling Q-mer. The decoder then take the levels as input for further processing and estimate the DNA sequence. In Fig.~\ref{fig:model}, we see that $G_1,..,G_m$ is sent through a dwelling process to obtain current trace $c(1),..,c(T)$ which after level-finding yields $\hat{G}_1,..,\hat{G}_m$. In this paper, we will assume that level-finding is perfect and hence $\hat{G}_i = G_i$. In practice, the effect of level-finding errors can be folded into the backtracking / skipping model built for the enzyme so that the end-to-end modeling is still quite accurate. 

\begin{figure}[!t]
\centering
\includegraphics[width=0.8\textwidth]{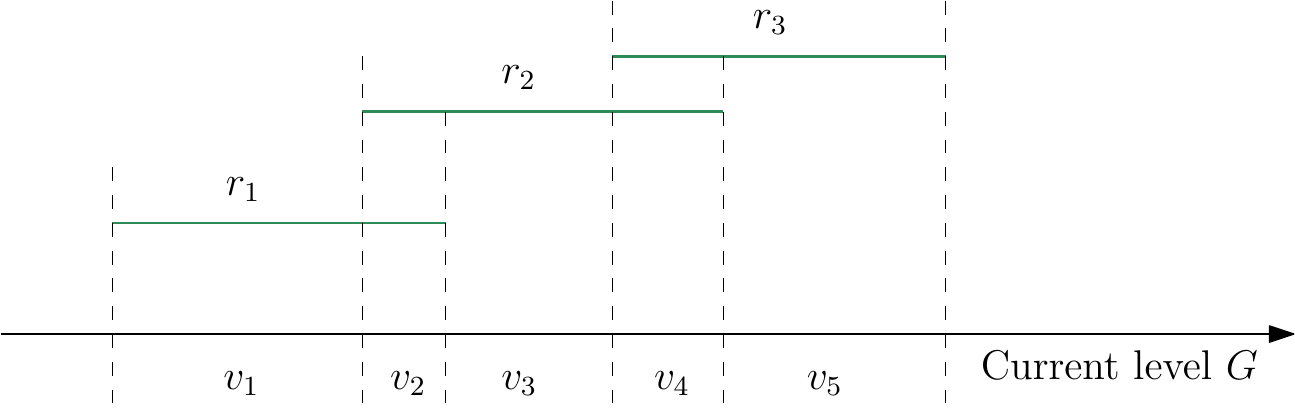}
\caption{Partition by intersections of Q-mer map ranges.}
\label{fig:interval-intersection}
\end{figure}

Thus in our mathematical model, we assume that the channel in the box in Fig.~\ref{fig:model} from $G_1,..,G_m$ to $\hat{G}_1,...,\hat{G}_m$ is an identity channel, thus modeling the nanopore purely at the DNA symbol level. We also make two further simplifications to facilitate the analysis of the model, while still trying to capture the most important effects for the nanopore experiment. First, in the experiment, skipping happens much more frequently than backtracking. Thus we ignore the backtracking effect and model skipping as an i.i.d. deletion process.\footnote{Post-processing of data can detect and remove backstepping by examining exact repeats of previously measured currents\cite{GundlachNature14}. Moreover, these backsteps are seen to occur more rarely than deletions, motivating this simplification.} Second, we model the Q-mer map fading as an independent random process, with each Q-mer having its own fading distribution. The nanopore experiment measurement results give statistics of these distributions, such as means and standard deviations (see Fig.~\ref{fig:q-mer}). In addition, usually these measured levels for the same Q-mer fall in a certain range. Thus a natural model to capture these characteristics is the uniform distribution\footnote{The uniform model captures the interactions within 1--2 standard deviation of the Q-mer map variations, as shown in Fig.~\ref{fig:q-mer} and~\ref{fig:interval-intersection}.}: we assume each Q-mer has a uniform fading distribution that conforms with its measured mean and standard deviation. The support of each uniform distribution is a bounded interval, with the center being the mean value of the corresponding Q-mer map and the width being a constant multiple of the standard deviation. It can also be called the range of a Q-mer map. As shown in Fig.~\ref{fig:q-mer}, the Q-mer map ranges have a lot of intersections and thus induce a partition on the set of all possible current level values (see Fig.~\ref{fig:interval-intersection} for an example, where $\{v_{1},\ldots,v_{5}\}$ partitions the union of the ranges $r_{1}$--$r_{3}$). Such intersections cause ambiguity in the inference of the dwelling Q-mer: for example, if the measured current level falls in the region $v_{2}$ in Fig.~\ref{fig:interval-intersection}, then there would be two possible Q-mers, $z_{1}$ and $z_{2}$, which correspond to ranges $r_{1}$ and $r_{2}$, respectively, causing such a measurement.

\begin{figure}[!t]
\centering
\includegraphics{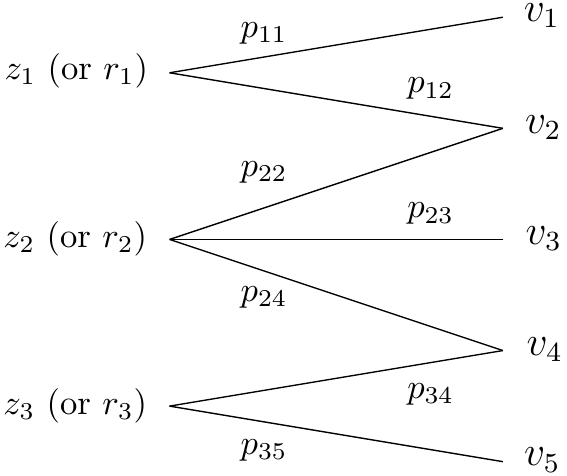}
\caption{DMC arising from the partition in Fig.~\ref{fig:interval-intersection}, transition probability $p_{ij} = l(v_{j})/l(r_{i})$.}
\label{fig:DMC}
\end{figure}

\begin{figure}[!t]
\centering
\includegraphics[width=0.7\textwidth]{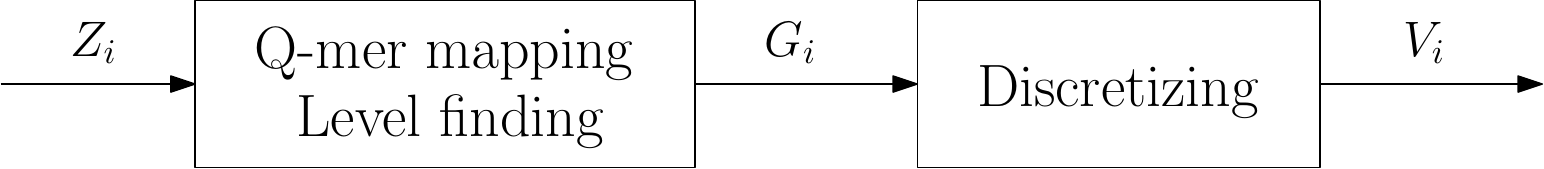}
\caption{Level-finding and discretization.}
\label{fig:fading-discretization}
\end{figure}

Such a partition provides a set of sufficient statistics for the channel output. For example, in Fig.~\ref{fig:interval-intersection}, if we observe a current level $G\in v_{2}$, then since the fading distribution is uniform, knowing the exact place in $v_{2}$ does not give us more information about which one of the two possible Q-mers, $z_{1}$ or $z_{2}$, is more likely to be the dwelling Q-mer. Thus we can just consider the part $v_{2}$ as the output of the channel, and call it the discretized level corresponding to $G$. Since the fading process is memoryless, the discretized random Q-mer map can be modeled as a DMC, as shown in Fig.~\ref{fig:DMC}. The channel transition probability is determined by
\[ p(v\cn z) = \begin{cases}
l(v)/l(r(z)) & \text{if } v\subseteq r(z), \\
0 & \text{o.w.}
\end{cases} \]
for any discretized interval $v$ and any Q-mer $z$, where $r(z)$ denotes the Q-mer map range of $z$ and $l(\cdot)$ denote the length of an interval. The process of level-finding and discretization is depicted in Fig.~\ref{fig:fading-discretization} (corresponding to the lower-right block in Fig.~\ref{fig:model}), which amounts to a DMC as described above.

\begin{figure}[!t]
\centering
\includegraphics[width=0.7\textwidth]{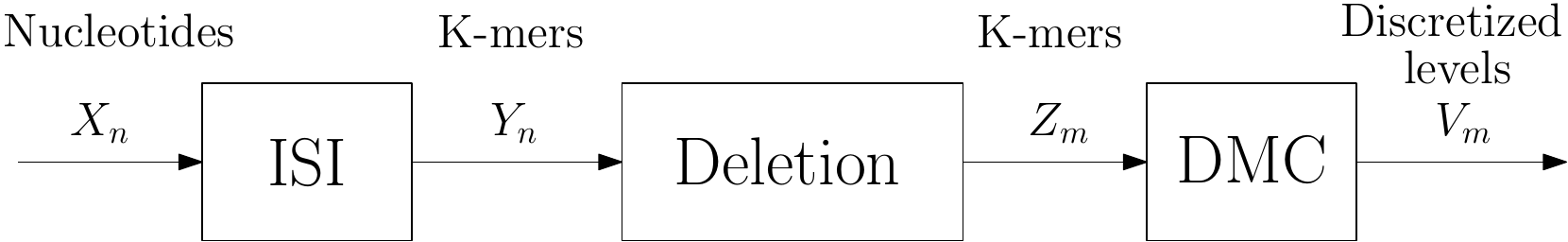}
\caption{Simplified nanopore channel.}
\label{fig:nanopore-channel}
\end{figure}

Hence we arrive at a simplification of the nanopore channel model. In this simplified channel (depicted in Fig.~\ref{fig:nanopore-channel}), the input (corresponding to the bases in the DNA sequence) symbol at time $n$ is denoted by $X_{n}$, which has a finite alphabet $\X$. $X_{n}$ is passed through an ISI channel with memory $K-1$, to form the $K$-mer $Y_{n} = X_{n-K+1}^{n} = (S_{n-1},X_{n})$,
where $S_{n-1} \triangleq X_{n-K+1}^{n-1}$ denotes the ISI channel state at time $n-1$. The alphabets for the $K$-mers and ISI states are denoted by $\Y = \X^{K}$ and $\St = \X^{K-1}$, respectively. The initial state $s_{0}$ is known (they are the prefixed adaptors). The $K$-mers $\{Y_{n}\}$ are then sent to a deletion channel, where each $K$-mer is independently deleted with probability $p_{d}$. We can represent the deletion process by an i.i.d. Bernoulli($p_{d}$) process $\{D_{i}\}_{i\geq1}$, where $D_{i} = 1$ denotes a deletion at time $i$. When the input of the deletion channel is a length-$n$ vector $Y^{n}$, its output is a $K$-mer vector of random length, denoted by $\dzs{n}$. The corresponding alphabet is
$\Y^{(n)}\triangleq\emptyset\cup \Y \cup \Y^{2} \cup \cdots \cup \Y^{n}$.
Similarly when a segment of $K$-mers $Y_{n_1}^{n_2}$ is input to the deletion channel, we use $\dz{n_1}{n_2}$ to denote the output.\footnote{The sub/superscripts in these notations have parentheses around them to denote the effect of deletion.} Finally, after deletion, a $K$-mer $Z_{m}$ is sent to a DMC to produce the channel output $V_{m}$ (corresponding to the discretized levels, see Fig.~\ref{fig:fading-discretization}), whose alphabet is denoted by $\V$. We also use $\dv{n_1}{n_2}$ to denote the output of DMC corresponding to the random-length $K$-mer vector $\dz{n_1}{n_2}$, and denote its alphabet by $\V^{(n_{2}-n_{1}+1)}$.

For readers' convenience, we summarize our notations in Table~\ref{table:notations}. Also note that we use the terms ``string'' and ``vector'' interchangeably in this paper.

\begin{table}[!t]
\caption{Notations}
\label{table:notations}
\centering
\begin{tabular}{c|p{0.8\textwidth}}
  \hline
$K$ & the input ISI length (the channel memory is $K-1$.) \\
$X_n$ & the $n$-th input symbol (nucleotide). \\
$S_n$ & the channel state generated at time $n$, $S_n = X_{n-K+2}^{n} = (X_{n-K+2},\cdots,X_{n})$. \\
$Y_n$ & the $n$-th $K$-mer: $Y_n = X_{n-K+1}^{n} = (S_{n-1},X_{n})$. \\
$D_{n}$ & the binary deletion r.v. at time $n$. We use $D_{n} = 1$ to denote the deletion of the $n$-th $K$-mer. \\
$Z_{m}$ & the $m$-th remaining $K$-mer after deletion. \\
$V_{m}$ & the $m$-th output symbol (observed discretized level). \\
$\dvl{n}$ & the output block that corresponds to the input symbol $X_n$. Note that the length of this block might not be 1 because of deletion (or more general synchronization errors, including insertion, etc.). \\
$\dv{n_1}{n_2}$ & the output block that corresponds to the input segment $X_{n_1}^{n_2}$. When $m=1$, we omit the subscript as usual (the sub/superscripts for these symbols have parentheses around them to denote the effect of random output length). \\
$\X,\St,\V$ & the (finite) alphabets of the input, ISI state, and output symbols, respectively. \\
$\Y^{(n)}$ & the alphabet of all strings of length at most $n$ on $\Y$, i.e., $\Y^{(n)} = \emptyset\cup \Y \cup \Y^{2} \cup \cdots \cup \Y^{n}$. \\
$\V^{*}$ & the alphabet of all strings of finite length on $\V$, i.e., $\V^{*} = \emptyset\cup \V \cup \V^{2} \cup \cdots$. \\
$\|a\|$ & the length of the string $a$. \\
$a\circ b$ & the concatenation of the strings $a$ and $b$. \\
  \hline
\end{tabular}
\end{table}

\section{Main Results}\label{sec:main-results}

As argued in Section \ref{sec:intro}, a performance
  metric for decoding capability of a nanopore sequencer, is the
  number of plausible input sequences given that the output is
  observed. This is quantified by the exponent of the average list
  size, which can be upper bounded as $\log|\mathcal{X}|-R$, where $R$
  is an achievable reliable decoding rate for i.i.d. uniform inputs. If we
  optimize the input distribution, then we obtain an achievable rate
  that can be used to measure the capability of the nanopore sequencer
  as a reader for an information storage system using DNA. Given this
  motivation, we next summarize the technical results presented in the
  paper. Also note that several of the techniques and results could
  also be of independent interest, beyond their application to nanopore
  sequencing.

In this paper we obtain a multi-letter capacity formula, and its lower and upper bounds for the nanopore channel given in Fig.~\ref{fig:nanopore-channel}. We first prove the multi-letter capacity formula for a class of channels with both synchronization errors and inter-symbol interference, which generalizes Dobrushin's Theorem~1 in \cite{Dobrushin-channel-sync-error} (see Section~\ref{sec:multi-letter-capacity} for the detailed definition of the channel) to channels with input memory:

\begin{theorem}\label{thm:multiletter-capacity}
The capacity of the channel \eqref{eq:ch-prob} exists and is given by $C = \lim_{n\to\infty}C_{n}$, where
\begin{equation}\label{eq:C-n}
C_{n} \triangleq \frac{1}{n}\max_{P_{X^{n}}}I\lt(X^{n};\dvs{n}\rt).
\end{equation}
\end{theorem}

Despite its generality, however, the multi-letter capacity formula is not computable. In Section~\ref{sec:ach-rate} we derive computable achievable rates for a cascade of a deletion channel with a DMC, i.e., the channel without the first block in Fig.~\ref{fig:nanopore-channel}, then apply the result to the nanopore channel. The main result for the lower bound is the following theorem (see Section~\ref{sec:ach-rate} for the notations and definitions):

\begin{theorem}\label{thm:ach-rate}
For the cascade of a deletion channel with a DMC, the following is an achievable rate for each irreducible Markov transition matrix $P$ on the input alphabet $\Y$:
\begin{equation}
\uC(P) = - \infg[(1-\theta)\gamma +  \theta E(\gamma)]/\ln2 - \theta H(\Z|\V).
\end{equation}
Therefore, if $\Ps$ is a collection of irreducible Markov transition matrices, then $\sup_{P\in\Ps}\uC(P)$ is also achievable.
\end{theorem}

For the achievable rates of the nanopore channel, we can apply the results above, with a constraint on the choice of Markov transition matrix $P$: since the current $K$-mer takes the form $Y_{n}= X_{n-K+1}^{n}$, the next $K$-mer $Y_{n+1}$ can only be obtained from left-shifting $X_{n-K+1}^{n}$ by one position and attaching a new nucleotide $X_{n+1}$ to the right. Thus for the nanopore channel the Markov transition matrix can only be supported on such legal transitions.

\begin{figure}[!t]
\centering
\includegraphics[width=0.8\textwidth]{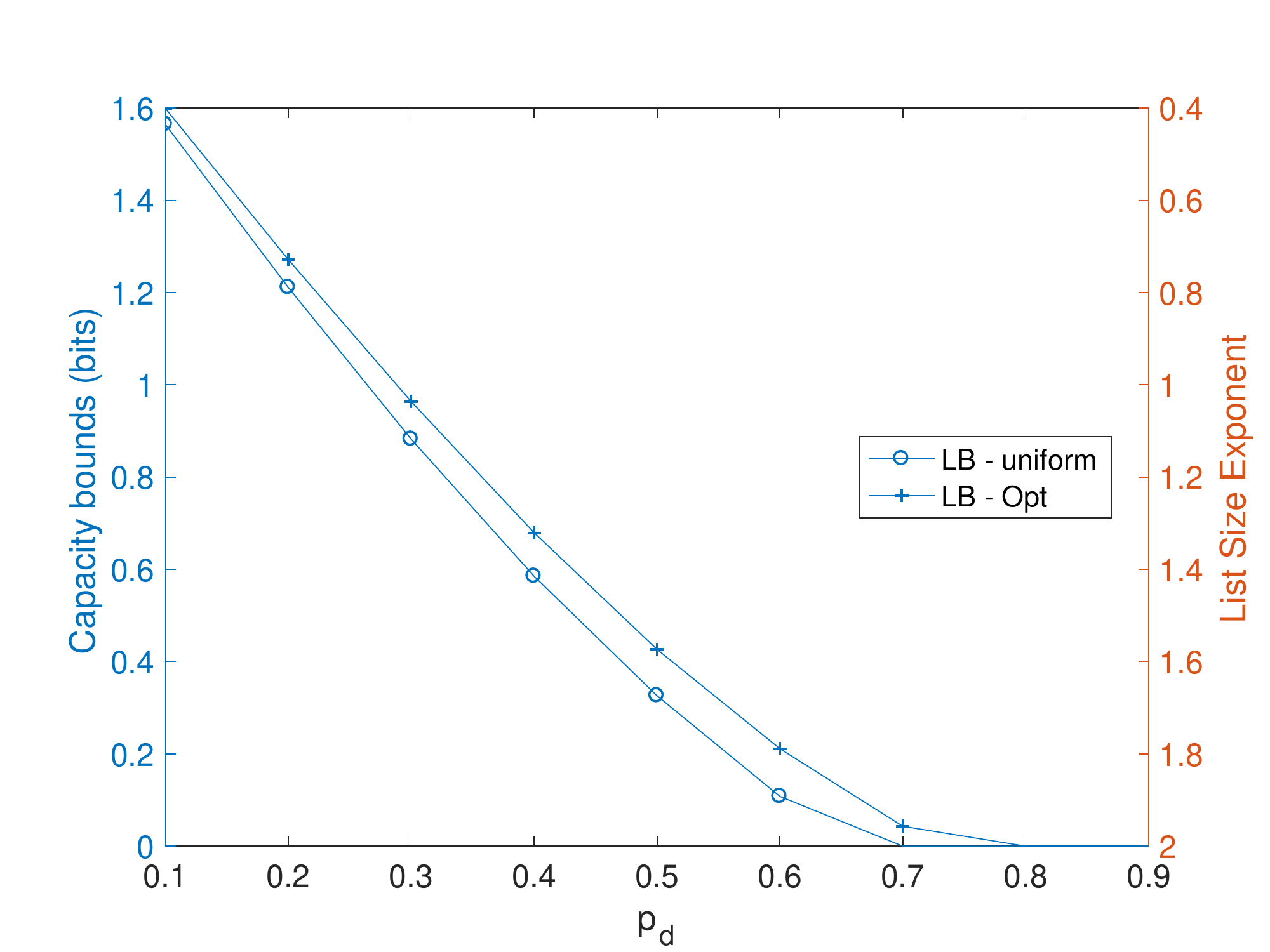}
\caption{Achievable rates using Q-mer map data from experiments \cite{GundlachNature14}.}
\label{fig:forward-pore-linear}
\end{figure}

\begin{figure}[!t]
\centering
\includegraphics[width=0.8\textwidth]{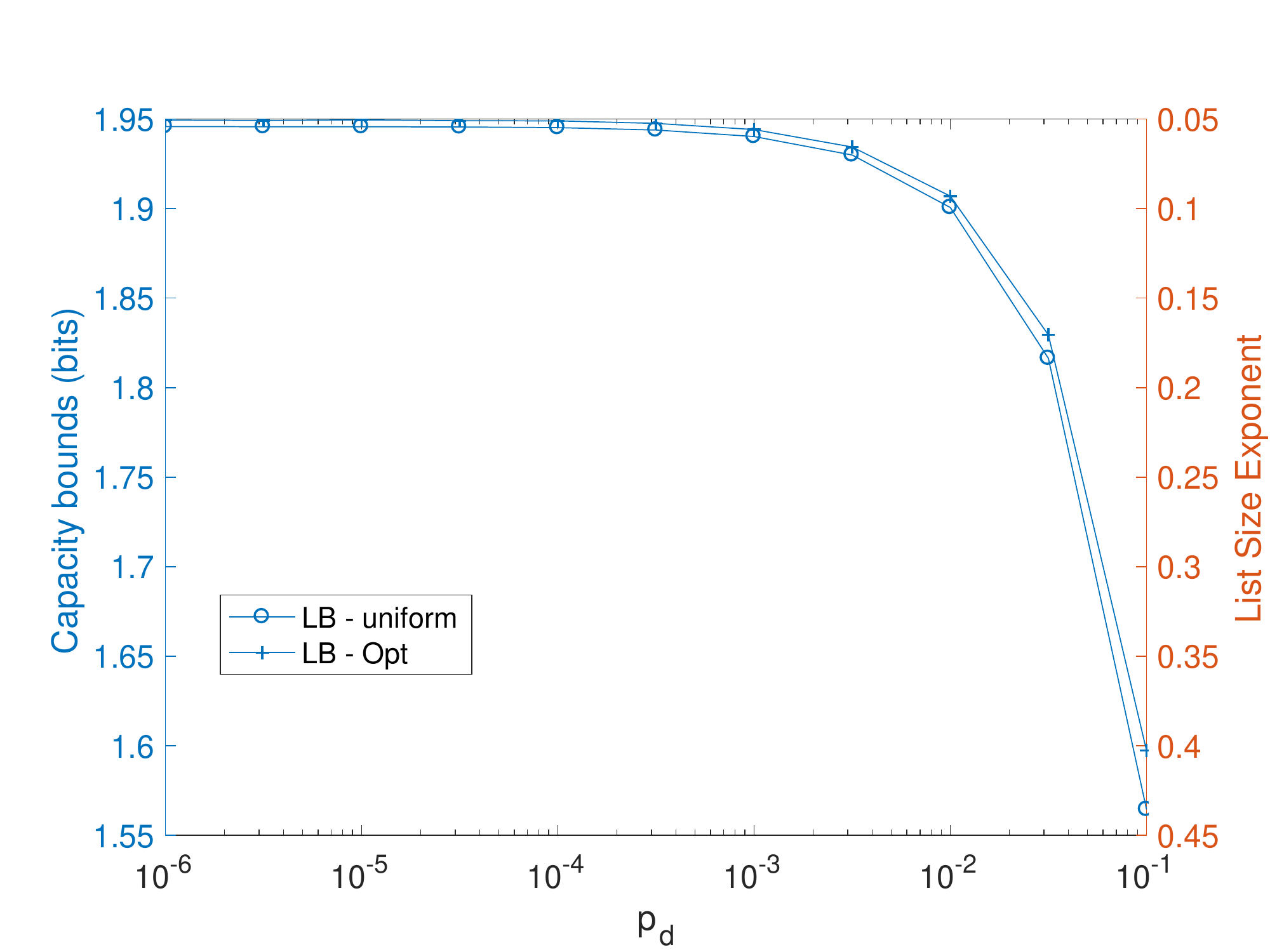}
\caption{Achievable rates using Q-mer map data from experiments \cite{GundlachNature14}.}
\label{fig:forward-pore-log}
\end{figure}

In Fig.~\ref{fig:forward-pore-linear} and~\ref{fig:forward-pore-log}, we plot the achievable rates in Theorem~\ref{thm:ach-rate}, both when the input distribution is i.i.d. uniform and when $P$ is optimized. The channel used is constructed from \emph{experimental data} using the Q-mer map data from experiments in\cite{GundlachNature14} with level discretization. We can see that optimized transition matrices can achieve higher rates than the uniformly generated one. When the deletion probability $p_{d}$ is small, the difference between the two cases become small.
As mentioned earlier, to connect these results
  to the performance metric for nanopore sequencers, we also evaluate
  the \emph{exponent of average list size}\footnote{We give a very
    coarse and imprecise connection between average list-size and
    error probability, by evaluating the cross-over probability for a
    corresponding to the average list size for a symmetric DMC in
    Section \ref{sec:Numerics}, Table \ref{table:equiv-sym-ch-prob}
    and Figure \ref{fig:equiv-sym-ch}.}. On the right vertical axes of
  these figures we mark the corresponding \emph{exponents of average
    list size}, which is defined to be $\log|\X| - R$ for a reliable
  decoding rate $R$. These exponents make the interpretation of the
  achievable rates clearer: the average number of plausible inputs
  given the output is approximately $2^{n(\log|\X| - R)}$, for length-$n$ input sequences.

The computable capacity upper bounds are derived in Section~\ref{sec:upper-bounds}. Our starting point is the following upper bounds provided by Theorem~\ref{thm:exisitence-limit}, when we prove the multi-letter capacity formula in Section~\ref{sec:multi-letter-capacity}: for each $n$, define
\[ \C_{n} = \frac{1}{n}\lt[\max_{s_{0}}\max_{P_{X^{n}}}I\big(X^{n};\dvs{n}\big|s_{0}\big) + \log|\St|\rt], \]
then each $\C_{n}$ is a capacity upper bound for the nanopore channel. These upper bounds are in the form of finite block mutual information and can be computed using the classic Blahut-Arimoto algorithm. However, since the extra term $\log|\St|$ above considerably affects the effectiveness of the upper bound for small to moderate $n$, while for large $n$ the computation is not practical due to the exponential complexity of the first term, we seek a relaxation that has much less complexity. For that purpose we introduce periodical synchronization symbols to the channel output (as in \cite{Fertonani-Duman-Novel-bounds}), and convert the resulting channel across synchronization periods to a finite state channel \cite{Gallager-IT_Reliable} with only ISI memory. In this form the upper bounds can be relaxed (see Theorem~\ref{thm:UBs-relaxation-C-tilde} in Section~\ref{sec:upper-bounds}), which upper bounds mutual information by the relative entropy between the ``worst-case'' block conditional probability and a stationary Markov distribution on the output. The relaxation can then be computed using the Viterbi algorithm with linear complexity, and so we can compute it for very large $n$, which suppresses the term $\log|\St|$ and yields practically more effective upper bounds.

\begin{figure}[!t]
\centering
\includegraphics[width=0.8\textwidth]{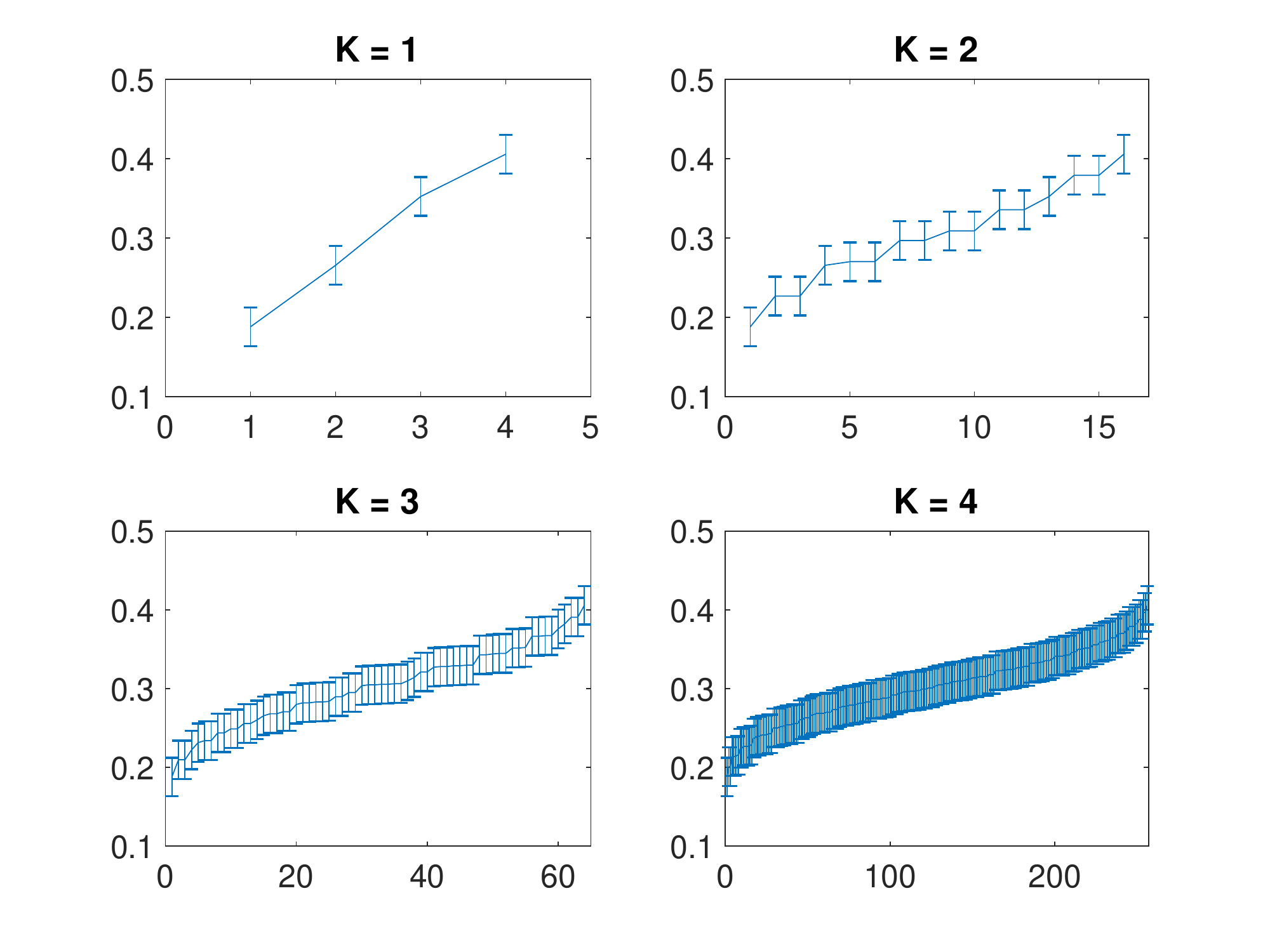}
\caption{$K$-mer maps for different $K$. The axes are similar to Fig.~\ref{fig:q-mer}.}
\label{fig:qmer-synth}
\end{figure}

We further investigate the trade-off of the effects of random $K$-mer map responses and deletion errors when increasing the nanopore length $K$. Since the experiment data only allows for the study of $K=4$, we construct a nanopore model as follows. In this model the mean value of the $K$-mer map is assumed to be a linear combination of the single-nucleotide responses, and the variance is a constant. To be specific, let $f(X)$ be the current value when all the nucleotides in the dwelling $K$-mer are uniformly $X$, we assume the mean value of the response of any $K$-mer $Y_{n} = X_{n-K+1}^{n}$ can be expressed as
\[ g(Y_{n}) = \sum_{i=1}^{K}\alpha_{i}f(X_{n-i+1}), \]
where $\{\alpha_{i}\}$ are all positive and sum to 1. Supposedly the nucleotides dwelling in the center of the nanopore have a larger influence in the $K$-mer response, so $\alpha_{i}$ is larger when $i$ is closer to $(K+1)/2$. Furthermore, assume the fading distribution for $Y_{n}$ takes the form
\[ g(Y_{n}) + U_{n}, \]
where $\{U_{n}\}$ are i.i.d. uniform on some fixed interval $[-A,A]$ (which does not change with $K$).
We generate such a fading model for different ISI length $K$. See Fig.~\ref{fig:qmer-synth} for an example of $K$-mer map means and standard deviations ($|\X| = 4$). The coefficients $\{\alpha_{i}\}$ are sampled from a scaled Cauchy distribution, i.e.
\[ \alpha_{i} = \frac{c}{[i-(K+1)/2]^{2} + \gamma} \]
for some positive constants $c$ and $\gamma$ (which are fit from the experimental data when $K=4$). Furthermore, $A$ is taken to be the average standard deviation in Fig.~\ref{fig:q-mer}, multiplied by a scaling factor $w$.

\begin{figure}[!t]
\centering
\includegraphics[width=0.8\textwidth]{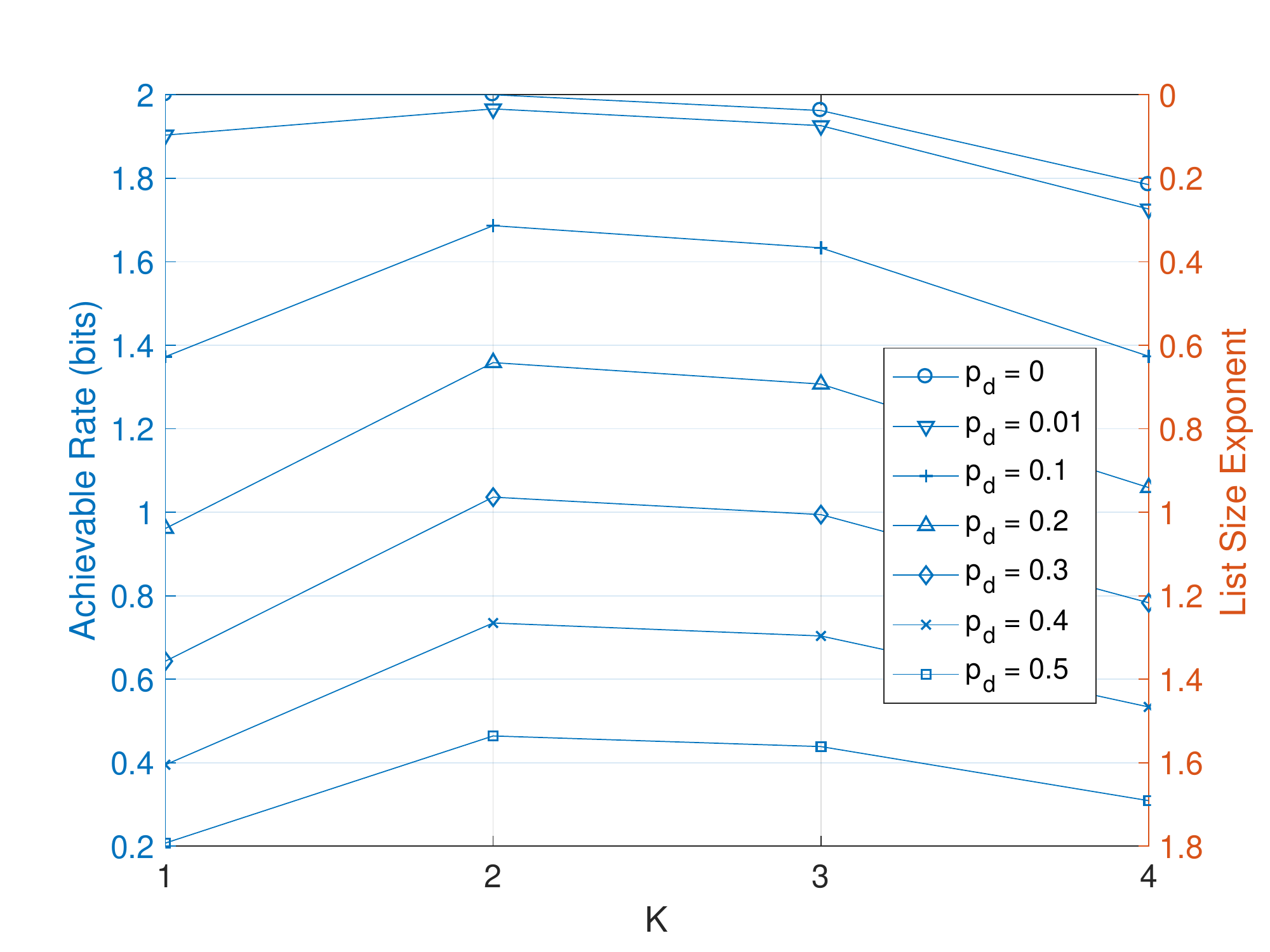}
\caption{Achievable rates for a model with linear ISI. $|\X| = 4$, $w=1$, uniform input.}
\label{fig:ach-rate-synth-nx4-sc-unf}
\end{figure}

\begin{figure}[!t]
\centering
\includegraphics[width=0.8\textwidth]{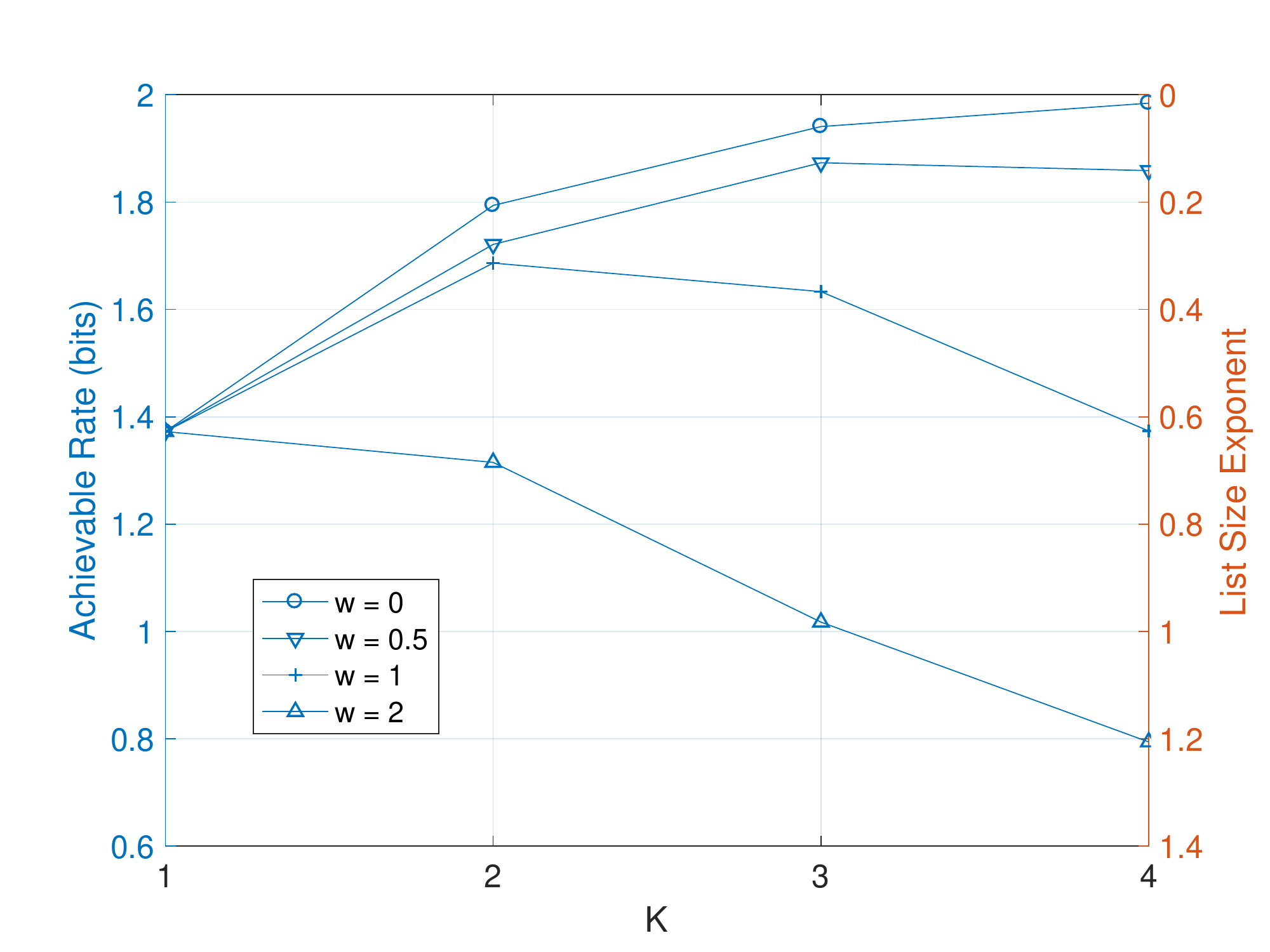}
\caption{Achievable rates for a model with linear ISI. $|\X| = 4$, $p_{d}=0.1$, uniform input.}
\label{fig:ach-rate-synth-nx4-pd-unf}
\end{figure}

We compute the achievable rates for the corresponding nanopore channels, which are generated using the interval-discretization recipe in Section~\ref{sec:model}. In Fig.~\ref{fig:ach-rate-synth-nx4-sc-unf} and~\ref{fig:ach-rate-synth-nx4-pd-unf}, for different deletion probability $p_{d}$ and for different scaling factor $w$, respectively, we plot the achievable rates for increasing ISI length $K$. The input is i.i.d. uniform with alphabet size $|\X| = 4$. Intuitively, the larger $K$ is, the more correlation the input symbols have, hence the transmitted message is better protected against deletions. This is reflected by the increasing achievable rate curve for the no-fading case ($w=0$) in Fig.~\ref{fig:ach-rate-synth-nx4-pd-unf}. On the other hand, the larger $K$ is, the more noisy the DMC is, since there are more intersections for the $K$-mer map intervals (cf. Fig.~\ref{fig:qmer-synth}). This effect is demonstrated by the decreasing achievable rate curve for the no-deletion case ($p_{d}=0$) in Fig.~\ref{fig:ach-rate-synth-nx4-sc-unf}. In general, there is a trade-off between these two effects: as shown in Fig.~\ref{fig:ach-rate-synth-nx4-sc-unf}, when $w=1$, for different non-zero $p_{d}$ the achievable rates first increase and then decrease, all peaking at $K = 2$. When the DMC quality improves/deteriorates, this peak shifts its position, as demonstrated in Fig.~\ref{fig:ach-rate-synth-nx4-pd-unf} when $p_{d}=0.1$. In particualr, larger $w$ corresponds to more intersections for the $K$-mer map intervals (cf. Fig.~\ref{fig:qmer-synth}), which means a poorer DMC. In this case for an increasing $K$ the second effect becomes more important, so the rate peak happens at a smaller value of $K$.

\section{Multi-letter capacity formula for channels with synchronization errors and ISI}
\label{sec:multi-letter-capacity}

We consider a channel model generalized from Dobrushin's channel with synchronization errors \cite{Dobrushin-channel-sync-error}, which includes the nanopore channel in Fig.~\ref{fig:nanopore-channel} as a special case. In Dobrushin's model, due to the synchronization errors, the number of output symbols is not fixed for each input symbol. In addition to that, we also consider channel memory, in the form of inter-symbols interference (ISI). We define the notations of this channel in Table~\ref{table:notations}. The channel has finite input and output alphabets $\X$ and $\V$, repsectively, and an ISI length $K$---thus there is a finite state $S_{n} = X_{n-K+2}^{n}$. Assume the initial state distribution $P_{S_{0}}$ is independent from the input (for the nanopore channel the initial state $s_{0}$ is fixed). At time $n$, when the previous ISI state is $s_{n-1}$, each transmission of an input symbol $x_{n}\in\X$ gives rise to an output string $v_{(n)}\in\V^{*}$ with transition probability $p(v_{(n)}\cn x_{n},s_{n-1})$ that is independent of time $n$. Also, the distribution of $V_{(n)}$ is determined from the system only through $s_{n-1}$ and $x_{n}$, that is, conditioned on $s_{n-1}$ and $x_{n}$, the output string $V_{(n)}$ is independent of all the other input, output, and state symbols. Furthermore, we assume that the average length of the output string is always a nonzero finite number, i.e., the transition distribution satisfies
\begin{equation}\label{eq:avg-length}
b\leq \sum_{v_{(n)}\in\V^{*}}\|v_{(n)}\|\cdot p(v_{(n)}\cn x_{n},s_{n-1}) \leq a,\qquad \forall x_{n}\in\X,\ \forall s_{n-1}\in\St
\end{equation}
for some $b>0$ and $a<\infty$, where $\|\cdot\|$ denotes the length of a string.

According to the channel law above, for the group of output strings $(v_{(1)},\cdots,v_{(n)})$ we have the group transition probability
\begin{equation}\label{eq:ch-transition-law}
p(v_{(1)},\cdots,v_{(n)}\cn x^{n}s_{0}) = \prod_{k=1}^{n}p(v_{(k)}\cn x_{k}s_{k-1}),
\end{equation}
where $s_{k-1} = x_{k-K+1}^{k-1}$. Therefore, the $n$-block channel transition probability can be written as
\begin{equation}\label{eq:ch-prob-s0}
p(\lvs{n}\cn x^{n}s_{0}) = \sum_{v_{(1)}\circ\,\cdots\,\circ\, v_{(n)} = \lvs{n}}p(v_{(1)},\cdots,v_{(n)}\cn x^{n}s_{0}),
\end{equation}
\begin{equation}\label{eq:ch-prob}
p(\lvs{n}\cn x^{n}) = \sum_{s_{0}}p(s_{0})p(\lvs{n}\cn x^{n}s_{0}),
\end{equation}
where $v_{(1)}\circ\cdots\circ v_{(n)}$ denotes the concatenation of the strings $v_{(1)},\cdots,v_{(n)}$ (see Table~\ref{table:notations}).

In this section we prove the multi-letter capacity formula (see Theorem~\ref{thm:multiletter-capacity} in Section~\ref{sec:main-results}) for the channel \eqref{eq:ch-prob}. The proof is obtained by adapting the proof of \cite[Theorem~1]{Dobrushin-channel-sync-error} to our generalized channel model. In the adaptation, we have to deal with several technical difficulties caused by the ISI channel (which is a finite state channel). The original proof in \cite{Dobrushin-channel-sync-error} consists of two parts: i) in Section~3 of \cite{Dobrushin-channel-sync-error} the block-capacity (maximized block mutual information, corresponding to $nC_{n}$ in our paper) is shown to be subadditive, and so the limit of the average capacity ($C_{n}$ in our paper) exists by Fekete's lemma; ii) in Section~4 of \cite{Dobrushin-channel-sync-error} the information stability for the channel is proved by constructing a sequence of input random variables, each of which is (mostly) composed of i.i.d. input blocks. Such input blocks induce i.i.d. channel blocks whose mutual information achieves the corresponding block-capacity. In our new model, the subadditivity proof does not work for the block-capacity any more.
To prove the existence of the limit of $C_{n}$, we demonstrate that an auxiliary term $\C_{n}$ (defined in \eqref{eq:UB-C-bar} below) satisfies the subadditive property, and then we show that $\C_{n}$ approximates $C_{n}$ asymptotically.
Furthermore, due to the input memory we cannot obtain i.i.d. channel blocks using the original construction; instead, we need to append a ``guard interval'' (consisting of deterministic symbols) at the end of each input block to eliminate the dependence across the blocks. Therefore we need a new proof to handle this situation, and we combine techniques from the classical finite state channels \cite{Gallager-IT_Reliable} and from Dobrushin \cite{Dobrushin-channel-sync-error}, along with a few of our own ingredients.

We first give an outline of the proof, and then show the technical details in the subsequent two subsections. As stated in \cite[Section~4]{Dobrushin-channel-sync-error}, to prove Theorem~\ref{thm:multiletter-capacity} it suffices to show the information stability\footnote{In \cite{Zeng-information-stability}, the information stability is also studied for a channel with ISI and timing errors. However, the ISI is linear (which comes from an FIR filter), and the timing error model is different from the synchronization errors considered here: no output symbols get deleted or inserted, but the linear coefficients (tap weights for the filter) change as a result of the timing errors. Hence we cannot directly apply those techniques and results to our problem.} of the sequences of transition probabilities \eqref{eq:ch-prob} for each block length $n$. Define the information density of the pair of jointly distributed discrete random variables $X$ and $Y$ by
\[ i_{XY} = \log\frac{p(X,Y)}{p(X)p(Y)}. \]
Note that for an $n$-block i.i.d. channel, the information capacity is $nC_{n}$ (defined in \eqref{eq:C-n}), which is associated with the length-$n$ block transition probability. If the limit of $C_{n}$ exists and is denoted by $C$, then information stability means the existence of a sequence of random variables $\bxi_{n}$ on the alphabet $\X^{n}$, such that if $\bta_{n}$ is the output of the length-$n$ block channel \eqref{eq:ch-prob} when the channel input is $\bxi_{n}$, then for any $\delta>0$
\begin{equation}\label{eq:information-stability}
\Pr\lt(|i_{\bxi_{n},\bta_{n}}-nC|>\delta n\rt) \to 0\qquad \text{as}\quad n\to\infty.
\end{equation}
Thus to prove Theorem~\ref{thm:multiletter-capacity} we need to 1) show the limit of $C_{n}$ exists, and 2) construct the random variables $\{\bxi_{n}\}$ that satisfy \eqref{eq:information-stability}. The former is achieved in Section~\ref{subsec:existence-of-limit}, while the latter is finished in Section~\ref{subsec:information-statbility}.

\subsection{Existence of limit }
\label{subsec:existence-of-limit}

As mentioned earlier, the original subadditivity proof in \cite[Section~3]{Dobrushin-channel-sync-error} does not work for the block-capacity $nC_{n}$ due to the ISI. To make the proof work, we introduce an auxiliary sequence $\{\C_{n}\}$ in the spirit of Gallager's finite state channels \cite{Gallager-IT_Reliable}, which satisfies subadditivity, and so its limit exists. Then we prove that $C_{n}$ defined in \eqref{eq:C-n} and $\C_{n}$ actually converge to the same limit $C$. Incidentally, we also show that $\C_{n}$ is an upper bound of $C$ for any finite $n$. The detailed proofs of the following results are deferred to Section~\ref{subsec:proof-existence-of-limit} in the Appendices.

\begin{lemma}\label{lem:subadditivity}
For each $n$ define
\begin{equation}\label{eq:UB-C-bar}
\C_{n} = \frac{1}{n}\lt[\max_{s_{0}}\max_{P_{X^{n}}}I\big(X^{n};\dvs{n}\big|s_{0}\big) + \log|\St|\rt].
\end{equation}
Let $N$ be arbitrary and let $m,n$ be positive integers satisfying $m+n=N$. Then we have
\begin{equation}\label{eq:subadditivity}
N\C_{N} \leq  n\C_{n} + m\C_{m},
\end{equation}
and so by using Fekete's lemma (Lemma~\ref{lem:Fekete} in Appendix), we can see that $\lim_{n \to \infty} \C_{n}$ exists and is equal to $\inf_{n} \C_{n}$.
\end{lemma}

\begin{theorem}\label{thm:exisitence-limit}
The limit of $C_{n}$ exists when $n\to\infty$ and is equal to $C \triangleq \inf_{n}\C_{n} = \lim_{n\to\infty}\C_{n}$. Hence for any finite $n$, $\C_{n}$ is an upper bound of $C$.
\end{theorem}

\subsection{Information stability}
\label{subsec:information-statbility}

For this part of the proof we develop on the ideas in \cite[Section~4]{Dobrushin-channel-sync-error} to construct a random sequence $\{\bxi_{n}\}$ that satisfies \eqref{eq:information-stability}, by creating a new construction suitable for the ISI channel. The construction uses i.i.d. input blocks of the same length $k$ that are drawn from a particular distribution, so that the information density can be approximated by the sum of information densities for (mostly) i.i.d. channel $k$-blocks, for which law of large numbers applies. As mentioned above, to guarantee independence among the blocks we need to introduce a guard interval. Moreover, for the information density approximation to work, two other random variables, $\cta_{n,k}$ and $\eta_{n,k}'$, are also constructed to approximate the channel output. With these approximations \eqref{eq:information-stability} is shown to hold using the tools of union bound and triangle inequality. These ideas are elaborated below.

Fix an integer $k$. For each integer $n$, we split $n$ symbols into blocks of length $k$ with a trailing remainder, i.e., we write
\begin{equation}\label{eq:expansion-n}
n = gk+l,
\end{equation}
where $g$ is an integer and $0\leq l<k$. Without loss of generality assume $k\geq K$, where $K$ is the ISI length, and write $k' = k - K + 1$. Note that $k-k'=K-1$, which is the length of channel memory. To eliminate the dependence across the channel $k$-blocks, we form an input $k$-block by constructing a random variable of length $k'$, and append a deterministic guard interval of length $K-1$ to it.

Assume $s_{0} = s^{*}\in\St = \X^{K-1}$ and $P_{X^{k'}} = P^{*}$ achieve the maximum of $\C_{k'}$ defined in \eqref{eq:UB-C-bar}. Define a random vector $\tX^{k}$ on $\X^{k}$ as follows: $\tX_{1}^{k'}$ has the distribution $P^{*}$ and $\tX_{k'+1}^{k} = s^{*}$, which has a deterministic value. The full distribution of $\tX^{k}$ is denoted by $\tP$.\footnote{Note that $s^{*}$, $P^{*}$ and $\tP$ all depend on $k$, that is, they may change their values when $k$ changes.} Let $\tV^{(k)}$ be the corresponding channel output of \eqref{eq:ch-prob} and define
\begin{equation}\label{eq:I-k}
I_{k} = \frac{1}{k}I\big(\tX^{k};\tV^{(k)}\big| S_{0} = s^{*}\big),
\end{equation}
then the following lemma shows that $I_{k}$ converges to the same limit $C$ as in Theorem~\ref{thm:exisitence-limit}. The proof is deferred to Section~\ref{subsec:proof-information-statbility} in the Appendices.

\begin{lemma}\label{lem:I-k-limit}
$\lim_{k\to\infty}I_{k} = C$.
\end{lemma}

\begin{figure}[!t]
\centering
\includegraphics[width=0.9\textwidth]{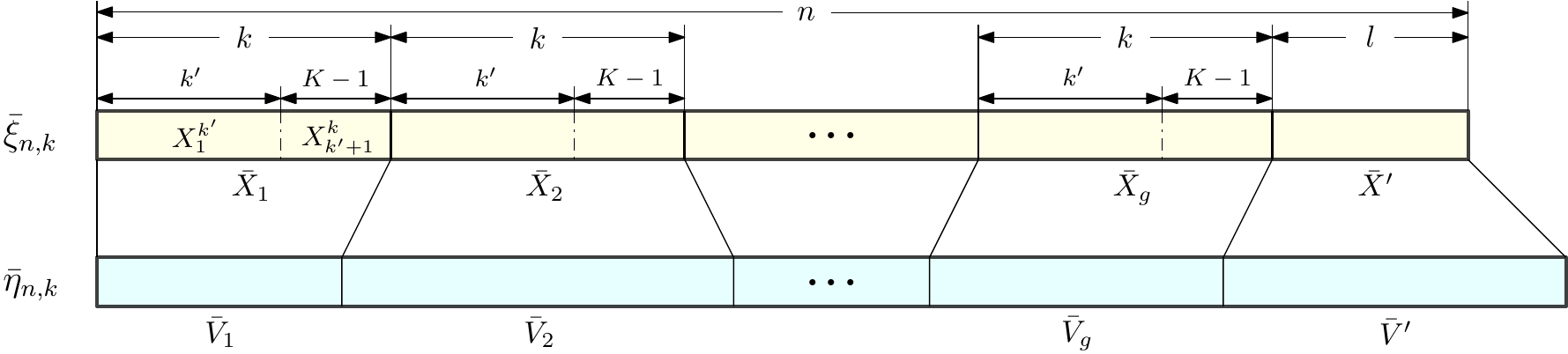}
\caption{The structures of $\bxi_{n,k}$ and $\bta_{n,k}$.}
\label{fig:input-random-vector}
\end{figure}

With these $k$-blocks we build a family of input random variables $\{\bxi_{n,k}\}_{n,k}$, within which $\{\bxi_{n}\}$ will be selected. See Fig.~\ref{fig:input-random-vector}. Let $x_{0}$ be an arbitrary fixed element of $\X$. Define
\begin{equation}\label{eq:bxi}
\bxi_{n,k} = X^{n} = \bX_{1}\bX_{2}\cdots\bX_{g}\bX',
\end{equation}
where we denote
$\bX_{i} = X_{(i-1)k+1}^{ik}$ for $i = 1,\ldots,g$, and
$\bX' = X_{gk+1}^{n}$ (cf. \eqref{eq:expansion-n}).
In this vector $\bxi_{n,k}$, all $\bX_{i}$, $i = 1,\ldots,g$, are i.i.d. $k$-dimensional random vectors in $\X^{k}$ having the distribution $\tP$. Furthermore, $\bX'$ is the fixed vector $\bx_{0} = (x_{0},x_{0},\ldots,x_{0})\in\X^{l}$. Let $P_{\xi}$ denote the distribution of $\bxi_{n,k}$ and let $\bta_{n,k}$ denote the corresponding channel output from \eqref{eq:ch-prob}. Let
$\bV_{i} = \dv{(i-1)k+1}{ik}$
denote the output symbols corresponding to the block $\bX_{i}$ for $i = 1,\ldots,g$, and let
$\bV' = \dv{gk+1}{n}$
be the output corresponding to $\bX'$. Then we can write
\[ \bta_{n,k} = \bV_{1}\circ\bV_{2}\circ\cdots\circ\bV_{g}\circ\bV', \]
which is the concatenation of the random vectors on the right side (cf. Table~\ref{table:notations}). Furthermore, define the tuple
\begin{equation}\label{eq:cta}
\cta_{n,k} = \big(\bV_{1},\bV_{2},\cdots,\bV_{g},\bV'\big).
\end{equation}

The structures of the input and output vectors $\bxi_{n,k}$ and $\bta_{n,k}$ are depicted in Fig.~\ref{fig:input-random-vector}. With such a construction we can show the following results, whose proofs are deferred to Section~\ref{subsec:proof-information-statbility} in the Appendices. We first show that the input and output are block-independent under input distribution $P_{\xi}$.
\begin{lemma}\label{lem:XV-indep}
When the channel input distribution is $P_{\xi}$, the pairs of random variables $(\bX_{1},\bV_{1})$, $(\bX_{2},\bV_{2})$, \ldots, $(\bX_{g}\bV_{g})$, and $(\bX',\bV')$ are independent. Moreover, $(\bX_{2},\bV_{2})$, \ldots, $(\bX_{g}\bV_{g})$ are i.i.d., with distribution
\[ \Pr(\bX_{i} = \bx_{i},\bV_{i} = \bv_{i}) = \tP(\bx_{i})p(\bv_{i}\cn \bx_{i}s^{*}). \]
\end{lemma}

With such a block-independence and a (modified) weak law of large numbers (Lemma~\ref{lem:WLLN} in Appendix~\ref{subsec:technical-lemmas}), we then prove two convergence results.

\begin{lemma}\label{lem:long-length-prob-vanishes}
When the channel input distribution is $P_{\xi}$, we have
$\Pr\big(\|\bta_{n,k}\| > 2an\big) \to 0$
as $n\to \infty$.
\end{lemma}

\begin{lemma}\label{lem:avg-info-density-converge}
For a fixed $k$, the average information density $i_{\bxi_{n,k},\,\cta_{n,k}}/g$ converges to $kI_{k}$ in probability when $g\to\infty$. Thus
\begin{equation} \label{eq:delta-ineq-3}
\Pr\lt(|i_{\bxi_{n,k},\,\cta_{n,k}}-kgI_{k}|>\delta n/4\rt) \to 0
\end{equation}
as $n\to\infty$.
\end{lemma}

The following lemma summarizes two results from \cite{Dobrushin-channel-sync-error}.
\begin{lemma}\label{lem:Dobrushin-aux}
Let $X,Y$ be jointly distributed random variables and let $\varphi$ be a mapping from $\Y$ to $\Z$. Define $S_{\varphi}(z) = |\varphi^{-1}(z)|$ for $z\in\Z$, i.e., the number of solutions of the equation $\varphi(y) = z$ in $\Y$. Then we have
\begin{equation}\label{eq:Dobrushin-expectation}
\Ex{|i_{XY} - i_{X,\,\varphi(Y)}|} \leq \max_{z\in\Z}\log S_{\varphi}(z),
\end{equation}
which is (4.3) in \cite{Dobrushin-channel-sync-error}, and
\begin{equation}\label{eq:Dobrushin-probability}
\Pr\lt(i_{XY} \neq i_{X,\,\varphi(Y)}\rt) \leq \Pr(S_{\varphi}(\varphi(Y))\neq 1),
\end{equation}
which is (4.7) in \cite{Dobrushin-channel-sync-error}. The first equation can be interpreted as the additional information needed to reconcile between $Y$ and $\varphi(Y)$.
\end{lemma}

Next we define another random variable $\eta_{n,k}'$ (in \eqref{eq:eta-prime} below) and bound the differences between $\bta_{n,k}$, $\cta_{n,k}$ and $\eta_{n,k}'$ through the functions relating them. First, the output $\bta_{n,k}$ of the channel \eqref{eq:ch-prob} (when the input is $\bxi_{n,k}$) can be written as
\[ \bta_{n,k} = \varphi(\cta_{n,k}), \]
where $\cta_{n,k}$ is defined in \eqref{eq:cta} and the function $\varphi:(\V^{*})^{g+1}\to\V^{*}$ concatenates its arguments:
\[ \varphi(\bv_{1},\bv_{2},\cdots,\bv_{g},\bv') = \bv_{1}\circ\bv_{2}\circ\cdots\circ\bv_{g}\circ\bv', \]
for all $\bv_{1},\bv_{2},\cdots,\bv_{g},\bv' \in \V^{*}$. The size of the inverse image $S_{\varphi}(\bv) = |\varphi^{-1}(\bv)|$ for $\bv\in\V^{*}$ is
\[ \binom{\|\bv\|+g}{g}, \]
which is the number of ways to partition $\|\bv\|$ symbols into $g+1$ blocks, including the empty ones. Next define
\begin{equation}\label{eq:eta-prime}
\eta_{n,k}' = \begin{cases}
\cta_{n,k} & \text{if } \|\bta_{n,k}\| \leq 2an, \\
\bta_{n,k} & \text{if } \|\bta_{n,k}\| > 2an.
\end{cases}
\end{equation}
Then $\eta_{n,k}' = \rho(\cta_{n,k})$ for some function $\rho$. By \eqref{eq:Dobrushin-probability} in Lemma~\ref{lem:Dobrushin-aux} and Lemma~\ref{lem:long-length-prob-vanishes} we have
\begin{align}
\Pr\lt(i_{\bxi_{n,k},\,\eta_{n,k}'} \neq i_{\bxi_{n,k},\,\cta_{n,k}}\rt) 
&= \Pr\lt(i_{\bxi_{n,k},\,\rho(\cta_{n,k})} \neq i_{\bxi_{n,k},\,\cta_{n,k}}\rt) \nonumber\\
&\leq \Pr(S_{\rho}(\rho(\cta_{n,k}))\neq 1)\nonumber\\
&\leq \Pr\big(\|\bta_{n,k}\| > 2an\big) \to 0 \label{eq:delta-ineq-2}
\end{align}
as $n\to\infty$. Moreover, $\bta_{n,k}$ can also be viewed as a function of $\eta_{n,k}'$, i.e., $\bta_{n,k} = \tau(\eta_{n,k}')$ for some function $\tau$. Note that $\tau(\eta_{n,k}') \neq \eta_{n,k}'$ only when $\|\bta_{n,k}\| \leq 2an$, in which case $\eta_{n,k}' = \cta_{n,k}$ and 
\[ \bta_{n,k} = \tau(\eta_{n,k}') = \varphi(\cta_{n,k}). \]
Then by \eqref{eq:Dobrushin-expectation} in Lemma~\ref{lem:Dobrushin-aux} we have
\begin{align}
\Ex{\lt|i_{\bxi_{n,k},\,\eta_{n,k}'} - i_{\bxi_{n,k},\,\bta_{n,k}}\rt|}
&= \Ex{\lt|i_{\bxi_{n,k},\,\eta_{n,k}'} - i_{\bxi_{n,k},\,\tau(\eta_{n,k}')}\rt|} \nonumber\\
&\leq \max_{\bv\in\V^{*}}\log S_{\tau}(\bv) \nonumber\\
&= \max_{\bv\in\V^{*}}\log S_{\varphi}(\bv) \nonumber\\
&= \max_{\|\bv\|\leq2an}\log\binom{\|\bv\|+g}{g} \nonumber\\
&= \log\binom{2an+g}{g} \label{eq:log-combin} \\
&\leq nr(k) \label{eq:stirling-approx}
\end{align}
with some function $r(k)$ of $k$ when $n$ is large enough. \eqref{eq:stirling-approx} is obtained through Stirling's formula, and $r(k) \to 0$ as $k\to\infty$ (see \cite[Section~4]{Dobrushin-channel-sync-error}). \eqref{eq:log-combin} can be viewed as an upper bound of the amount of side information needed to go from $\bta_{n,k}$ to $\eta_{n,k}'$. By Markov's inequality, when $n$ is large enough
\begin{align}
\Pr\lt(\lt|i_{\bxi_{n,k},\,\eta_{n,k}'} - i_{\bxi_{n,k},\,\bta_{n,k}}\rt|> \delta n/4 \rt) 
&\leq \frac{4}{\delta n}\Ex{\lt|i_{\bxi_{n,k},\,\eta_{n,k}'} - i_{\bxi_{n,k},\,\bta_{n,k}}\rt|} \nonumber\\
&\leq \frac{4}{\delta n}\cdot nr(k) \nonumber\\
& = \frac{4r(k)}{\delta}. \label{eq:delta-ineq-1}
\end{align}

Finally we have all the ingredients to tackle \eqref{eq:information-stability}. Note that by the triangle inequality, we have
\begin{IEEEeqnarray*}{r/C/l}
\lt|i_{\bxi_{n,k},\,\bta_{n,k}} - nC\rt|
&\leq& \lt|i_{\bxi_{n,k},\,\bta_{n,k}} - i_{\bxi_{n,k},\,\eta_{n,k}'}\rt| + \lt|i_{\bxi_{n,k},\,\eta_{n,k}'} - i_{\bxi_{n,k},\,\cta_{n,k}}\rt| \\
&& +\: \lt|i_{\bxi_{n,k},\,\cta_{n,k}} - kgI_{k}\rt| + \lt|kgI_{k} -nC\rt| \\
&=& \Delta_{1} + \Delta_{2} + \Delta_{3} + \Delta_{4},
\end{IEEEeqnarray*}
where $\Delta_{1}$, $\Delta_{2}$, $\Delta_{3}$, and $\Delta_{4}$ are respectively defined as the first to fourth terms on the right side of the inequality. By union bound,
\begin{align}
\Pr\lt(\lt|i_{\bxi_{n,k},\,\bta_{n,k}}-nC\rt|>\delta n\rt) 
& \leq \Pr\lt(\Delta_{1} + \Delta_{2} + \Delta_{3} + \Delta_{4} > \delta n\rt) \nonumber\\
& \leq \Pr\lt(\cup_{j=1}^{4}\{\Delta_{j} > \delta n/4\}\rt) \nonumber\\
& \leq \sum_{j=1}^{4}\Pr\lt(\Delta_{j} > \delta n/4\rt). \label{eq:prob-triangle-ineq}
\end{align}
Now by \eqref{eq:delta-ineq-1},
\begin{equation}\label{eq:delta-bound-1}
\Pr\lt(\Delta_{1} > \delta n/4\rt) = \Pr\lt(\lt|i_{\bxi_{n,k},\,\bta_{n,k}} - i_{\bxi_{n,k},\,\eta_{n,k}'}\rt|> \delta n/4 \rt)
\leq \frac{4r(k)}{\delta}
\end{equation}
when $n$ is large enough; by \eqref{eq:delta-ineq-2},
\begin{equation}\label{eq:delta-bound-2}
\Pr\lt(\Delta_{2} > \delta n/4\rt) \leq \Pr\lt(i_{\bxi_{n,k},\,\eta_{n,k}'} \neq i_{\bxi_{n,k},\,\cta_{n,k}}\rt) \to 0 
\end{equation}
as $n\to\infty$; by \eqref{eq:delta-ineq-3},
\begin{equation}\label{eq:delta-bound-3}
\Pr\lt(\Delta_{3} > \delta n/4\rt) = \Pr\lt(|i_{\bxi_{n,k},\,\cta_{n,k}}-kgI_{k}|>\delta n/4\rt) \to 0
\end{equation}
as $n\to\infty$; and at last, by Lemma~\ref{lem:I-k-limit}, if $k$ is so large (but fixed) that $|I_{k} - C|<\delta/8$, then
\begin{align*}
|kgI_{k} -nC| &\leq kg|I_{k} - C| + lC \\
&\leq n|I_{k} - C| + kC \\
&< (\delta /8 + kC/n)n \\
&< (\delta /8 + \delta /8)n \\
&= \delta n/4
\end{align*}
when $n$ is large enough, which means
\begin{equation}\label{eq:delta-bound-4}
\Pr\lt(\Delta_{4} > \delta n/4\rt) = 0
\end{equation}
in this case.

Now we combine the results from \eqref{eq:delta-bound-1}--\eqref{eq:delta-bound-4}. Since $r(k) \to 0$ as $k\to\infty$, for any $\epsilon>0$, we can find an integer $k_{\epsilon}$ such that 
\[ \frac{4r(k_{\epsilon})}{\delta} \leq \frac{\epsilon}{3} \]
and $|I_{k_{\epsilon}} - C|<\delta/8$. With this $k_{\epsilon}$ we can further find $N_{\epsilon}$, such that if $k = k_{\epsilon}$ and $n\geq N_{\epsilon}$, we have
\begin{align*}
\Pr\lt(\Delta_{1} > \delta n/4\rt) &\leq \frac{4r(k_{\epsilon})}{\delta} \leq \frac{\epsilon}{3}, \\
\Pr\lt(\Delta_{2} > \delta n/4\rt) &\leq \frac{\epsilon}{3}, \\
\Pr\lt(\Delta_{3} > \delta n/4\rt) &\leq \frac{\epsilon}{3}, \\
\Pr\lt(\Delta_{4} > \delta n/4\rt) &= 0.
\end{align*}
Therefore, by \eqref{eq:prob-triangle-ineq}
\[ \Pr\lt(\lt|i_{\bxi_{n,k_{\epsilon}},\,\bta_{n,k_{\epsilon}}}-nC\rt|>\delta n\rt) \leq \sum_{j=1}^{4}\Pr\lt(\Delta_{j} > \delta n/4\rt) \leq \epsilon \] 
when $n\geq N_{\epsilon}$.

With such a result we can construct the required $\bxi_{n}$ from the family $\{\bxi_{n,k}\}_{n,k}$. Let $\epsilon_{0} = 1$ and let $\{\epsilon_{m}\}_{m\geq1}$ be a decreasing sequence with $\lim_{m\to\infty}\epsilon_{m} = 0$. Then by the results above, for each $m\geq1$, we can find $k_{m}'$ and $N_{m}$, defined as
\[ k_{m}' = k_{\epsilon_{m}}, \]
\[ N_{m} = N_{\epsilon_{m}}, \]
such that when $n\geq N_{m}$,
\[ \Pr\lt(\lt|i_{\bxi_{n,k_{m}'},\,\bta_{n,k_{m}'}}-nC\rt|>\delta n\rt) \leq \epsilon_{m}. \] 
Without loss of generality, we can choose $N_{m+1}\geq N_{m}+1$. Also for notational convenience we define $k_{0}' = 1$ and $N_{0}=1$. Furthermore, for each $n>0$ define
\[ m(n) = \sup\{m\geq0: n\geq N_{m}\}, \]
\[ k(n) = k_{m(n)}'. \]
Note that $m(n)\to\infty$ as $n\to\infty$. For each $n$, we have $n\geq N_{m(n)}$ and so
\[ \Pr\lt(\lt|i_{\bxi_{n,k(n)},\,\bta_{n,k(n)}}-nC\rt|>\delta n\rt) 
= \Pr\lt(\lt|i_{\bxi_{n,k_{m(n)}'},\,\bta_{n,k_{m(n)}'}}-nC\rt|>\delta n\rt) \leq \epsilon_{m(n)}. \] 
(Note that this inequality holds trivially when $m(n)=0$.) Now define
\[ \bxi_{n} = \bxi_{n,k(n)}. \]
By definition $\bta_{n} = \bta_{n,k(n)}$, and so
\[ \Pr\lt(|i_{\bxi_{n},\bta_{n}}-nC|>\delta n\rt) 
= \Pr\lt(\lt|i_{\bxi_{n,k(n)},\,\bta_{n,k(n)}}-nC\rt|>\delta n\rt) \leq \epsilon_{m(n)} \to 0 \]
as $n\to\infty$, since $m(n)\to\infty$ as well. Hence \eqref{eq:information-stability} is true for such choices of $\bxi_{n}$ and information stability holds for the channel \eqref{eq:ch-prob}.

\section{Achievable Rates}
\label{sec:ach-rate}

In this section we derive computable achievable rates for a cascade of a deletion channel with a DMC, which can be applied to the nanopore channel. Our approach is based on and generalizes the lower bound ideas for (noisy) deletion channels in \cite{Diggavi-Grossglauser-finite-buffer-channel}. The codebook is generated randomly using a stationary ergodic Markov process.\footnote{In the nanopore channel, the ISI channel naturally provides such a Markov structure.} The decoder outputs an estimated codeword if, roughly speaking, the output length is typical and it is the unique codeword which is jointly typical with the output. With such a coding system we derive an achievable rate by analyzing the average error probability, which utilizes techniques for watched Markov chains in \cite{Diggavi-Grossglauser-finite-buffer-channel}, the general AEP \cite{Cover-Thomas-ElemIT}, and ergodic theory \cite{Walters-Ergodic-Theory}.

Consider an i.i.d. deletion channel with input alphabet $\Y$, connected to a DMC with output alphabet $\V$. We assume all alphabets are finite and the deletion probability $p_{d}$ satisfies $0\leq p_{d} < 1$. Let a stationary Markov process $\{Y_{i}\}_{i\geq1}$ on $\Y$ with an \emph{irreducible} transition matrix $P$ be input to the deletion channel, and let $\{Z_{j}\}_{j\geq1}$ denote the corresponding output of the deletion channel. Let $\pi$ be the steady state distribution of $P$. Then $\{Z_{j}\}$ is a watched Markov chain\footnote{A watched Markov chain, a.k.a. a censored Markov chain, is formed by a regular Markov chain $\{U_{i}\}$ and a subset $\U'$ of its state space $\U$. If we only observe the outcome of the Markov chain when $U_{i}\in\U'$, then the result is called a watched Markov chain with watched set $\U'$. In our case, $U_{i} = (Y_{i},D_{i})$ and $\U' = \Y\times\{0\}$, i.e., we only observe $U_{i}$ when there is no deletion.}\cite{Diggavi-Grossglauser-finite-buffer-channel}, which is also a stationary Markov process \cite[Lemma~6-6]{Kemeny-Denumerable-Markov-chains}. Define
$\theta = 1-p_{d} > 0$,
then the watched Markov chain $\{Z_{j}\}$ has a transition matrix
\[ \bP = \frac{\theta}{1-\theta}\sum_{k=1}^{\infty}(1-\theta)^{k}P^{k} = \theta P[I - (1-\theta)P]^{-1}, \]
which also has $\pi$ as the steady state distribution. Note that as $P$ is irreducible, $\bP$ is primitive, i.e., all of its entries are positive. Hence by \cite[Thm~1.19]{Walters-Ergodic-Theory}, $\{Z_{j}\}$ is a stationary ergodic process. The corresponding output process $\{V_{j}\}_{j\geq1}$ of the DMC is a hidden Markov process, and the joint process $\{Z_{j},V_{j}\}_{j\geq1}$ is also stationary and ergodic (see \cite{Gray-EIT}). As in \cite{Cover-Thomas-ElemIT} we use $H(\V)$ to denote the entropy rate of the stationary ergodic process $\{V_{j}\}$ and define the conditional entropy rate $H(\Z|\V) = H(\Z,\V) - H(\V)$.

Let $a,b\in\Y$. Define $\tP_{b}$ to be the submatrix of $P$ obtained from removing the $b$-th row and $b$-th column, $\tq_{b}$ to be the column vector obtained from removing the $b$-th entry of the $b$-th column of $P$, and $\tp_{ab}$ to be the row vector obtained from removing the $b$-th entry of the $a$-th row of $P$. Let
\[ E_{ab}(\gamma) = \ln\lt[P_{ab} + \tp_{ab}(e^{\gamma}I- \tP_{b})^{-1}\tq_{b}\rt] \]
for $\gamma>0$ and define
\[ E(\gamma) = \sum_{a,b\in\Y}\pi_{a}\bP_{ab}E_{ab}(\gamma). \]
With all these definitions we have the achievable rate theorem (Theorem~\ref{thm:ach-rate} in Section~\ref{sec:main-results}), whose proof is presented below.

\begin{IEEEproof}[Proof of Theorem~\ref{thm:ach-rate}]
First we need to give a few more definitions. If we transmit $n$ symbols $Y_{1}^{n}$ over the deletion channel, then the number of received symbols
\[ T = T(n) = \sum_{i=0}^{n}\1{D_{i} = 0} \]
is a random variable. For $\epsilon > 0$ and $m>0$, we define typical sequences for the process $\{Z_{j}\}$ as usual \cite{Cover-Thomas-ElemIT} and denote the corresponding typical set by $\Ae = \Ae(Z)$. Furthermore, for a sequence $z^{m}\in\Ae$, if $\forall a,b\in\Y$ we have
\[ \lt| \frac{1}{m}\sum_{l=2}^{m}\1{(z_{l-1},z_{l}) = (a,b)}\ -\ \pi_{a}\bP_{ab} \rt| \leq \epsilon, \]
then we say $z^{m}$ is \emph{doubly-strongly typical}, and the doubly-strongly typical set is denoted by $\sAe$. Almost all (with probability approaching 1) sequences of $\{Z_{j}\}$ are typical, according to the general AEP\cite[Thm 16.8.1]{Cover-Thomas-ElemIT}); in fact, by the ergodic theorem \cite[Thm 1.14]{Walters-Ergodic-Theory}, almost all sequences are doubly-strongly typical.

Now consider a (random) codebook of $2^{nR}$ length-$n$ block codes $\lt\{Y^{n}(1),\cdots,Y^{n}(2^{nR})\rt\}$, generated from $2^{nR}$ independent copies of the stationary Markov process $\{Y_{i}\}$. The decoder is defined as follows. Without loss of generality, assume $Y^{n}(1)$ is the transmitted codeword and let $V^{T}$ denote the received symbols (from the DMC). Also denote the (intermediate) output symbols of the deletion channel by $Z^{T}$. They can be viewed as the first $T$ entries of the processes $\{Z_{j}\}$ and $\{V_{j}\}$ corresponding to the input process that generates $Y^{n}(1)$.

We write $Z^{m}\sqsubset Y^{n}$ if $Z^{m}$ is a subsequence of some $Y^{n}$. Let $0<\epsilon<\theta$ and define
\[ m = (\theta - \epsilon)n. \]
The decoder claims an error if one of the following happens:
\begin{enumerate}
\item[(1)] $T < m$.
\item[(2)] $T \geq m$, but $V^{m}$ is not jointly typical with any doubly-strongly typical subsequence $z^{m}$ of $Y^{n}(1)$ (w.r.t. the stationary ergodic joint process $\{Z_{j},V_{j}\}$).
\item[(3)] $T \geq m$, $V^{m}$ is jointly typical with some doubly-strongly typical subsequence $z^{m}$ of $Y^{n}(k)$, for  some $k>1$.
\end{enumerate}
If none of the above holds, then $V^{m}$ is correctly decoded to $Y^{n}(1)$.

We study the average probability of error $\Pe{i}$, $i=1,2,3$, induced by the three error patterns above, respectively, when $n$ is large. First, by WLLN of the deletion process, $\Pe{1}\to0$ when $n\to\infty$. Second, in Appendix~\ref{subsec:Pe2} we prove that $\Pe{2}\to0$ as well. Finally, for $\Pe{3}$, by union bound and as $Y^{n}(2)$, \ldots, $Y^{n}(2^{nR})$ are independent, we can write
\begin{equation}\label{eq:Pe3}
\Pe{3} \leq 2^{nR}\cdot \Pr\lt(\{T\geq m\}\cap F \rt) \leq 2^{nR}\cdot\Pr\lt( F \rt),
\end{equation}
where the event $F$ is defined as
\begin{align*}
F &=  \lt\{ (z^{m},V^{m})\in\Ae(Z,V) \text{ for some } z^{m} \text{ satisfying } z^{m} \sqsubset Y^{n}(2) \text{ and } z^{m} \in \sAe(Z)\rt\} \\
&=  \bigcup_{z^{m} \in \sAe(Z)}\lt(\lt\{ (z^{m},V^{m})\in\Ae(Z,V)\rt\}\cap\lt\{z^{m} \sqsubset Y^{n}(2)\rt\}\rt).
\end{align*}
Now by union bound again and by the independence of $V^{m}$ and $Y^{n}(2)$, we have
\begin{IEEEeqnarray}{rCl}
\Pr\lt( F \rt) &\leq \sum_{z^{m} \in \sAe(Z)}\Pr&\lt(\lt\{ (z^{m},V^{m})\in\Ae(Z,V)\rt\}\cap\lt\{z^{m} \sqsubset Y^{n}(2)\rt\}\rt) \nonumber\\
&= \sum_{z^{m} \in \sAe(Z)} \Pr&\lt((z^{m},V^{m})\in\Ae(Z,V)\rt)\cdot \Pr\lt(z^{m} \sqsubset Y^{n}(2)\rt). \label{eq:Prob-F}
\end{IEEEeqnarray}
Let $\Ae(V|z^{m})$ denote the set of sequences $v^{m}$ that are jointly typical with $z^{m}$. We know from \cite[Thm 15.2.2]{Cover-Thomas-ElemIT} that for $z^{m}$ typical
\[ |\Ae(V|z^{m})| \leq 2^{m(H(\V|\Z) + 2\epsilon)} \]
and for any typical sequence $v^{m}$
\[ \Pr\lt(V^{m}=v^{m}\rt) \leq 2^{-m(H(\V) - \epsilon)}. \]
(As mentioned earlier, we use $H(\V)$ to denote the entropy rate of the non-i.i.d. stationary ergodic process $\{V_{j}\}$ and define the conditional entropy rate $H(\V|\Z) = H(\Z,\V) - H(\Z)$. In what follows $H(\Z)$ and $H(\Z|\V)$ are defined similarly.) Thus we can express
\begin{align}
\Pr\big((z^{m},V^{m}\in&\Ae(Z,V)\big) \nonumber\\
&= \sum_{v^{m}\in\Ae(V|z^{m})}\Pr\lt(V^{m}=v^{m}\rt) \nonumber\\
&\leq |\Ae(V|z^{m})|\cdot 2^{-m(H(\V) - \epsilon)} \nonumber\\
&\leq 2^{m(H(\V|\Z) - H(\V) + 3\epsilon)}. \label{eq:Pr-cond-typical-seq}
\end{align}
Furthermore, in Appendix~\ref{subsec:upper-bound-subseq-prob} we show that 
\begin{equation}\label{eq:Pr-subseq}
\beta_{m} = \infg \exp\lt[(n-m)\gamma + E_{1}(\gamma) +  mE(\gamma) + m\epsilon E_{0}(\gamma)\rt]
\end{equation}
is an upper bound for $\Pr\lt(z^{m} \sqsubset Y^{n}(2)\rt)$ when $z^{m} \in \sAe(Z)$. Since both bounds \eqref{eq:Pr-cond-typical-seq} and \eqref{eq:Pr-subseq} are independent of $z^{m}$ and
\[|\sAe(Z)|\leq|\Ae|\leq2^{m(H(\Z)+\epsilon)}, \]
we have
\begin{align*}
\Pr\lt( F \rt) &\leq \sum_{z^{m} \in \sAe(Z)} 2^{m(H(\V|\Z) - H(\V) + 3\epsilon)}\cdot\beta_{m} \\
&= |\sAe(Z)|\cdot 2^{m(H(\V|\Z) - H(\V) + 3\epsilon)}\cdot\beta_{m} \\
&\leq 2^{m(H(\Z)+\epsilon)}\cdot 2^{m(H(\V|\Z) - H(\V) + 3\epsilon)}\cdot\beta_{m} \\
&= 2^{m(H(\Z|\V)+4\epsilon)}\cdot\beta_{m}.
\end{align*}
Plugging into \eqref{eq:Pe3}:
\begin{IEEEeqnarray}{rCl}
\Pe{3}& \leq& 2^{nR}\cdot 2^{m(H(\Z|\V)+4\epsilon)}\cdot\beta_{m} \nonumber\\
&\leq& \exp\big[\ (nR+mH(\Z|\V)+4m\epsilon)\ln2+(n-m)\gamma \nonumber\\
&&\hspace{10ex}+\: E_{1}(\gamma) +  mE(\gamma) + m\epsilon E_{0}(\gamma)\ \big] \nonumber\\
&=& \exp\bigg\{n\cdot\bigg[\ \lt(R+\frac{m}{n}H(\Z|\V)+4\frac{m}{n}\epsilon\rt)\ln2 \nonumber\\
&&\hspace{10ex}+\lt(1-\frac{m}{n}\rt)\gamma + E_{1}(\gamma)/n \nonumber\\
&&\hspace{10ex} +\: \frac{m}{n}E(\gamma) + \frac{m}{n}\epsilon E_{0}(\gamma)\ \bigg]\bigg\}
\end{IEEEeqnarray}
for any $\gamma>0$. Now fix $\gamma$. If we want $\Pe{3}\to0$ as $n\to\infty$, we need the term inside the square brackets above to be less than zero when $n$ is large enough. Since $\frac{m}{n} = \theta - \epsilon$, this condition requires that
\begin{equation}\label{eq:R-cond-Pe3}
(R+ \theta H(\Z|\V))\cdot\ln2+(1-\theta)\gamma + \theta E(\gamma) + c\epsilon < 0,
\end{equation}
where $c$ is a finite constant. Since $\epsilon$ is arbitrary, for any
\[ R < - [(1-\theta)\gamma +  \theta E(\gamma)]/\ln2 - \theta H(\Z|\V), \]
we can always find $\epsilon >0$ such that \eqref{eq:R-cond-Pe3} is satisfied, and so all average error probabilities $\Pe{i}$, $i=1,2,3$ vanishes when $n$ is large. Thus using standard achievability argument \cite{Cover-Thomas-ElemIT} the rate $R$ is achievable. Taking supremum over all $\gamma>0$, we have
\begin{align*}
R &< \sup_{\gamma>0} - [(1-\theta)\gamma +  \theta E(\gamma)]/\ln2 - \theta H(\Z|\V) \\
&= - \infg[(1-\theta)\gamma +  \theta E(\gamma)]/\ln2 - \theta H(\Z|\V) 
\end{align*}
is also an achievable rate.
\end{IEEEproof}

\begin{corollary}\label{cor:del-ach-rate}
For a deletion channel, the following is an achievable rate for each irreducible Markov transition matrix $P$:
\[ \uC(P) = - \infg[(1-\theta)\gamma +  \theta E(\gamma)]/\ln2. \]
If $\Ps$ is a collection of irreducible Markov transition matrices, then $\sup_{P\in\Ps}\uC(P)$ is also achievable.
\end{corollary}

The corollary above is obtained by setting the DMC in Theorem~\ref{thm:ach-rate} to be an identity channel. These results are generalizations and improvements of the Markov lower bound results for (noisy) deletion channels in \cite{Diggavi-Grossglauser-finite-buffer-channel}. First, Theorem~\ref{thm:ach-rate} applies to any DMC, while the noisy deletion channel result in \cite[Theorem~5.1]{Diggavi-Grossglauser-finite-buffer-channel} is only valid for symmetric DMC. Second, the result in \cite[Theorems~4.3]{Diggavi-Grossglauser-finite-buffer-channel} only considers a special subset of irreducible Markov transition matrices as the optimization space, while the results above can optimize over any collection of irreducible transition matrices. Finally, even over the same optimization space, for each transition matrix the results above can still provide a better lower bound than \cite{Diggavi-Grossglauser-finite-buffer-channel}. For example, specializing to the binary deletion channel, the achievable rate (in nats) in \cite[Corollary~4.2]{Diggavi-Grossglauser-finite-buffer-channel} is the supremum of
\[ -\theta\ln[(1-q)A + qB] - \gamma \]
over a certain set and for some parameters $q$, $A$ and $B$. Corollary~\ref{cor:del-ach-rate}, however, in this case gives an achievable rate whose corresponding objective function takes the form
\[ -\theta[(1-q)\ln A + q\ln B] - \gamma, \]
which is an improvement to the previous result by Jensen's inequality. Note that the same formula for this special case is also obtained  in \cite[Corollary~1]{DM06} using a different method.

The codewords designed on the $K$-mers naturally translate to the codewords on the nucleotides using the bijective ISI mapping, but the length is increased by $K-1$. If $X_{1}^{n}$ denotes the codeword on the nucleotides, then the corresponding codeword on the $K$-mers is $Y_{K}^{n}$. The resulting code rate is reduced by a factor of $(n-K+1)/n$, which approaches 1 when $n\to\infty$, and so the achievable rate results from last section still apply. There is yet another potential issue with this scheme. Since the transmission starts from time 1, the first few received symbols may be influenced by the initial state $s_{0}$ (the prefixed adaptors in the nanopore experiments). However, as $n\to\infty$, such influence is negligible and the decoder in the previous section still works with the same asymptotic error performance.

\section{Capacity upper bounds}
\label{sec:upper-bounds}

In this section we derive computable capacity upper bounds for the nanopore channel. First, from Theorem~\ref{thm:exisitence-limit}, we know for each $n$, $\C_{n}$ defined in \eqref{eq:UB-C-bar} is an upper bound for the nanopore channel capacity, which can be computed using the classic Blahut-Arimoto algorithm. However, since these bounds are in the form of finite block mutual information plus an extra term, they suffers from two issues:
i) the computational complexity grows exponentially, since the optimization space is $|\X|^{n}$-dimensional;
ii) for smaller $n$, the extra term $\frac{1}{n}\log|\St| = \frac{K-1}{n}\cdot\log|\X|$ is relatively large and decays very slowly with $n$, which greatly reduces the effectiveness of the upper bounds.

To address these issues, we seek a relaxation of $\C_{n}$ that allows for an easy computation with large $n$, so that the second term in \eqref{eq:UB-C-bar} is negligible after division by $n$, while the growth of the first term (the block mutual information) is still tracked in a certain way. For that purpose we add periodic synchronization symbols to the output (as in \cite{Fertonani-Duman-Novel-bounds}), and use the formulation developed in \cite{Vontobel-Arnold-ISI-Upperbound}.

Let $M$ denote the period of synchronization and consider block length $N = nM$. Let $B_{k}$ be the random variable that denotes the number of output symbols of the deletion channel that corresponds to the chunk of input symbols $X_{(k-1)M+1}^{kM}$, for $1\leq k\leq n$. Then $0\leq B_{k}\leq M$ and we have the one-to-one correspondence
\[ \big(B^{n},\dvs{N}\big) \longleftrightarrow \big(\dvs{M},\dv{M+1}{2M},\ldots,\dv{(n-1)M+1}{nM}\big).  \]
Let $\bX_{k}$ and $\bV_{k}$ denote $X_{(k-1)M+1}^{kM}$ and $\dv{(k-1)M+1}{kM}$, respectively. Then we have
\[ I\big(X^{N};\dvs{N}\big|s_{0}\big) \leq I\big(X^{N};\dvs{N},B^{n}\big|s_{0}\big) = I\big(\bX^{n};\bV^{n}\big|s_{0}\big). \]

Note that if we group every $M$ channel uses together, we can define a new channel with input $\bX_{n}$ and output $\bV_{n}$, whose input and output alphabets are $\X^{M}$ and $\V^{(M)}$, respectively. This is a finite state channel\cite{Gallager-IT_Reliable} with state
\[ \bS_{n} \triangleq S_{nM} = X_{nM-K+2}^{nM}, \]
since the channel transition probability satisfies
\begin{align*}
p\big(\bv_{n}\bs_{n}\cn \bx^{n}\bs^{n-1}\bv^{n-1}\big) &= p\big(\bv_{n}\cn \bx_{n}\bs_{n-1}\big)p\big(\bs_{n}\cn \bx_{n}\bs_{n-1}\big) \\
&= p\big(\bv_{n}\bs_{n}\cn \bx_{n}\bs_{n-1}\big).
\end{align*}
Observe that since $\bs_{0} = s_{0}$ and $\St$ is also the alphabet for $\bS_{n}$, we can rewrite $I\big(\bX^{n};\bV^{n}\big|s_{0}\big) = I\big(\bX^{n};\bV^{n}\big|\bs_{0}\big)$ and thus obtain the following relaxation of \eqref{eq:UB-C-bar}:
\[ \C_{N} \leq \frac{1}{M}\C'_{n}, \]
\begin{equation}\label{eq:UB-C'-bar}
\C'_{n} \triangleq \frac{1}{n}\lt[\max_{\bs_{0}}\max_{P_{\bX^{n}}}I\big(\bX^{n};\bV^{n}\big|\bs_{0}\big) + \log|\St|\rt].
\end{equation}
Note that $\C'_{n}$ also coincides exactly with the upper bound in \cite[Thm 4.6.1]{Gallager-IT_Reliable} for the new finite state channel. Moreover, this is indeed a channel with only inter-symbol interference (ISI) memory, since $\bS_{n}$ is fully determined by $\bX_{n}$ and $\bS_{n-1}$. As a result, we can apply methods in \cite{Vontobel-Arnold-ISI-Upperbound} to this channel to simplify the computation of $\C'_{n}$.

Let $\bU_{k} = \big(\bS_{k-1},\bX_{k}\big)$ denote the input branch of the new channel at time $k$, then we have an one-to-one correspondence $(\bs_{0},\bx^{n}) \longleftrightarrow \bu^{n}$. Following \cite{Vontobel-Arnold-ISI-Upperbound} we have
\[ \max_{\bs_{0}}\max_{P_{\bX^{n}}}I\big(\bX^{n};\bV^{n}\big|\bs_{0}\big) \leq \max_{\bu^{n}}D(\,W(\cdot|\bu^{n})\|R(\cdot)\,), \]
where $W$ denotes the transition probability $p(\bv^{n}\cn\bu^{n})$ and $R$ can be any probability distribution on sequences $\bv^{n}$. Hence we have a relaxed capacity upper bound $\tC'_{n}$ for the new channel
\begin{equation}\label{eq:UB-C-tilde}
\C'_{n}\leq \tC'_{n} \triangleq \frac{1}{n}\lt[\max_{\bu^{n}}D(\,W(\cdot|\bu^{n})\|R(\cdot)\,) + \log|\St|\rt],
\end{equation}
and for the original channel we have the relaxation
\begin{theorem}\label{thm:UBs-relaxation-C-tilde}
$\C_{N} \leq \frac{1}{M}\tC'_{n}$ for all $N = nM$.
\end{theorem}

This new upper bound admits a linear complexity algorithm, if we make further assumptions on the distribution $R$, as in \cite{Vontobel-Arnold-ISI-Upperbound}. First take $R$ to be the probability measure for an $L$-th order Markov chain. Since $W(\bv^{n}\cn\bu^{n}) = \prod_{k=1}^{n}W(\bv_{k}\cn\bu_{k})$ and $R(\bv^{n}) = R(\bv^{L})\cdot\!\! \prod_{k=L+1}^{n}R(\bv_{k}\cn\bv_{k-L}^{k-1})$, we can establish the decompostion (following \cite{Vontobel-Arnold-ISI-Upperbound}, but notation changed slightly)
\[ D(\,W(\cdot|\bu^{n})\|R(\cdot)\,) = \tphi_{L}(\ts_{L}) + \sum_{k=L+1}^{n}\tphi_{k}(\ts_{k-1},\bu_{k}), \]
where we define $\ts_{k} = \bu_{k-L+1}^{k}$ for $k\geq L$, define
\begin{IEEEeqnarray*}{rCl}
\tphi_{k}(\ts_{k-1},\bu_{k}) &=& E_{W(\cdot\cn\bu_{k-L}^{k})}\lt[\log\frac{W(\bV_{k}\cn\bu_{k})}{R(\bV_{k}\cn\bV_{k-L}^{k-1})}\rt] \\
&=& - H\big(\bV_{k}\cn\bu_{k}\big) - E_{W(\cdot\cn\bu_{k-L}^{k})}\log R(\bV_{k-L}^{k}) \\
&& +\: E_{W(\cdot\cn\bu_{k-L}^{k-1})}\log R(\bV_{k-L}^{k-1})
\end{IEEEeqnarray*}
for $k> L$, and define
\begin{align*}
\tphi_{L}(\ts_{L}) &= E_{W(\cdot\cn\bu^{L})}\lt[\log\frac{\prod_{k=1}^{L}W(\bV_{k}\cn\bu_{k})}{R(\bV^{L})}\rt] \\
&= - \sum_{k=1}^{L} H\big(\bV_{k}\cn\bu_{k}\big) - E_{W(\cdot\cn\bu^{L})}\log R(\bV^{L}).
\end{align*}
If we further restrict $R$ to be a stationary measure of an $L$-th order homogeneous Markov chain, then $\tphi_{k}(\cdot)$ is time invariant for $k>L$, and so we can drop the subscript and write $\tphi = \tphi_{k}$. Consequently, we can build a trellis with
\begin{equation}\label{eq:alphabet-size-s-tilde}
\big|\St\times(\X^{M})^{L}\big| = |\X|^{L\cdot M + K-1}
\end{equation}
states to compute $D(\,W(\cdot|\bu^{n})\|R(\cdot)\,)$, starting from time index $L$. The state of the trellis is exactly $\ts_{k}$ at time $k$, each branch $(\ts_{k-1},\bu_{k})$ carrying a weight $\tphi(\ts_{k-1},\bu_{k})$, and the initial weight for the state $\ts_{L}$ is $\tphi_{L}(\ts_{L})$. For a specific $\bu^{n}$, $D(\,W(\cdot|\bu^{n})\|R(\cdot)\,)$ can be computed as the cumulative weight along the corresponding path, and so we can use the Viterbi algorithm to compute the maximum path weight for \eqref{eq:UB-C-tilde}.

Since the Viterbi algorithm has linear complexity, we can compute \eqref{eq:UB-C-tilde} for a large $n$, so that the influence of the second term $\log|\St|$ is negligible. The corresponding trellis path length is $n-L+1$. To make this bound as tight as possible, we also need to choose the probability distribution $R$ carefully. As mentioned in \cite{Vontobel-Arnold-ISI-Upperbound}, a good choice of $R$ can be obtained from the $(L+1)$-dimensional output distribution of the finite state channel $p\big(\bv_{n}\bs_{n}\cn \bx_{n}\bs_{n-1}\big)$ when the input distribution is optimized with respect to the achievable rate (e.g., using the generalized Blahut-Arimoto algorithm \cite{Vontobel-GBAA}). The computational complexity of this algorithm is mainly limited by $M$ and $L$: the number of states grows exponentially with $M\cdot L$, and so does the output alphabet size, which is $|\V^{(M)}|^{L}$. The latter turns out to be the major bottleneck for the computation in the case of the nanopore channel.

\section{Numerical Results}
\label{sec:Numerics}

\begin{figure}[!t]
\centering
\includegraphics[width=0.8\textwidth]{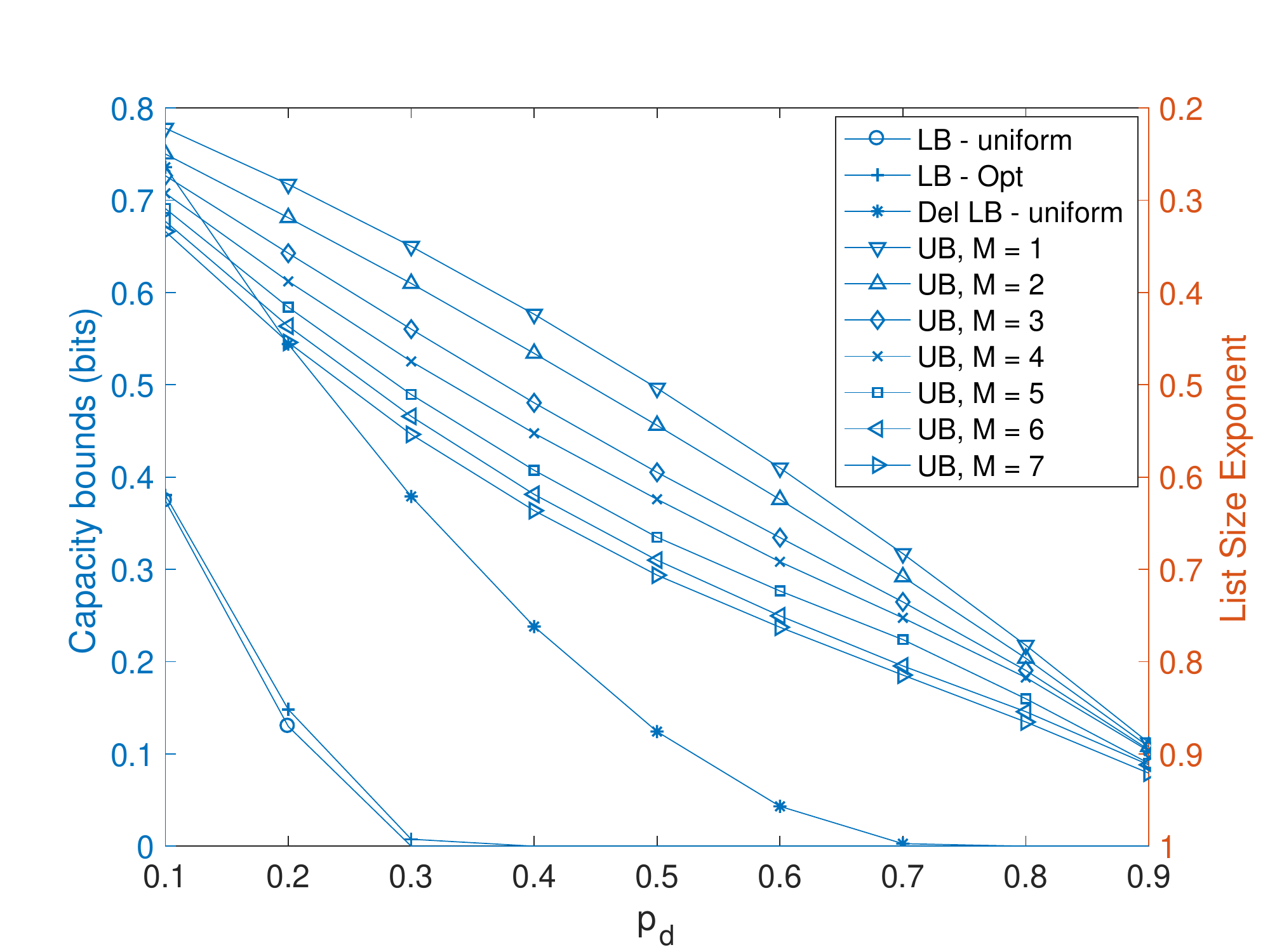}
\caption{$K = 2$, $|\X| = 2$,  symmetric DMC.}
\label{fig:sym-ch-linear}
\end{figure}
\begin{figure}[!t]
\centering
\includegraphics[width=0.8\textwidth]{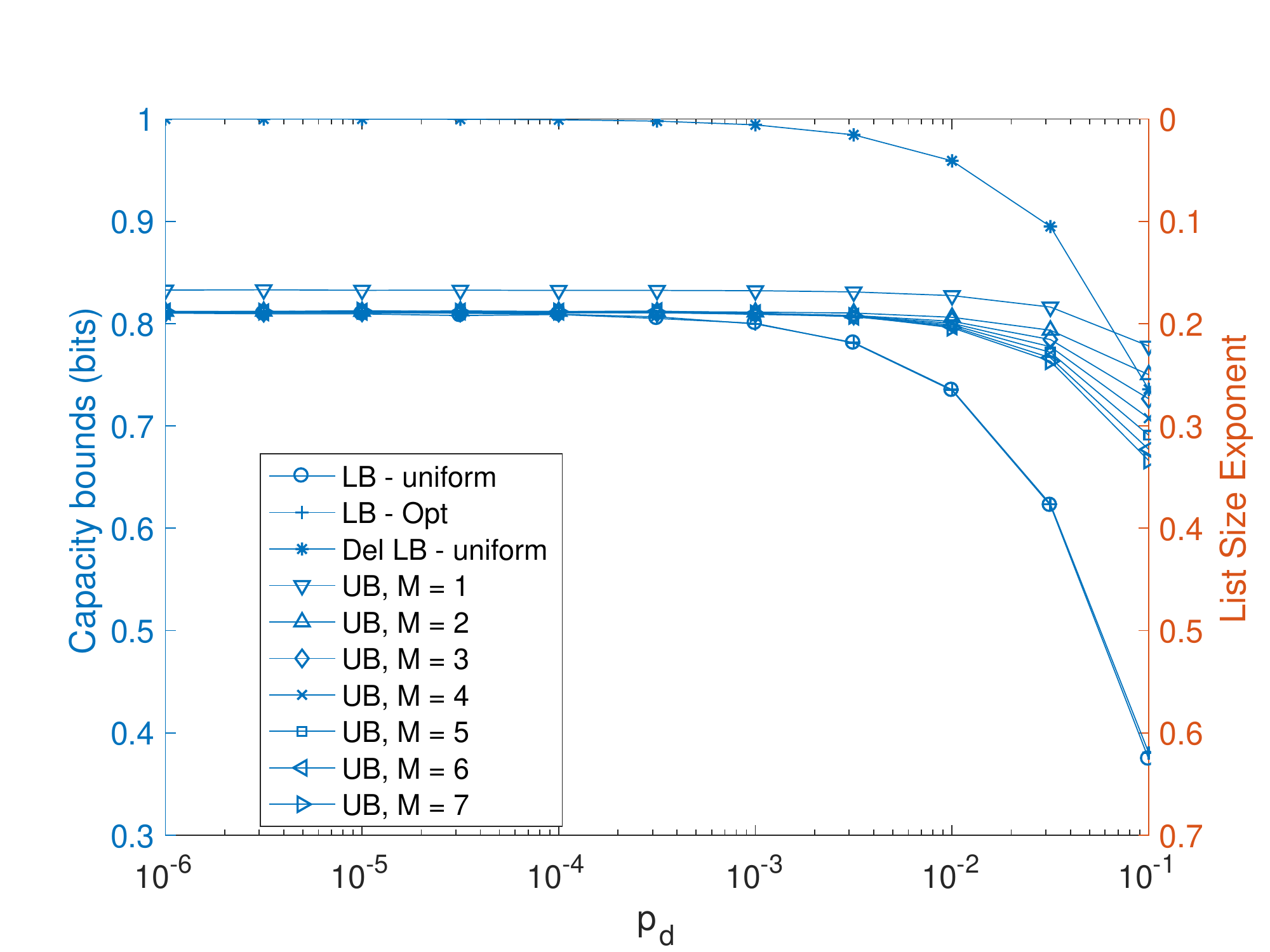}
\caption{$K = 2$, $|\X| = 2$,  symmetric DMC.}
\label{fig:sym-ch-log}
\end{figure}
\begin{figure}[!t]
\centering
\includegraphics[width=0.8\textwidth]{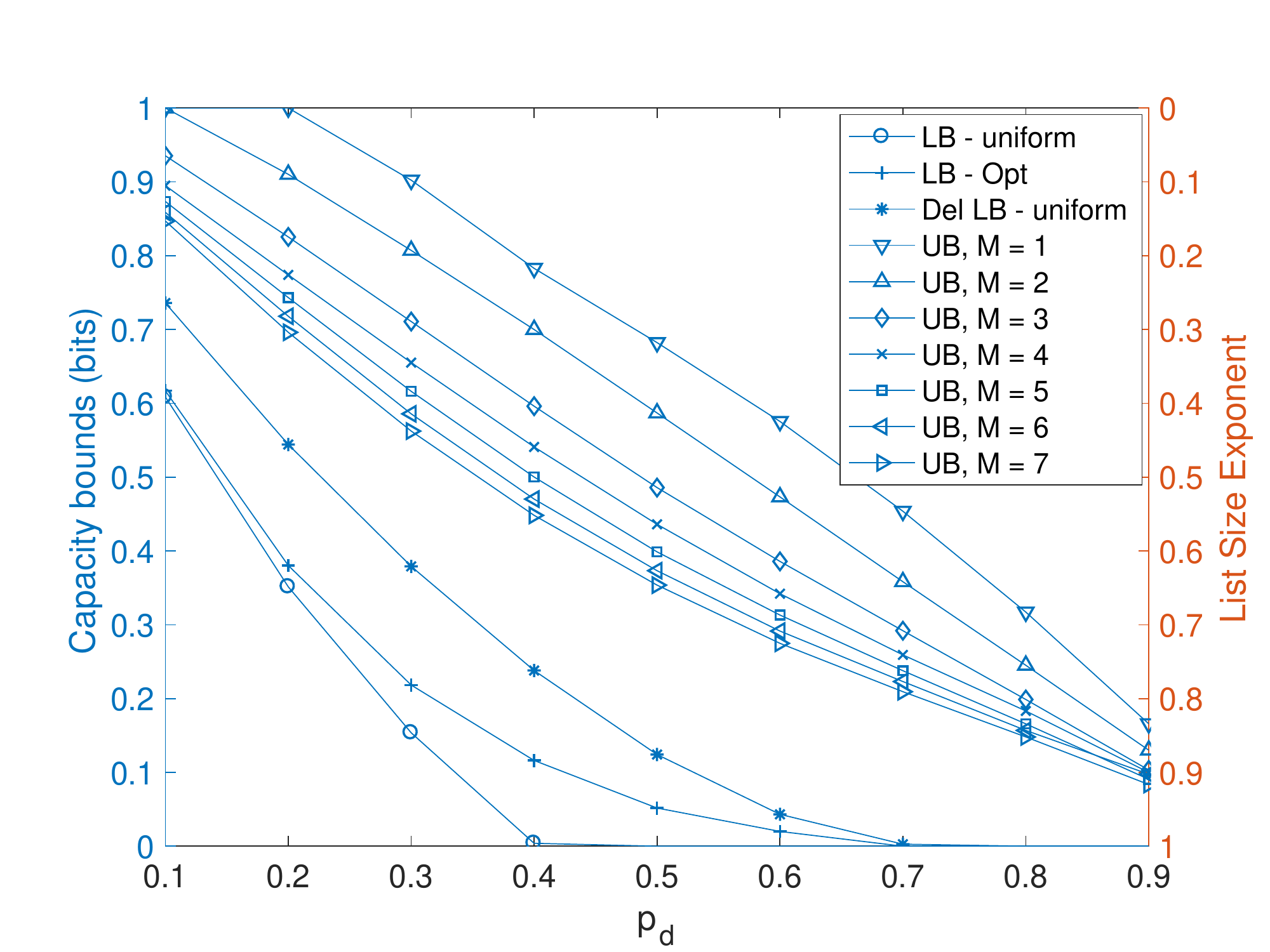}
\caption{$K = 2$, $|\X| = 2$,  non-symmetric DMC.}
\label{fig:synz-param-linear}
\end{figure}
\begin{figure}[!t]
\centering
\includegraphics[width=0.8\textwidth]{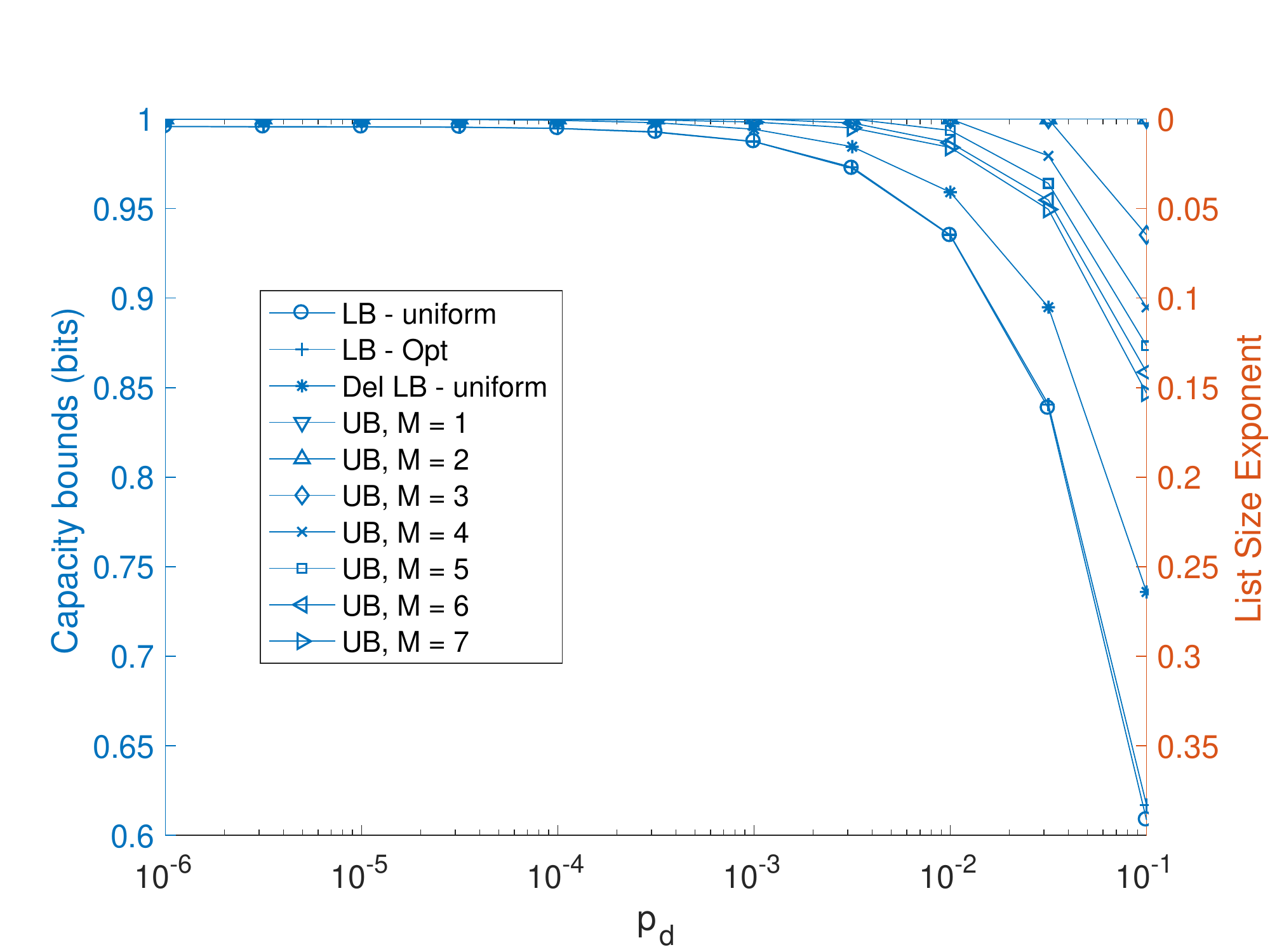}
\caption{$K = 2$, $|\X| = 2$,  non-symmetric DMC.}
\label{fig:synz-param-log}
\end{figure}

First we use two more examples to illustrate the computation of the capacity bounds for the nanopore channel. The first example has ISI memory length $K=2$, input alphabet size $|\X| = 2$, and uses a symmetric DMC: in this case $\V = \Y$, $p(y|y) = c$ for all $y$ and $p(y'|y) = c'$ for all $y'\neq y$, for some constants $c$ and $c'$. The bounds are plotted in Fig.~\ref{fig:sym-ch-linear} and~\ref{fig:sym-ch-log} for different deletion probabilities. The second example has the same ISI memory length and input alphabet size, but the DMC is not symmetric: it is constructed similarly to the example in Section~\ref{sec:main-results}, where a part of the Q-mer map data in Fig.~\ref{fig:q-mer} (corresponding to the first $2^{2} = 4$ Q-mers) is extracted to simulate the real world situation. The corresponding bounds are plotted in Fig.~\ref{fig:synz-param-linear} and~\ref{fig:synz-param-log}.

In both examples, we plot the achievable rates in Theorem~\ref{thm:ach-rate} as the capacity lower bounds, both when the input distribution is i.i.d. uniform and when $P$ is optimized. We also plot the achievable rate of pure deletion channel from Corollary~\ref{cor:del-ach-rate} when the input distribution is i.i.d. uniform, to illustrate the rate loss due to the signal degradation caused the DMC. From Fig.~\ref{fig:sym-ch-linear} and~\ref{fig:sym-ch-log} we can see that for the channel with a symmetric DMC, i.i.d. uniformly generated codewords already have a performance very close to the optimal coding scheme in Section~\ref{sec:ach-rate}. From Fig.~\ref{fig:synz-param-linear} and \ref{fig:synz-param-log} (and also Fig.~\ref{fig:forward-pore-linear} and \ref{fig:forward-pore-log} in Section~\ref{sec:main-results}), we can see that when the DMC is not symmetric, non-uniformly generated codewords can achieve higher rates than the uniform ones. But when the deletion probability $p_{d}$ is small, the uniform case is still close to optimal.

We also plot the capacity upper bounds in Theorem~\ref{thm:UBs-relaxation-C-tilde} with different parameters for these two examples. From the computation results we found that either increasing the synchronization period $M$ or the output Markov order $L$ (see Section~\ref{sec:upper-bounds} for the definitions) yields a tighter upper bound, but when $M$ is not too small the bounds become very close to each other for different $L$. Hence in Fig.~\ref{fig:sym-ch-linear}, \ref{fig:sym-ch-log}, \ref{fig:synz-param-linear} and \ref{fig:synz-param-log} we only plot the upper bounds for different $M$ (with $L = 1$). We can see that in the case with a symmetric DMC, the upper and lower bounds almost coincide for small $p_{d}$ and so we can bound the capacity tightly. With a non-symmetric DMC, the gap between the upper and lower bounds is also small for small $p_{d}$. We note that further computational optimization is needed to calculate the upper bounds for the example in Section~\ref{sec:main-results}.

\begin{figure}[!t]
\centering
\includegraphics[width=0.8\textwidth]{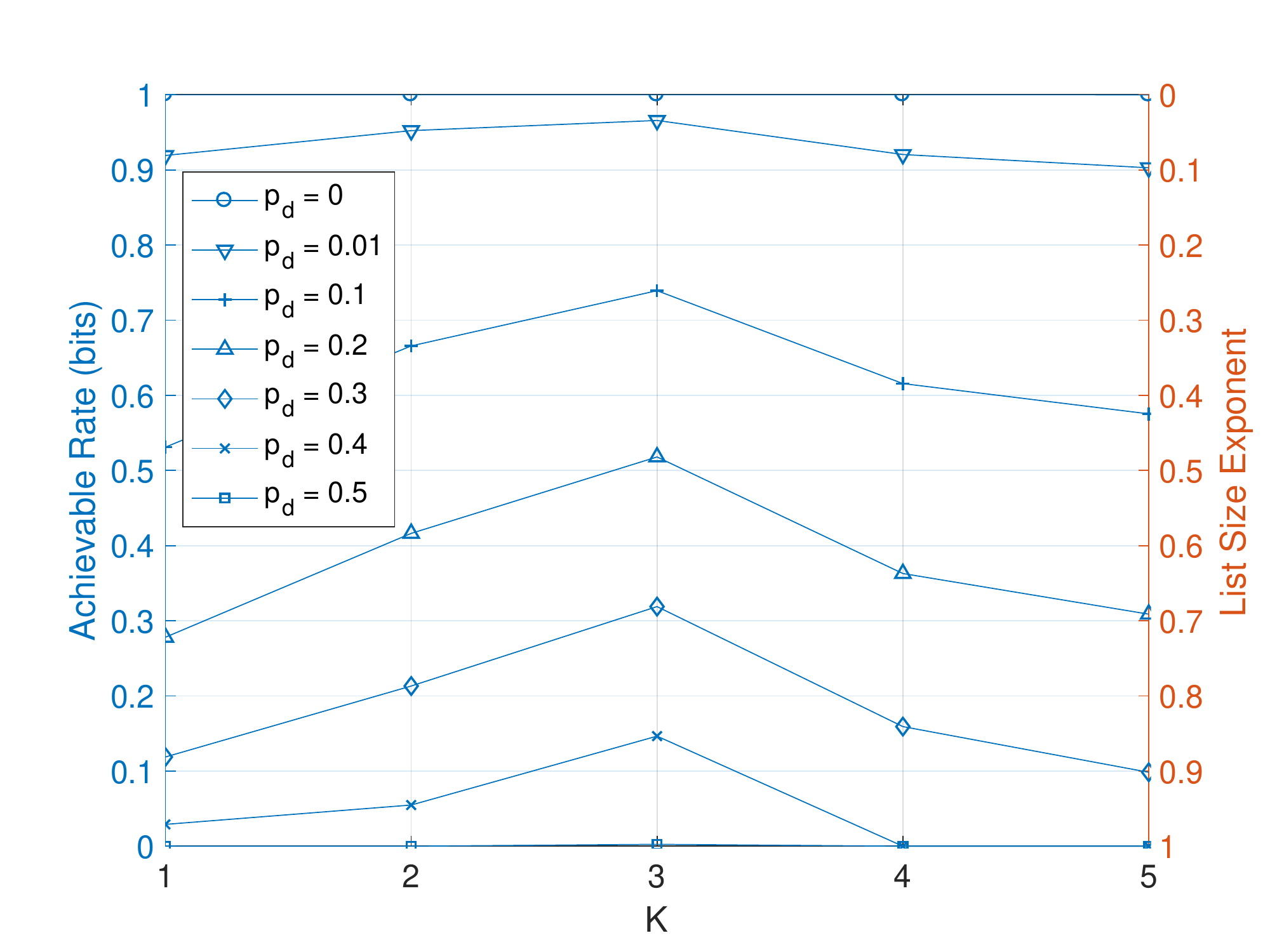}
\caption{$|\X| = 2$, $w=2$, uniform input.}
\label{fig:ach-rate-synth-nx2-sc-unf}
\end{figure}

\begin{figure}[!t]
\centering
\includegraphics[width=0.8\textwidth]{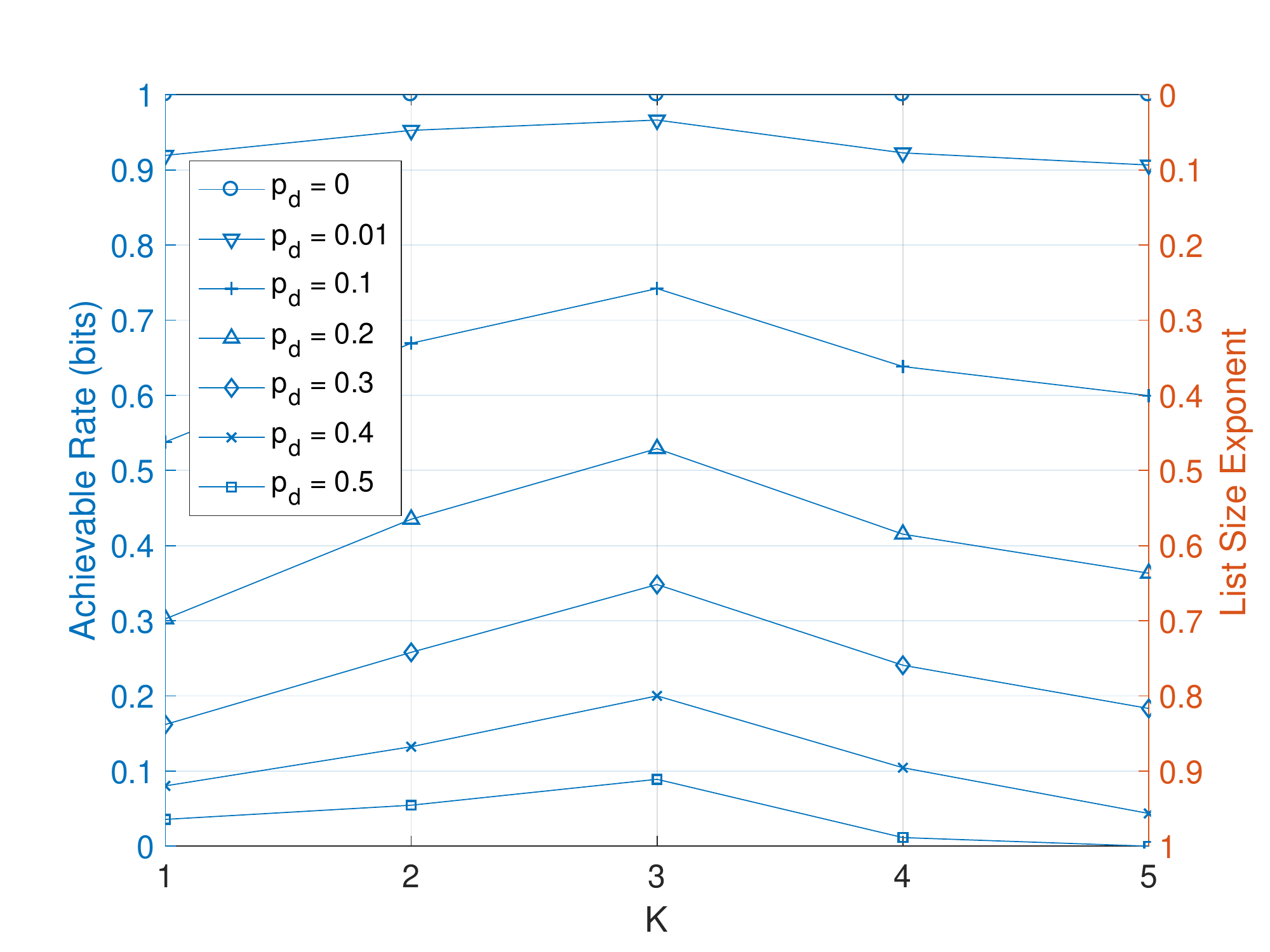}
\caption{$|\X| = 2$, $w=2$, optimized input.}
\label{fig:ach-rate-synth-nx2-sc-opt}
\end{figure}

\begin{figure}[!t]
\centering
\includegraphics[width=0.8\textwidth]{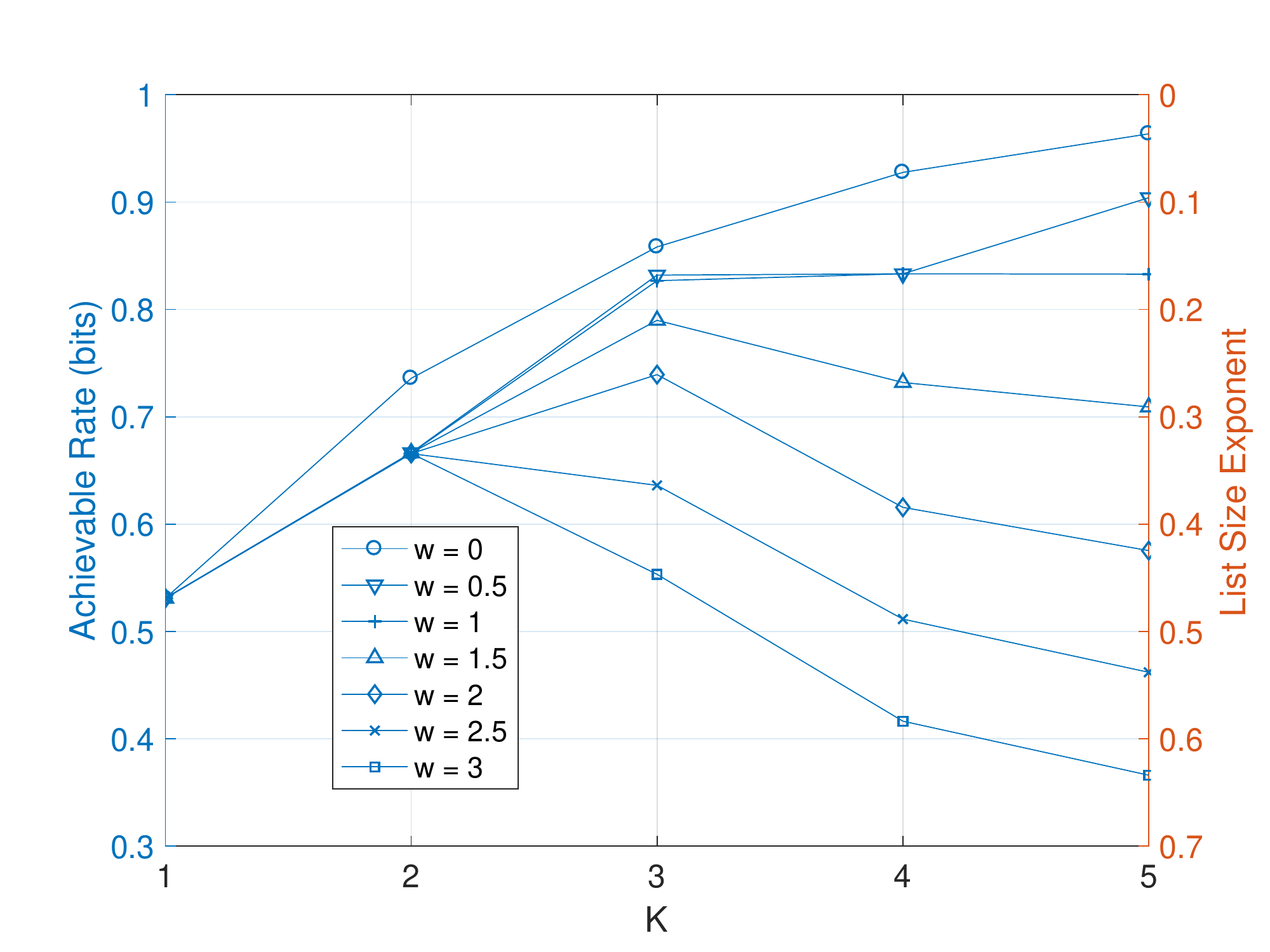}
\caption{$|\X| = 2$, $p_{d}=0.1$, uniform input.}
\label{fig:ach-rate-synth-nx2-pd-unf}
\end{figure}

\begin{figure}[!t]
\centering
\includegraphics[width=0.8\textwidth]{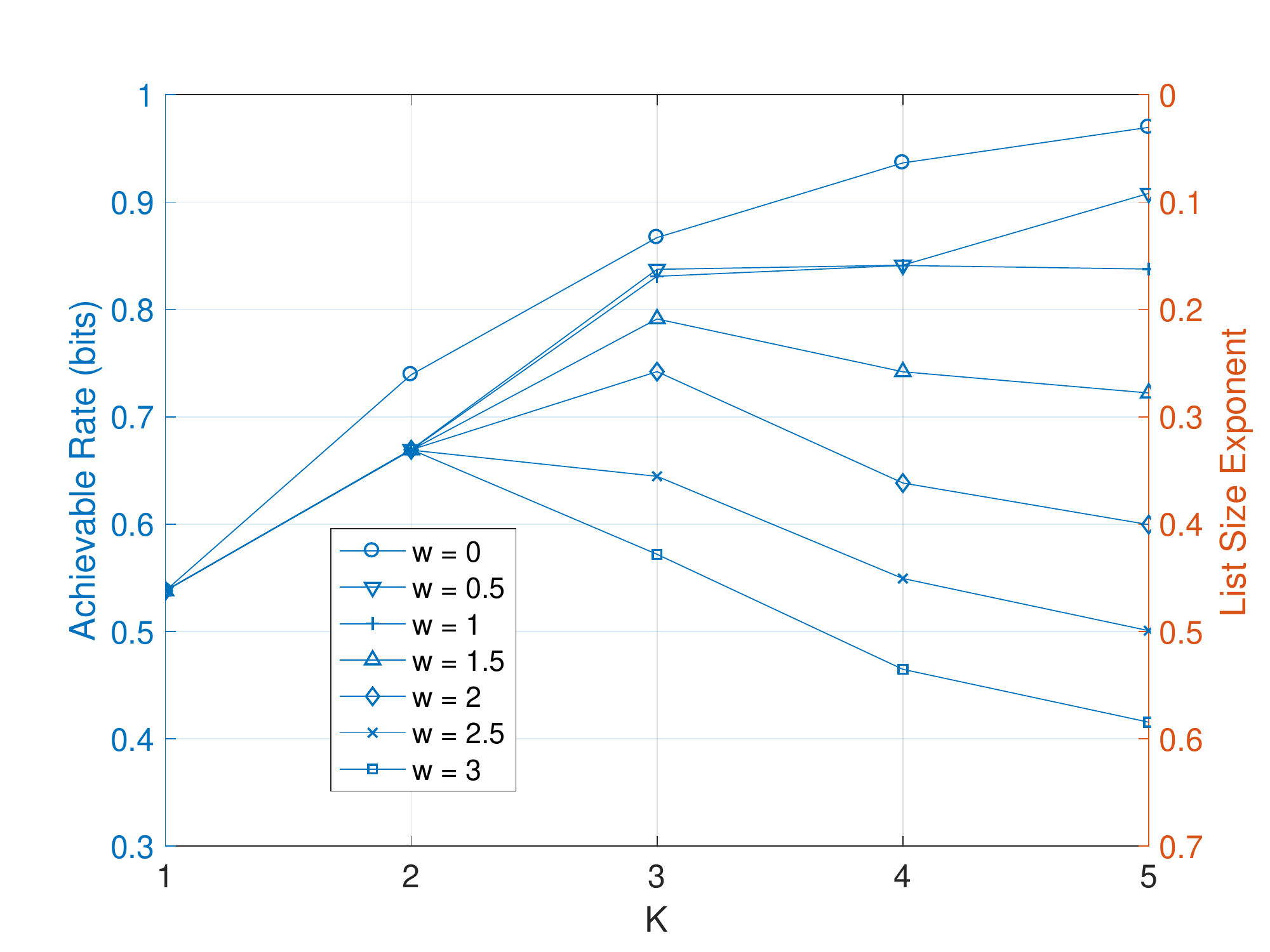}
\caption{$|\X| = 2$, $p_{d}=0.1$, optimized input.}
\label{fig:ach-rate-synth-nx2-pd-opt}
\end{figure}

In Section~\ref{sec:main-results} we discussed a nanopore model with linear ISI and uniform fading, to illustrate the influence of ISI length on the performance of nanopore sequencers. In particular, increasing the ISI length $K$ brings in better protection against deletion errors, but it also introduces more noise in the DMC. In what follows we provide more numerical results illustrating this trade-off. We compute the achievable rates for the nanopore channels when the input alphabet size is $|\X| = 2$ (we use the measured data for nucleotides A and C). In Fig.~\ref{fig:ach-rate-synth-nx2-sc-unf}, \ref{fig:ach-rate-synth-nx2-sc-opt}, \ref{fig:ach-rate-synth-nx2-pd-unf}  and~\ref{fig:ach-rate-synth-nx2-pd-opt}, these achievable rates for different ISI length $K$ are plotted, both when the input is i.i.d. uniform and when the input Markov chain is optimized. (Recall that $w$ is the scaling factor defined in Section~\ref{sec:main-results}.) Again, in Fig.~\ref{fig:ach-rate-synth-nx2-sc-unf} and~\ref{fig:ach-rate-synth-nx2-sc-opt} we see the peaks of the achievable rates, resulting from the opposing effects in the trade-off, and in Fig.~\ref{fig:ach-rate-synth-nx2-pd-unf}  and~\ref{fig:ach-rate-synth-nx2-pd-opt} we see the shifting of these peaks due to different noise levels in the DMC.

\begin{figure}[!t]
\centering
\includegraphics[width=\textwidth]{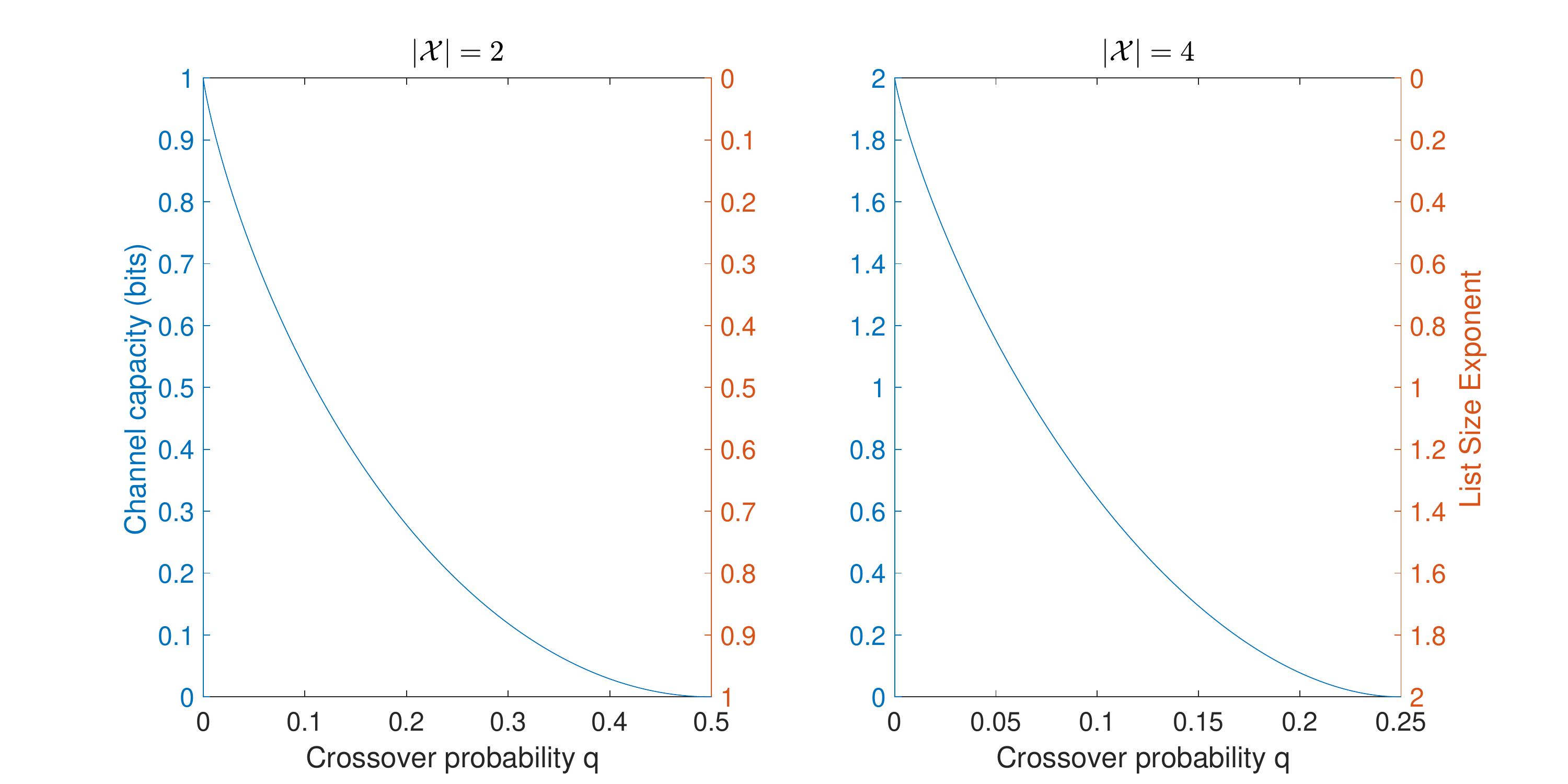}
\caption{Capacity of symmetric DMCs.}
\label{fig:equiv-sym-ch}
\end{figure}

Finally, we want to discuss an alternative point of view for the capacity bounds. As pointed out earlier, achievable rates under the uniform distribution can be translated into equivalent uncertainty list sizes for base-calling. An alternative measure is the decoding error-rate, which is the preferred metric when decoding to a single choice. The two measures, decoding error-rate and list-size are related since the uncertainty list is not usually comprised of arbitrary sequences but of sequences {\em around} the true sequence. While it is hard to make this formal, we can get a sense of the relationship between the two by using the following back-of-the-envelope calculation. Consider an ``equivalent'' 4-ary symmetric DMC with crossover probability $q$, whose capacity is equal to the achievable rate of the nanopore. The input of this channel can be viewed as the DNA sequence to be measured, $X^{n}$, while the output can be viewed as a certain estimate of the sequence, $\hX^{n}$. The transition probability of this DMC is defined as
\[ p(b|a) = \begin{cases}
q & \text{if } a \neq b,\\
1-3q & \text{if } a = b,
\end{cases} \]
where $a,b\in\X$. The channel capacity and the corresponding list size exponent (defined in Section~\ref{sec:main-results}) are ploted in Fig.~\ref{fig:equiv-sym-ch} (the right figure). As a rule of thumb, from this figure, we can read off the corresponding crossover probability $q$ from the achievable rates in Fig.~\ref{fig:forward-pore-linear} and~\ref{fig:forward-pore-log}. For example, for a set of typical deletion probabilities $p_{d}$, the corresponding $q$ are listed in Table~\ref{table:equiv-sym-ch-prob}. We can also treat the cases for alphabet size 2 similarly, using a binary symmetric DMC and the left figure of Fig.~\ref{fig:equiv-sym-ch}.  


\begin{table}[!t]
\caption{Equivalent symmetric channel crossover probability}
\label{table:equiv-sym-ch-prob}
\centering
\begin{tabular}{c|c|c|c|c}
  \hline
  $p_{d}$ & 0.01 & 0.1 & 0.2 & 0.3 \\
  \hline
  $q$ & 0.0032 & 0.0189 & 0.0407 & 0.0664 \\
  \hline
\end{tabular}
\end{table}

\section{Discussion}

This paper introduced a signal-level mathematical model for the nanopore sequencer, and developed an information-theoretical analysis of the nanopore channel. The model incorporates various impairments to the nanopore read: ISI, deletions, and random Q-mer responses. We first generalized Dobrushin's multi-letter capacity formula to the channels with both ISI and synchronization errors, which includes the nanopore channel as a special case. Then we derived a single-letter computable lower bound of nanopore channel capacity. Furthermore, we developed novel computable upper bounds for the capacity. In addition, we also analyzed the trade-off of the achievable rates when increasing the nanopore length, through a linear ISI model. These results could enable benchmarking nanopore sequencer performances and provide some insight in designing better nanopore sequencers.

There are also several open questions. The information theoretical analysis gives bounds on the list size, but does not directly  translate to error probability. One open question is how to develop bounds on error probability for the nanopore channel, for errors in the sense of edit distances. To improve the error rate performance of the nanopore sequencer, one may want to combine the results from multiple reads/measurements of the same DNA sequence. For example, current technology might enable the read of both a DNA sequence and its reverse complement. Such experiments give rise to a parallel nanopore channel, whose capacity analysis is quite challenging---it involves the capacity of multiple deletion channels, which is not well understood. Another interesting but also challenging problem is the assembly of the multiple read results. For the current technology, multiple reads may come from different random sections from the same DNA. If each section is passed through a nanopore sequencer to produce a read, then how to decode, align and assemble them to give the correct decoding results for the original DNA is an important, but very challenging problem. Distinct from the classical assembly problem for the shotgun sequencing technology where short reads are precisely sequenced, here each read is long and is subject to some error (cf. \cite{kannan2016shannon,lam_near-optimal_2014,tse-2016isit}, etc.). Finally, the computation of upper bounds developed in this paper is still quite expensive. Developing more computationally efficient upper and lower bounds to the decoding capability of the nanopore channel is an important open question.

\appendices
\section{Proofs for Section~\ref{sec:multi-letter-capacity}}

\subsection{Technical lemmas}
\label{subsec:technical-lemmas}

We introduce several technical lemmas to be used in the proofs.

\begin{lemma}[Fekete's lemma]\label{lem:Fekete}
If the sequence $\{ a_n\}_{n=1}^\infty$ is subadditive, i.e., $a_{m+n}\leq a_{m}+a_{n}$ for all $m$ and $n$, then the limit $\lim_{n \to \infty} \frac{a_n}{n}$ exists and is equal to $\inf_{n} \frac{a_n}{n}$. Similarly, if the sequence is superadditive, then $\lim_{n \to \infty} \frac{a_n}{n} = \sup_{n}\frac{a_n}{n}$.
\end{lemma}

\begin{IEEEproof}
See, e.g., \cite[Lemma~2, Appendix~4A]{Gallager-IT_Reliable}.
\end{IEEEproof}

\begin{lemma}\label{lem:entropy-mean-value-bound}
Assume $B$ is a discrete random variable on the nonnegative integers. If the mean value $\Ex{B}$ is finite, then its entropy $H(B)$ is also finite. In particular, $H(B) \leq 1+ \Ex{B}$.
\end{lemma}

\begin{IEEEproof}
When $\Ex{B} = 0$, $B=0$ with probability 1 and so $H(B)=0$. Assume $\Ex{B}>0$. We can follow the derivation for the maximum entropy theorem (Theorem~12.1.1) in \cite{Cover-Thomas-ElemIT} to show that
\[ H(B) \leq H(G),  \]
where $G$ is a geometric random variable with the same mean value as $B$:
\[ \Pr(G = k) = (1-q)^{k}q,\quad \forall k\geq 0 \]
\[ q = \frac{1}{1+\Ex{B}}. \]
Note that $0<q <1$ as $\Ex{B}$ is finite non-zero. Let $H(q)$ denote the binary entropy function of $q$, the entropy of $G$ is
\[ H(G) = \frac{1}{q}H(q) <\infty, \]
Hence
\[ H(B) \leq H(G) \leq \frac{1}{q} = 1+\Ex{B}. \]
\end{IEEEproof}

\begin{lemma}\label{lem:entropy-finite-avg-length-bound}
Let $V$ be a random variable on $\V^{*}$ with finite expected length $\Ex{\|V\|}$. If the underlying alphabet $\V$ for $\V^{*}$ is finite, then the entropy of $V$ is also finite, and is bounded by
\[ H(V) \leq 1 + \Ex{\|V\|}(\log|\V|+1). \]
\end{lemma}

\begin{IEEEproof}
By Lemma~\ref{lem:entropy-mean-value-bound}, $H(\|V\|)$ is finite and
\[ H(\|V\|) \leq 1+ \Ex{\|V\|}. \]
Denote $\V^{0} = \emptyset$ and write
\[ p_{n} = \Pr(\|V\| = n) = \sum_{v\in\V^{n}}p(v) \]
for $n\geq0$. For $v\in\V^{n}$ define the notation
\[ q_{n}(v) = \begin{cases}
 \Pr(V = v\cn \|V\| = n) = p(v)/p_{n} & \text{if } p_{n} > 0, \\
 0 & \text{otherwise},
\end{cases} \]
we have $p(v) = p_{n}q_{n}(v)$ and
\begin{align*}
H(V) &= - \sum_{v\in\V^{*}}p(v)\log p(v) \\
&= - \sum_{n = 0}^{\infty}\ \sum_{v\in\V^{n}}p(v)\log p(v) \\
&= - \sum_{n = 0}^{\infty}p_{n} \sum_{v\in\V^{n}}q_{n}(v)\lt[\log p_{n} + \log q_{n}(v)\rt] \\
&= - \sum_{n = 0}^{\infty}p_{n} \lt[\log p_{n} + \sum_{v\in\V^{n}}q_{n}(v)\log q_{n}(v)\rt].
\end{align*}
Note that all the terms in the series are negative. Since
\[ \sum_{v\in\V^{n}}q_{n}(v)\log q_{n}(v)  = - H(V\cn \|V\| = n) \geq - n\log|\V|, \]
\[ p_{n} \lt[\log p_{n} + \sum_{v\in\V^{n}}q_{n}(v)\log q_{n}(v)\rt] \geq p_{n} \lt(\log p_{n} - n\log|\V| \rt). \]
Summing over $n$ we have
\[ \sum_{n = 0}^{\infty}p_{n} \lt[\log p_{n} + \sum_{v\in\V^{n}}q_{n}(v)\log q_{n}(v)\rt] \geq -H(\|V\|) -\Ex{\|V\|}\log|\V|, \]
and so
\begin{align*}
H(V) &\leq H(\|V\|) + \Ex{\|V\|}\log|\V| \\
&\leq 1+ \Ex{\|V\|} + \Ex{\|V\|}\log|\V| \\
&= 1 + \Ex{\|V\|}(\log|\V| + 1) < \infty.
\end{align*}
\end{IEEEproof}

The following is a small modification of the standard law of large numbers.
\begin{lemma}\label{lem:WLLN}
Let $U_{1}$, $U_{2}$, \ldots be independent random variables, among which $U_{2}$, $U_{3}$, \ldots are identically distributed. Assume $U_{1}$ is finite almost surely, and assume $\Ex{|U_{2}|} < \infty$. Write $S_{n} = \sum_{k=1}^{n}U_{k}$ and $\mu = \Ex{U_{2}}$. Then $S_{n}/n\to\mu$ in probability as $n\to\infty$.
\end{lemma}

\begin{IEEEproof}
This lemma is a direct consequence of Kolmogorov's strong law of large numbers (see, e.g., \cite[Sec.~4.3, Theorem~3]{Shiryaev-Probability}.) Define $S'_{n} = \sum_{k=2}^{n}U_{k}$. Since $U_{2}$, $U_{3}$, \ldots are i.i.d.,  by SLLN, when $n\to\infty$,
\begin{equation}\label{eq:SLLN}
\frac{S'_{n}}{n-1} \to\mu
\end{equation}
almost surely. Now on a set $E$ with probability 1 we both have $U_{1}$ being finite and \eqref{eq:SLLN}, thus at any sample point $\omega\in E$
\[ \frac{S_{n}(\omega)}{n} = \frac{U_{1}(\omega)}{n}  + \frac{n-1}{n}\cdot\frac{S'_{n}(\omega)}{n-1} \to 0+\mu = \mu \] 
as $n\to\infty$. Hence $S_{n}/n\to\mu$ almost surely, and the lemma follows as almost sure convergence implies convergence in probability.
\end{IEEEproof}

\subsection{Proofs for Section~\ref{subsec:existence-of-limit}}
\label{subsec:proof-existence-of-limit}

Before proving Lemma~\ref{lem:subadditivity} and Theorem~\ref{thm:exisitence-limit}, we first prove another lemma for the channel \eqref{eq:ch-prob}, which shows that the influence of the initial state (per channel use) diminishes as the block length increases.
\begin{lemma}\label{lem:initial-state-does-not-matter}
$\forall \epsilon>0$, $\exists N>0$ such that $\forall n\geq N$, $\forall P_{X^{n}}$, and $\forall s_{0},s_{0}' \in \St$, for the channel \eqref{eq:ch-prob} we have
\begin{equation}\label{eq:info-diff-s0}
\lt| I\big(X^{n};\dvs{n}\big|s_{0}\big) - I\big(X^{n};\dvs{n}\big|s_{0}'\big) \rt| < n\epsilon. 
\end{equation}
\end{lemma}

\begin{IEEEproof}
Without loss of generality assume $n>K$. Let $P_{X^{n}}$, and $s_{0},s_{0}'$ be arbitrary but fixed. Let $B = \|\dvs{K}\|$ be the length of output string induced by the first $K$ input symbols $X^{K}$, where $K$ is the ISI length. Then we have the one-to-one correspondence
\begin{equation}\label{eq:BVK-bijection}
\big(B,\dvs{n}\big) \longleftrightarrow \big(\dvs{K},\dv{K+1}{n}\big),
\end{equation}
i.e., we can partition the output $\dvs{n}$ into two chunks that correspond to $X^{K}$ and $X_{K+1}^{n}$, respectively.\footnote{We can also work with any other partition, as long as the first input block has a constant length that is at least $K-1$.}

First we want to bound the difference in \eqref{eq:info-diff-s0} by the corresponding mutual information expression with $B$ inserted in the output. We have
\begin{align}
I\big(X^{n};\dvs{n}\big|s_{0}\big) &\leq I\big(X^{n};\dvs{n},B\big|s_{0}\big) \nonumber\\
&\leq I\big(X^{n};\dvs{n}\big|s_{0}\big) + I\big(X^{n};B\big|\dvs{n},s_{0}\big) \nonumber\\
&\leq I\big(X^{n};\dvs{n}\big|s_{0}\big) + H(B). \label{eq:I-ineq-sandwich1}
\end{align}
Since
\[ \Ex{B} = \Ex{\|\dvs{K}\|} = \sum_{k=1}^{K}\Ex{\|\dvl{k}\|}, \]
by \eqref{eq:avg-length} we have
\[ 0 < Kb \leq \Ex{B} \leq Ka < \infty. \]
As $B$ is a distribution on the non-negative integers with finite mean value, by Lemma~\ref{lem:entropy-mean-value-bound} we have
\[ H(B) \leq 1+\Ex{B} \leq 1+Ka < \infty, \]
and so by \eqref{eq:I-ineq-sandwich1}
\begin{equation}\label{eq:I-ineq-sandwich2}
I\big(X^{n};\dvs{n}\big|s_{0}\big) \leq I\big(X^{n};\dvs{n},B\big|s_{0}\big) \leq I\big(X^{n};\dvs{n}\big|s_{0}\big) + K',
\end{equation}
where $K' \triangleq1+Ka$ is a constant. We can also derive the same inequality for $s_{0}'$, and taking differences of them yields
\begin{align*}
I\big(X^{n};\dvs{n}\big|s_{0}\big) - I\big(X^{n};\dvs{n}\big|s_{0}'\big) - K'  &\leq I\big(X^{n};\dvs{n},B\big|s_{0}\big) - I\big(X^{n};\dvs{n},B\big|s_{0}'\big) \\
&\leq \lt| I\big(X^{n};\dvs{n},B\big|s_{0}\big) - I\big(X^{n};\dvs{n},B\big|s_{0}'\big) \rt|.
\end{align*}
Thus by symmetry, we have the bound
\begin{equation}\label{eq:I-B-bound}
\lt| I\big(X^{n};\dvs{n}\big|s_{0}\big) - I\big(X^{n};\dvs{n}\big|s_{0}'\big) \rt| \leq \lt| I\big(X^{n};\dvs{n},B\big|s_{0}\big) - I\big(X^{n};\dvs{n},B\big|s_{0}'\big) \rt| + K'.
\end{equation}

Next we want to analyze the difference of the mutual information on the right hand side of \eqref{eq:I-B-bound}. To simplify notations we denote $\bX_{1} = X^{K}$, $\bX_{2} = X_{K+1}^{n}$, $\bV_{1} = \dvs{K}$, and $\bV_{2} = \dv{K+1}{n}$. With the bijective relation \eqref{eq:BVK-bijection} we can write
\begin{align}
I\big(X^{n};\dvs{n},B\big|s_{0}\big)  &= I\big(X^{K}X_{K+1}^{n};\dvs{K}\dv{K+1}{n}\big|s_{0}\big)  \nonumber\\
&= I\big(\bX_{1}\bX_{2};\bV_{1}\bV_{2}\big|s_{0}\big) \nonumber\\
&= I\big(\bX_{1};\bV_{1}\bV_{2}\big|s_{0}\big) + I\big(\bX_{2};\bV_{1}\big|\bX_{1}s_{0}\big) + I\big(\bX_{2};\bV_{2}\big|\bX_{1}\bV_{1}s_{0}\big) \nonumber\\
&= I_{1} + I_{2} + I_{3}, \label{eq:I-BK-decomp}
\end{align}
where $I_{1}$, $I_{2}$, and $I_{3}$ are defined as the first, second, and the term in the line above them, respectively. This decomposition closely resembles that in \cite[Theorem~4.6.4]{Gallager-IT_Reliable} and will be treated similarly here. Note that by the channel transition law \eqref{eq:ch-transition-law} and \eqref{eq:ch-prob-s0}, we have
\begin{align*}
p(\bv_{2}\cn\bx_{1}\bx_{2}\bv_{1}s_{0}) 
&= p\big(\lv{K+1}{n}\cn x^{n}\lvs{K}s_{0}\big) \\
&= \sum_{v_{(K+1)}\circ\,\cdots\,\circ\,v_{(n)}=\lv{K+1}{n}}\!\!\!\!\!p(v_{(K+1)},\cdots,v_{(n)}\cn x^{n}\lvs{K}s_{0}) \\
&= \sum_{v_{(K+1)}\circ\,\cdots\,\circ\,v_{(n)}=\lv{K+1}{n}}\!\!\!\!\!p(v_{(K+1)},\cdots,v_{(n)}\cn x_{K+1}^{n}s_{K}) \\
&= p\big(\lv{K+1}{n}\cn x_{K+1}^{n}s_{K}\big) \\
&= p(\bv_{2}\cn\bx_{2}s_{K}),
\end{align*}
where $s_{K} = x_{2}^{K}$. Therefore we have the relation
\begin{equation}\label{eq:ISI-prob-decomp}
p(\bx_{1}\bx_{2}\bv_{1}\bv_{2}\cn s_{0}) = 
p(\bx_{1})\cdot p(\bx_{2}\cn\bx_{1})\cdot p(\bv_{1}\cn\bx_{1} s_{0})\cdot p(\bv_{2}\cn\bx_{2}s_{K}).
\end{equation}
With this relation we have the Markov chains $\bX_{2}\to \bX_{1}S_{0} \to \bV_{1}$ and $\bV_{1}S_{0} \to \bX_{1}\bX_{2}S_{K} \to \bV_{2}$. Thus the second term of \eqref{eq:I-BK-decomp} is zero:
\begin{equation}\label{eq:I-BK-2}
I_{2} = I\big(\bX_{2};\bV_{1}\big|\bX_{1}s_{0}\big) = 0.
\end{equation}
Furthermore, for $I_{3}$ we have
\begin{align*}
H(\bV_{2}\big|\bX_{2}\bX_{1}\bV_{1}s_{0}) & = H(\bV_{2}\big|\bX_{2}\bX_{1}S_{K}\bV_{1}s_{0}) \\
& = H(\bV_{2}\big|\bX_{2}\bX_{1}S_{K}) \\
& = H(\bV_{2}\big|\bX_{2}\bX_{1}),
\end{align*}
as $S_{K}$ is determined by $\bX_{1}$. Moreover, observing that by \eqref{eq:ISI-prob-decomp},
\begin{align*}
p(\bv_{2}\big|\bx_{1}\bv_{1}s_{0}) 
&= \sum_{\bx_{2}}p(\bx_{2}\cn \bx_{1}\bv_{1}s_{0})p(\bv_{2}\big|\bx_{2}\bx_{1}\bv_{1}s_{0}) \\
&= \sum_{\bx_{2}}p(\bx_{2}\cn \bx_{1})p(\bv_{2}\big|\bx_{2}\bx_{1}s_{K}) \\
&= \sum_{\bx_{2}}p(\bx_{2}\cn \bx_{1})p(\bv_{2}\big|\bx_{2}\bx_{1}) \\
&= p(\bv_{2}\big|\bx_{1}),
\end{align*}
we have $H(\bV_{2}\big|\bX_{1}\bV_{1}s_{0}) = H(\bV_{2}\big|\bX_{1})$. Therefore the third term of \eqref{eq:I-BK-decomp} is
\begin{equation}\label{eq:I-BK-3}
I_{3} = I\big(\bX_{2};\bV_{2}\big|\bX_{1}\bV_{1}s_{0}\big) = H(\bV_{2}\big|\bX_{1}) - H(\bV_{2}\big|\bX_{2}\bX_{1}) = I\big(\bX_{2};\bV_{2}\big|\bX_{1}\big).
\end{equation}
Finally, we can bound the first term of \eqref{eq:I-BK-decomp} by
\[ I_{1} = I\big(\bX_{1};\bV_{1}\bV_{2}\big|s_{0}\big) \leq H(\bX_{1}) \leq K\log|\X|, \]
and so from \eqref{eq:I-BK-decomp}, \eqref{eq:I-BK-2}, and \eqref{eq:I-BK-3}, we have
\[ I\big(\bX_{2};\bV_{2}\big|\bX_{1}\big) \leq I\big(X^{n};\dvs{n},B\big|s_{0}\big) \leq K\log|\X| + I\big(\bX_{2};\bV_{2}\big|\bX_{1}\big). \]
We can also derive the same bounds for $s_{0}'$, and taking differences of them yields
\[ I\big(X^{n};\dvs{n},B\big|s_{0}\big) - I\big(X^{n};\dvs{n},B\big|s_{0}'\big) \leq K\log|\X|. \]
By symmetry again
\begin{equation}\label{eq:I-B-diff}
\lt| I\big(X^{n};\dvs{n},B\big|s_{0}\big) - I\big(X^{n};\dvs{n},B\big|s_{0}'\big) \rt| \leq K\log|\X|.
\end{equation}

Combining \eqref{eq:I-B-bound} and \eqref{eq:I-B-diff}, we have
\[ \lt| I\big(X^{n};\dvs{n}\big|s_{0}\big) - I\big(X^{n};\dvs{n}\big|s_{0}'\big) \rt| \leq K'' \triangleq K\log|\X| + K'. \]
Since $K''$ is a constant that does not depend on $P_{X^{n}}$, $s_{0}$, or $s_{0}'$, we have shown that for any $\epsilon>0$, we can always find $N > K''/\epsilon$, such that $\forall n\geq N$, for any $P_{X^{n}}$, and any $s_{0},s_{0}'$,
\[ \lt| I\big(X^{n};\dvs{n}\big|s_{0}\big) - I\big(X^{n};\dvs{n}\big|s_{0}'\big) \rt| < n\epsilon. \]
\end{IEEEproof}

\begin{IEEEproof}[Proof of Lemma~\ref{lem:subadditivity}]
We use the same techniques as in the proof above. Consider the mutual information $I\big(X^{N};\dvs{N}\big|s_{0}\big)$. Let $B = \|\dvs{n}\|$ be the random variable that denotes the number of output symbols of the channel corresponding to the input $X^{n}$. Then we have the one-to-one correspondence
\[ \big(B,\dvs{N}\big) \longleftrightarrow \big(\dvs{n},\dv{n+1}{N}\big),  \]
i.e., we can partition the output $\dvs{N}$ into two chunks. Denote $\bX_{1} = X^{n}$, $\bX_{2} = X_{n+1}^{N}$, $\bV_{1} = \dvs{n}$, and $\bV_{2} = \dv{n+1}{N}$.

The rest of the proof closely parallels \cite[Thm~4.6.1]{Gallager-IT_Reliable}. For any $P_{X^{N}}$ and $s_{0}$ consider the decomposition
\begin{IEEEeqnarray}{r/C/l}
I\big(X^{N};\dvs{N}\big|s_{0}\big) &\leq& I\big(X^{N};B,\dvs{N}\big|s_{0}\big) \nonumber\\
&=& I\big(X^{N};\dvs{n},\dv{n+1}{N}\big|s_{0}\big) \nonumber\\
&=& I\big(\bX_{1}\bX_{2};\bV_{1}\bV_{2}\big|s_{0}\big) \nonumber\\
&=& I\big(\bX_{1};\bV_{1}| s_{0}\big) + I\big(\bX_{2};\bV_{1}\cn \bX_{1}s_{0}\big) + I\big(\bX_{1}\bX_{2};\bV_{2}\cn\bV_{1}s_{0}\big) \nonumber\\
&=& I_{1} + I_{2} + I_{3}, \label{eq:I-decomp-subadditivity}
\end{IEEEeqnarray}
where $I_{1}$, $I_{2}$, and $I_{3}$ are respectively defined as the first to third terms in the line above them. Similar to the proof above, we have the relation
\[ p(\bx_{1}\bx_{2}\bv_{1}\bv_{2}\cn s_{0}) = 
p(\bx_{1})\cdot p(\bx_{2}\cn\bx_{1})\cdot p(\bv_{1}\cn\bx_{1} s_{0})\cdot p(\bv_{2}\cn\bx_{2}s_{n}). \]
As a result, we have $I_{2} = 0$ and
\[ H\big(\bV_{2}\cn \bX_{1}\bX_{2}\bV_{1}s_{0}S_{n}\big) = H\big(\bV_{2}\cn \bX_{2}S_{n}\big). \]

Recall that $S_{n} = X_{n-K+2}^{n}$. We can write
\begin{align}
I_{3} &= I\big(\bX_{1}\bX_{2},S_{n};\bV_{2}\cn\bV_{1}s_{0}\big) \nonumber \\
&= I\big(\bX_{1}\bX_{2};\bV_{2}\cn\bV_{1}s_{0}S_{n}\big) + I\big(S_{n};\bV_{2}\cn\bV_{1}s_{0}\big) \nonumber \\
&\leq H\big(\bV_{2}\cn\bV_{1}s_{0}S_{n}\big) - H\big(\bV_{2}\cn \bX_{1}\bX_{2}\bV_{1}s_{0}S_{n}\big) + H(S_{n}) \nonumber\\
&\leq H\big(\bV_{2}\cn S_{n}\big) - H\big(\bV_{2}\cn \bX_{2}S_{n}\big) + \log |\St| \nonumber \\
&= I\big(\bX_{2};\bV_{2}\cn S_{n}\big) + \log |\St| \nonumber \\
&= I\big(X_{n+1}^{N};\dv{n+1}{N}\cn S_{n}\big) + \log |\St| \nonumber \\
&\leq \max_{s_{n}}\max_{P_{X_{n+1}^{N}}}I\big(X_{n+1}^{N};\dv{n+1}{N}| s_{n}\big) + \log|\St| \nonumber \\
&= \max_{s_{0}}\max_{P_{X^{m}}}I\big(X^{m};\dvs{m}\big|s_{0}\big) + \log|\St| \nonumber \\
&= m\C_{m}, \label{eq:I-3-final}
\end{align}
where we used the relation $m+n = N$ and the shift-invariance of the channel transition probability (when $s_{0} = s_{n}$). Now the expansion \eqref{eq:I-decomp-subadditivity} can be upper bounded by
\[ I\big(X^{N};\dvs{N}\big|s_{0}\big) \leq I_{1} + I_{3} \leq I\big(X^{n};\dvs{n}| s_{0}\big) + m\C_{m}. \]
Adding $\log |\St|$ to both sides we have
\begin{align*}
I\big(X^{N};\dvs{N}\big|s_{0}\big) + \log |\St| &\leq I\big(X^{n};\dvs{n}| s_{0}\big) + \log |\St| + m\C_{m} \\
&\leq \max_{s_{0}}\max_{P_{X^{n}}}I\big(X^{n};\dvs{n}| s_{0}\big) + \log |\St| + m\C_{m} \\
&\leq n\C_{n} + m\C_{m}.
\end{align*}
Since this inequality is true for all $P_{X^{N}}$ and $s_{0}$, it must be true for the maximization over them, and thus \eqref{eq:subadditivity} holds.
\end{IEEEproof}

\begin{IEEEproof}[Proof of Theorem~\ref{thm:exisitence-limit}]
For any $\epsilon>0$, we need to show that $\exists N>0$ such that $\forall n\geq N$,
\[ |C_{n}-C|<\epsilon. \]
We achieve this in three steps. First, by Lemma~\ref{lem:subadditivity}, $\exists N_{1}>0$ such that $\forall n\geq N_{1}$,
\begin{equation}\label{eq:C-limit-bound1}
|\C_{n}-C|<\epsilon/2.
\end{equation}

Second, by Lemma~\ref{lem:initial-state-does-not-matter}, $\exists N_{2}>0$ such that $\forall n\geq N_{2}$, $\forall P_{X^{n}}$, and $\forall s_{0},s_{0}' \in \St$, we have
\[ \lt| I\big(X^{n};\dvs{n}\big|s_{0}\big) - I\big(X^{n};\dvs{n}\big|s_{0}'\big) \rt| < n\epsilon/4. \]
Averaging over $s_{0}\in\St$ with respect to the initial distribution $P_{S_{0}}$, we have
\begin{align}
\lt| I\big(X^{n};\dvs{n}\big|S_{0}\big) - I\big(X^{n};\dvs{n}\big|s_{0}'\big) \rt| 
&\leq \sum_{s_{0}\in\St}p(s_{0})\lt| I\big(X^{n};\dvs{n}\big|s_{0}\big) - I\big(X^{n};\dvs{n}\big|s_{0}'\big) \rt| 
< n\epsilon/4. \label{eq:C-limit-I-s0-diff}
\end{align}
On the other hand, by \cite[Appendix~4A, Lemma~1]{Gallager-IT_Reliable}
\begin{equation}\label{eq:C-limit-I-cond-diff}
\lt| I\big(X^{n};\dvs{n}\big| S_{0}\big) - I\big(X^{n};\dvs{n}\big) \rt| \leq \log|\St|.
\end{equation}

To simplify notation and emphasize the dependence of mutual information on the input distribution, when $P_{X^{n}} = P$, we rewrite
\[ I(P) \triangleq I\big(X^{n};\dvs{n}\big), \]
\[ I_{S_{0}}(P) \triangleq I\big(X^{n};\dvs{n}\big| S_{0}\big), \]
\[ I_{s_{0}}(P) \triangleq I\big(X^{n};\dvs{n}\big|s_{0}\big). \]
Let $P_{X^{n}} = P^{*}$ and $s_{0} = s^{*}$ achieve the maximum of $\C_{n}$ in \eqref{eq:UB-C-bar} and let $P_{X^{n}} = \tP$ achieve the maximum of $C_{n}$ in \eqref{eq:C-n}. Then we can write
\[ C_{n} = \frac{1}{n}I(\tP),\qquad \C_{n} =  \frac{1}{n}\lt[ I_{s^{*}}(P^{*}) + \log|\St| \rt]. \]
By \eqref{eq:C-limit-I-s0-diff} and \eqref{eq:C-limit-I-cond-diff} with $s_{0}' = s^{*}$, we have
\begin{align*}
nC_{n} = I(\tP) &\leq I_{S_{0}}(\tP) + \log|\St| \\
&< I_{s^{*}}(\tP) + n\epsilon/4 + \log|\St| \\
&\leq I_{s^{*}}(P^{*}) + \log|\St| + n\epsilon/4 \\
&= n\C_{n} + n\epsilon/4;
\end{align*}
\begin{align*}
nC_{n} = I(\tP) &\geq I(P^{*}) \\
&\geq I_{S_{0}}(P^{*}) - \log|\St| \\
&> I_{s^{*}}(P^{*}) - n\epsilon/4 - \log|\St| \\
&= I_{s^{*}}(P^{*}) + \log|\St| - n\epsilon/4 - 2\log|\St| \\
&= n\C_{n} - n\epsilon/4 - 2\log|\St|.
\end{align*}
Therefore for $n\geq N_{2}$
\[ |C_{n}-\C_{n}| < \epsilon/4 + \frac{2\log|\St|}{n}. \]
If we choose $N_{3} >  8\log|\St|/\epsilon$, then for $n\geq \max\{N_{2}, N_{3}\}$,
\begin{equation}\label{eq:C-limit-bound2}
|C_{n}-\C_{n}| < \epsilon/2.
\end{equation}

Now let $N = \max\{N_{1},N_{2}, N_{3}\}$. Then by \eqref{eq:C-limit-bound1} and \eqref{eq:C-limit-bound2}, for $n\geq N$ we have
\[ |C_{n}-C| \leq |C_{n}-\C_{n}| + |\C_{n}-C| < \epsilon/2 + \epsilon/2 = \epsilon. \]
\end{IEEEproof}

\subsection{Proofs for Section~\ref{subsec:information-statbility}}
\label{subsec:proof-information-statbility}

\begin{IEEEproof}[Proof of Lemma~\ref{lem:I-k-limit}]
The proof of this lemma uses the same techniques as Lemmas~\ref{lem:initial-state-does-not-matter} and \ref{lem:subadditivity}. Consider the mutual information $I\big(X^{k};\dvs{k}\big|s_{0}\big)$. Let $B = \|\dv{k'+1}{k}\|$ be the length of the output string induced by the last $K-1$ input symbols. As always we have the one-to-one correspondence
\[ \big(B,\dvs{k}\big) \longleftrightarrow \big(\dvs{k'},\dv{k'+1}{k}\big).  \]
i.e., we can partition the output $\dvs{k}$ into two chunks. Denote $\bX_{1} = X^{k'}$, $\bX_{2} = X_{k'+1}^{k}$, $\bV_{1} = \dvs{k'}$, and $\bV_{2} = \dv{k'+1}{k}$.

First, similar to \eqref{eq:I-ineq-sandwich2} we can show that for any $s_{0}$ and $P_{X^{k}}$
\begin{equation}\label{eq:I-ineq-sandwich3}
I\big(X^{k};\dvs{k}\big|s_{0}\big) \leq I\big(X^{k};\dvs{k},B\big|s_{0}\big) \leq I\big(X^{k};\dvs{k}\big|s_{0}\big) + K',
\end{equation}
where $K' \triangleq1+(K-1)a$ is a constant. Second, with the bijection above, similar to \eqref{eq:I-decomp-subadditivity} we can write
\begin{IEEEeqnarray}{r/C/l}
I\big(X^{k};\dvs{k},B\big|s_{0}\big)  &=& I\big(X^{k'}X_{k'+1}^{k};\dvs{k'}\dv{k'+1}{k}\big|s_{0}\big) \nonumber \\
&=& I\big(\bX_{1}\bX_{2};\bV_{1}\bV_{2}\big|s_{0}\big)  \nonumber \\
&=& I\big(\bX_{1};\bV_{1}| s_{0}\big) + I\big(\bX_{2};\bV_{1}\cn \bX_{1}s_{0}\big)
+ I\big(\bX_{1}\bX_{2};\bV_{2}\cn\bV_{1}s_{0}\big)  \nonumber \\
&=& I_{1}' + I_{2}' + I_{3}', \label{eq::I-decomp-I-k-limit}
\end{IEEEeqnarray}
where $I_{1}'$, $I_{2}'$, and $I_{3}'$ are respectively defined as the first to third terms in the line above them. Since we have  the relation
\[ p(\bx_{1}\bx_{2}\bv_{1}\bv_{2}\cn s_{0}) = 
p(\bx_{1})\cdot p(\bx_{2}\cn\bx_{1})\cdot p(\bv_{1}\cn\bx_{1} s_{0})\cdot p(\bv_{2}\cn\bx_{2}s_{k'}), \]
we have $I_{2}' = 0$ and $H\big(\bV_{2}\cn \bX_{1}\bX_{2}\bV_{1}s_{0}S_{k'}\big) = H\big(\bV_{2}\cn \bX_{2}S_{k'}\big)$, so
\begin{align*}
I_{3}' &= I\big(\bX_{1}\bX_{2},S_{k'};\bV_{2}\cn\bV_{1}s_{0}\big) \\
&= I\big(\bX_{1}\bX_{2};\bV_{2}\cn\bV_{1}s_{0}S_{k'}\big) + I\big(S_{k'};\bV_{2}\cn\bV_{1}s_{0}\big) \\
&\leq H\big(\bV_{2}\cn\bV_{1}s_{0}S_{k'}\big) - H\big(\bV_{2}\cn \bX_{1}\bX_{2}\bV_{1}s_{0}S_{k'}\big) + H(S_{k'}) \\
&\leq H\big(\bV_{2}\cn S_{k'}\big) - H\big(\bV_{2}\cn \bX_{2}S_{k'}\big) + \log |\St| \\
&= I\big(\bX_{2};\bV_{2}\cn S_{k'}\big) + \log |\St| \\
&\leq H(\bX_{2}) + \log |\St| \\
&\leq (K-1)\log|\X| + \log |\St| \\
&= 2\log|\St|.
\end{align*}
Substituting the (in)equalities for $I_{2}'$ and $I_{3}'$ to \eqref{eq::I-decomp-I-k-limit}, we have
\[ I\big(X^{k};\dvs{k},B\big|s_{0}\big) \leq I_{1}' + 2\log|\St|. \]
Since $I_{1}' = I\big(X^{k'};\dvs{k'}\big|s_{0}\big)$, and by \eqref{eq::I-decomp-I-k-limit} $I_{1}' \leq I\big(X^{k};\dvs{k},B\big|s_{0}\big)$,
\begin{equation}\label{eq:I-ineq-sandwich4}
I\big(X^{k'};\dvs{k'}\big|s_{0}\big) \leq I\big(X^{k};\dvs{k},B\big|s_{0}\big) \leq I\big(X^{k'};\dvs{k'}\big|s_{0}\big) + 2\log|\St|.
\end{equation}
Combining \eqref{eq:I-ineq-sandwich3} and \eqref{eq:I-ineq-sandwich4}, we have
\[ \lt| I\big(X^{k};\dvs{k}\big|s_{0}\big) - I\big(X^{k'};\dvs{k'}\big|s_{0}\big) \rt| \leq K' + 2\log|\St|. \]

Finally, let $s_{0} = s^{*}$ and $P_{X^{k}} = \tP$. Then the inequality above becomes
\[ \lt| kI_{k} - (k'\C_{k'} - \log|\St|) \rt| \leq K' + 2\log|\St|, \]
so by the triangle inequality
\[ \lt| kI_{k} - k'\C_{k'} \rt| \leq K' + 3\log|\St|, \]
\begin{align*}
k|I_{k} - C| &\leq \lt|kI_{k} - k'\C_{k'}\rt| + \lt|k'\C_{k'} - k'C\rt| + (K-1)C \\
&\leq K' + 3\log|\St| + k'|\C_{k'} - C| + (K-1)C \\
&\leq K'' + k|\C_{k'} - C|,
\end{align*}
where $K'' = K' + 3\log|\St| + (K-1)C$ is a constant. Hence
\[ |I_{k} - C| \leq \frac{K''}{k} + \lt|\C_{k'} - C\rt|. \]
Since when $k\to\infty$ we also have $k'\to\infty$, by Theorem~\ref{thm:exisitence-limit}, the right side converges to 0 as $k\to\infty$, so does the left side.
\end{IEEEproof}

\begin{IEEEproof}[Proof of Lemma~\ref{lem:XV-indep}]
Similar to the channel transition laws \eqref{eq:ch-transition-law}, \eqref{eq:ch-prob-s0} and \eqref{eq:ch-prob}, we have
\[ p\big(v_{((i-1)k+1)},\cdots,v_{(ik)}\cn x_{(i-1)k+1}^{ik},s_{(i-1)k}\big) = \prod_{j=(i-1)k+1}^{ik}p(v_{(j)}\cn x_{j}s_{j-1}), \]
\begin{align*}
p(\bv_{i}\cn \bx_{i}s_{(i-1)k}) &= p\big(\lv{(i-1)k+1}{ik}\cn x_{(i-1)k+1}^{ik},s_{(i-1)k}\big) \\
&= \sum_{v_{((i-1)k+1)}\circ\,\cdots\,\circ\, v_{(ik)} = \bv_{i}}p\big(v_{((i-1)k+1)},\cdots,v_{(ik)}\cn x_{(i-1)k+1}^{ik},s_{(i-1)k}\big),
\end{align*}
and there are similar expressions for $\bx'$, $\bv'$ and $s_{gk}$. So the group transition probabilities can be expressed as
\[ p(\bv_{1},\cdots,\bv_{g},\bv'\cn \bx_{1},\cdots,\bx_{g},\bx',s_{0}) = \prod_{i=1}^{g}p\big(\bv_{i}\cn \bx_{i}s_{(i-1)k}\big) \cdot p\big(\bv'\cn \bx's_{gk}\big), \]
\begin{align*}
p(\bv_{1},\cdots,\bv_{g},\bv'\cn \bx_{1},\cdots,\bx_{g},\bx') &= \sum_{s_{0}}p(s_{0})\prod_{i=1}^{g}p(\bv_{i}\cn \bx_{i}s_{(i-1)k}) \cdot p\big(\bv'\cn \bx's_{gk}\big) \\
&= p(\bv_{1}\cn \bx_{1}) \cdot \prod_{i=2}^{g}p(\bv_{i}\cn \bx_{i}s_{(i-1)k}) \cdot p\big(\bv'\cn \bx's_{gk}\big).
\end{align*}
When the input distribution is $P_{\xi}$, we know all $\bX_{i}$, $i = 1,\ldots,g$, are i.i.d. with distribution $\tP$ and $\bX' = \bx_{0}$. Hence almost surely we have $S_{ik} = s^{*}$, $i = 1,\ldots,g$. When all $s_{ik} = s^{*}$, the equation above can be rewritten as
\[ p(\bv_{1},\cdots,\bv_{g},\bv'\cn \bx_{1},\cdots,\bx_{g},\bx') = p(\bv_{1}\cn \bx_{1}) \cdot \prod_{i=2}^{g}p(\bv_{i}\cn \bx_{i}s^{*}) \cdot p\big(\bv'\cn \bx's^{*}\big), \]
So for any $\bx_{1}\cdots\bx_{g}\bx'$ and $(\bv_{1},\cdots,\bv_{g},\bv')$ we have the expansion
\begin{align}
\Pr\big(\bxi_{n,k} &= \bx_{1}\cdots\bx_{g}\bx',\ \cta_{n,k} = (\bv_{1},\cdots,\bv_{g},\bv')\big) \nonumber\\
&= \tP(\bx_{1})p(\bv_{1}\cn \bx_{1}) \cdot \prod_{i=2}^{g}\tP(\bx_{i})p(\bv_{i}\cn \bx_{i}s^{*}) \cdot \1{\bx'=\bx_{0}}p\big(\bv'\cn \bx's^{*}\big). \label{eq:indep-xv}
\end{align}
On the other hand, when the input distribution is $P_{\xi}$, almost surely $S_{ik} = s^{*}$ and so we always have
\begin{equation}\label{eq:bxbv}
\Pr(\bX_{i} = \bx_{i},\bV_{i} = \bv_{i}) = \tP(\bx_{i})p(\bv_{i}\cn \bx_{i}s^{*})
\end{equation}
for $2\leq i\leq g$ and
\[ \Pr(\bX' = \bx',\bV' = \bv') = \1{\bx'=\bx_{0}}p(\bv'\cn \bx's^{*}). \]
Therefore, \eqref{eq:indep-xv} becomes
\begin{align*}
\Pr\big((\bX_{1},\bV_{1}) &= (\bx_{1},\bv_{1}),\ \cdots,\ (\bX_{g},\bV_{g}) = (\bx_{g},\bv_{g}),\ (\bX',\bV') = (\bx',\bv')\big)\\
&= \Pr(\bX_{1} = \bx_{1},\bV_{1} = \bv_{1}) \cdot \prod_{i=2}^{g}\Pr(\bX_{i} = \bx_{i},\bV_{i} = \bv_{i}) \cdot \Pr(\bX' = \bx',\bV' = \bv')
\end{align*}
and so we have the desired independence. The i.i.d. property of $(\bX_{2},\bV_{2})$, \ldots, $(\bX_{g}\bV_{g})$ follows from \eqref{eq:bxbv}.
\end{IEEEproof}

\begin{IEEEproof}[Proof of Lemma~\ref{lem:long-length-prob-vanishes}]
Fix $\epsilon>0$. In view of \eqref{eq:expansion-n} define $U_{i} = \|\bV_{i}\|$ for $i = 1,\ldots,g$ and $B = \|\bV'\|$. Then
\[ \|\bta_{n,k}\| = \sum_{i=1}^{g}U_{i} + B. \]
By lemma~\ref{lem:XV-indep} we know all $U_{i}$, $i = 1,2,\ldots,g$ are independent, and $U_{i}$, $i = 2,\ldots,g$ are i.i.d.. Note that by \eqref{eq:avg-length} we have
\[ \Ex{U_{1}},\ \Ex{U_{2}} \leq ka <\infty \]
and so $U_{1}<\infty$ almost surely. Then by Lemma~\ref{lem:WLLN}, there exists $G_{1}>0$ such that when $g\geq G_{1}$,
\[ \Pr\lt(\lt|\frac{1}{g}\sum_{i=1}^{g}U_{i} - \Ex{U_{2}}\rt| > \frac{ka}{2}\rt) < \epsilon/2, \]
which implies
\begin{align}
\Pr\lt(\sum_{i=1}^{g}U_{i} > \frac{3an}{2}\rt) &\stackrel{(*)}{\leq} \Pr\lt(\sum_{i=1}^{g}U_{i} > \frac{3akg}{2}\rt) \nonumber\\
&\leq \Pr\lt(\frac{1}{g}\sum_{i=1}^{g}U_{i} > \frac{ka}{2} + \Ex{U_{2}}\rt) \nonumber\\
&\leq \Pr\lt(\lt|\frac{1}{g}\sum_{i=1}^{g}U_{i} - \Ex{U_{2}}\rt| > \frac{ka}{2}\rt) \nonumber\\
&< \epsilon/2, \label{eq:prob-vanish1}
\end{align}
where for $(*)$ we used the relation $n = gk+l$.

On the other hand, since
\[ \Ex{B} \leq la \leq ka, \]
by Markov inequality we have
\[ \Pr\lt(B>\frac{an}{2}\rt) \leq \frac{2\Ex{B}}{an} \leq \frac{2k}{n}. \]
If we choose $G_{2} > 4/\epsilon$, then for $g\geq G_{2}$
\[ \Pr\lt(B>\frac{an}{2}\rt) \leq \frac{2k}{n} \leq \frac{2}{g} < \epsilon/2. \]
Now combining this inequality with \eqref{eq:prob-vanish1}, using union bound we have
\begin{align*}
\Pr\big(|\bta_{n,k}| > 2an\big)  &= \Pr\lt(\sum_{i=1}^{g}U_{i} + B > 2an\rt) \\
&\leq \Pr\lt(\lt\{\sum_{i=1}^{g}U_{i} > \frac{3an}{2} \rt\}\bigcup \lt\{B > \frac{an}{2}\rt\}\rt) \\
&\leq \Pr\lt(\sum_{i=1}^{g}U_{i} > \frac{3an}{2} \rt) + \Pr\lt(B > \frac{an}{2}\rt) \\
&< \epsilon/2 + \epsilon/2 = \epsilon,
\end{align*}
whenever $g \geq \max\{G_{1},G_{2}\}$, i.e., $n \geq \max\{G_{1}k,G_{2}k\}$.
\end{IEEEproof}

\begin{IEEEproof}[Proof of Lemma~\ref{lem:avg-info-density-converge}]
By Lemma~\ref{lem:XV-indep}, all the pairs $(\bX_{i},\bV_{1})$, $(\bX_{2},\bV_{2})$, \ldots, $(\bX_{g}\bV_{g})$, and $(\bX',\bV')$ are independent, and so we have the decomposition
\begin{align}
i_{\bxi_{n,k},\,\cta_{n,k}} & = \log\frac{p(\bX_{1},\bX_{2},\cdots,\bX_{g},\bX',\bV_{1},\bV_{2},\cdots,\bV_{g},\bV')}{p(\bX_{1},\bX_{2},\cdots,\bX_{g}\bX')p(\bV_{1},\bV_{2},\cdots,\bV_{g},\bV')} \nonumber\\
& = \sum_{i=1}^{g}\log\frac{p(\bX_{i}\bV_{i})}{p(\bX_{i})p(\bV_{i})} + \log\frac{p(\bX'\bV')}{p(\bX')p(\bV')} \nonumber\\
& = \sum_{i=1}^{g}\log\frac{p(\bX_{i}\bV_{i})}{p(\bX_{i})p(\bV_{i})}  \nonumber\\
& = \sum_{i=1}^{g}i_{\bX_{i}\bV_{i}}, \label{eq:info-density-expansion}
\end{align}
since $i_{\bX'\bV'} = 0$ by construction of $\bX'$.

We now use Lemma~\ref{lem:WLLN} to show the convergence. First, the random variables $i_{\bX_{i}\bV_{i}}$, $i = 1,2,\ldots,g$ are independent, and $i_{\bX_{i}\bV_{i}}$, $i = 2,\ldots,g$ are i.i.d.. Second, note that we can write
\[ i_{\bX_{i}\bV_{i}} = \log p(\bV_{i}\cn\bX_{i}) - \log p(\bV_{i}) \]
almost everywhere. Since
\[ \Ex{\|\bV_{i}\|}\leq ka <\infty, \]
\[ \Ex{\|\bV_{i}\|\cn \bX_{i} = \bx_{i}} \leq ka <\infty,\qquad \forall \bx_{i}\in\X^{k}, \]
by Lemma~\ref{lem:entropy-finite-avg-length-bound} we know
$H(\bV_{i}) <\infty$,
and for all $\bx_{i}\in\X^{k}$,
\begin{align*}
H(\bV_{i}\cn \bX_{i} = \bx_{i}) & \leq 1 + \Ex{\|\bV_{i}\|\cn \bX_{i} = \bx_{i}}(\log|\V| + 1) \\
&\leq 1 + ka(\log|\V| + 1) < \infty,
\end{align*}
\[ H(\bV_{i}\cn \bX_{i}) \leq 1 + ka(\log|\V| + 1) < \infty. \]
Therefore, for $i = 1,2,\ldots,g$ we can write
\begin{align*}
 \Ex{|i_{\bX_{i}\bV_{i}}|} &= \Ex{-\log p(\bV_{i}\cn\bX_{i})} + \Ex{-\log p(\bV_{i})} \\
 &= H(\bV_{i}\cn\bX_{i}) + H(\bV_{i}) < \infty.
\end{align*}
In particular, $i_{\bX_{1}\bV_{1}}$ is finite almost surely. Finally, by Lemma~\ref{lem:XV-indep},
\[ p(\bX_{2}\bV_{2}) = \tP(\bX_{2})p(\bV_{2}\cn \bX_{2}s^{*}) \]
and so
\[ \Ex{i_{\bX_{2}\bV_{2}}} = I(\bX_{2};\bV_{2}) = I\big(X^{k};\dvs{k}\big| S_{0} = s^{*}\big)\Big|_{P_{X^{k}} = \tP} = kI_{k}. \]

Therefore, by Lemma~\ref{lem:WLLN} and \eqref{eq:info-density-expansion} we have
\[ i_{\bxi_{n,k},\,\cta_{n,k}}/g \to kI_{k} \]
in probability when $g\to\infty$. In particular, for any $\epsilon>0$, there exist $G>0$ such that when $g\geq G$,
\[ \Pr\lt(|i_{\bxi_{n,k},\,\cta_{n,k}}-kgI_{k}|>\delta kg/4\rt) < \epsilon. \]
Consequently, if we choose $N = kG$, then for all $n\geq N$, we have $g\geq G$ and so
\[ \Pr\lt(|i_{\bxi_{n,k},\,\cta_{n,k}}-kgI_{k}|>\delta n/4\rt) \leq \Pr\lt(|i_{\bxi_{n,k},\,\cta_{n,k}}-kgI_{k}|>\delta kg/4\rt) < \epsilon. \]
\end{IEEEproof}

\section{Proofs for Section~\ref{sec:ach-rate}}

\subsection{Proof for $\Pe{2}\to0$}\label{subsec:Pe2}

Let $\Ae(Z,V)$ denote the jointly typical set for the process $\{Z_{j},V_{j}\}$. Define the event
\begin{align*}
F &=  \lt\{ (z^{m},V^{m})\notin\Ae(Z,V) \text{ for all } z^{m} \text{ satisfying } z^{m} \sqsubset Y^{n}(1) \text{ and } z^{m} \in \sAe(Z)\rt\} \\
&=  \bigcap_{z^{m} \in \sAe(Z)}\lt\{ (z^{m},V^{m})\notin\Ae(Z,V) \text{ if } z^{m} \sqsubset Y^{n}(1)\rt\},
\end{align*}
then we can decompose $\Pe{2}$ as follows:
\begin{IEEEeqnarray*}{rCl}
\Pe{2} &=& \Pr\lt(\{T\geq m\}\cap F \rt) \\
&=& \Pr\lt(\{T\geq m\}\cap \lt\{Z^{m}\notin \sAe(Z)\rt\}\cap F \rt) \\
&& +\: \Pr\lt(\{T\geq m\}\cap \lt\{Z^{m}\in \sAe(Z)\rt\}\cap F \rt).
\end{IEEEeqnarray*}
The first term is bounded by
$\Pr\lt(\lt\{Z^{m}\notin \sAe(Z)\rt\} \rt)$,
which goes to 0 as $n\to\infty$ by the general AEP and the ergodic theorem. For the second term, define the event
\[ G = \{T\geq m\}\cap \lt\{Z^{m}\in \sAe(Z)\rt\}. \]
In this event, $Z^{m}$ is clearly a subsequence of $Y^{n}(1)$ since $T\geq m$, and it is doubly-strongly typical. Thus the event
\[ G \cap \lt\{(Z^{m},V^{m})\notin\Ae(Z,V)\rt\} \]
contains $G\cap F$ as a subset, since the latter requires that for all doubly-strongly typical subsequence $z^{m}$ of $Y^{n}(1)$,
\[ (z^{m},V^{m})\notin\Ae(Z,V), \]
while the former just requires this to hold for one of them (i.e., for $Z^{m}$). As a result, the second term in the decomposition of $\Pe{2}$ can be bounded by
\begin{align*}
\Pr\lt(G\cap F \rt) &\leq \Pr\lt(G \cap \lt\{(Z^{m},V^{m})\notin\Ae(Z,V)\rt\} \rt) \\
 &\leq \Pr\lt((Z^{m},V^{m})\notin\Ae(Z,V) \rt)
 \end{align*}
which also goes to 0 as $n\to\infty$ by the general AEP for the stationary ergodic joint process $\{Z_{j},V_{j}\}$. Therefore, we conclude that $\Pe{2}\to0$.

\subsection{Upper bound for $\Pr\lt(z^{m} \sqsubset Y^{n}(2)\rt)$}\label{subsec:upper-bound-subseq-prob}

Let $z^{m} \in \sAe(Z)$. We generalize the idea in \cite{Diggavi-Grossglauser-finite-buffer-channel} to compute and upper bound for $\Pr\lt(z^{m} \sqsubset Y^{n}(2)\rt)$. For $l\leq m$ define $\tau_{l}$ to be the first time that $z^{l}$ appears as a subsequence in $\{Y_{i}\}_{i\geq1}$, i.e.,
\[ \tau_{l} = \inf\lt\{t: z^{l}\sqsubset Y^{t}\rt\}, \]
then $\tau_{l}$ is a stopping time. Let $\tau_{0} = 0$ and for $l>0$ define $N_{l}$ to be the first time (hitting time) that $z_{l}$ is visited in $\{Y_{i}\}_{i\geq1}$ after time $\tau_{l-1}$:
\[ N_{l} \triangleq \inf\lt\{i>\tau_{l-1}:Y_{i} = z_{l}\rt\} - \tau_{l-1}, \]
and so
\[ \tau_{l} = \sum_{i\leq l}N_{i}. \]
Note that for $k>0$ we have
\[ \Pr(N_{1} = k) = \Pr\lt(Y_{1},\cdots,Y_{k-1}\neq z_{1}; Y_{k} = z_{1}\rt). \]
Furthermore, when $l>1$, by the strong Markov property, for $k>0$
\begin{align}
\Pr(N_{l} = k\cn N_{1}=n_{1},&\cdots,N_{l-1} = n_{l-1};\tau_{l-1}<\infty) = \Pr(N_{l} = k\cn \tau_{l-1}<\infty) \nonumber \\
&=\Pr\lt(Y_{2},\cdots,Y_{k}\neq z_{l}; Y_{k+1} = z_{l}\cn Y_{1} = z_{l-1}\rt). \label{eq:N-l-prob}
\end{align}
Thus we have the following lemma:

\begin{lemma}
For all $l$, $\Pr(N_{l} < \infty) = 1$ and so $\Pr(\tau_{l} < \infty) = 1$. Hence
\[ \Pr(N_{l} = k\cn\tau_{l-1}<\infty) = \Pr(N_{l} = k), \]
and all $N_{l}$, $1\leq l\leq m$ are independent.
\end{lemma}
\begin{IEEEproof}
Since $P$ is a finite irreducible Markov chain, it is a finite closed equivalent class, and thus is recurrent by \cite[Proposition~4-30]{Kemeny-Denumerable-Markov-chains}. Following the notations of \cite[Sections~4.6-4.7]{Kemeny-Denumerable-Markov-chains}, define the hitting probability matrices $H$ and $\bH$, where the $(a,b)$-th entry of $H$ and $\bH$ denotes the probability that $b$ is reached in finite time from $a$ since time 1 and after time 1, respectively, then both $H$ and $\bH$ are all-1 matrices, by Lemma~4-23 and Proposition~4-14 in \cite{Kemeny-Denumerable-Markov-chains}.

As a consequence, we have
\[ \Pr(\tau_{1} < \infty) = \Pr(N_{1} < \infty) = \pi H = 1. \]
Also, simple conditional probability calculation gives us
\[ \Pr(N_{2} = k) = \Pr(N_{2} = k\cn\tau_{1}<\infty), \]
and using strong Markov property as above we conclude that $N_{1},N_{2}$ are independent. Now by \eqref{eq:N-l-prob},
\[ \Pr(N_{2}<\infty) = \sum_{k>0}\Pr(N_{2} = k) = \sum_{k>0}\Pr(N_{2} = k\cn\tau_{1}<\infty) = \bH_{z_{1}z_{2}} = 1. \]
Therefore, since the intersection of two measure-1 events still has measure 1, we have
\[ \Pr(\tau_{2} < \infty) = \Pr(N_{1} + N_{2} < \infty) = \Pr(N_{1} < \infty, N_{2}< \infty) = 1. \]
Continuing in this way we can show that for all $l$
\[ \Pr(N_{l} = k\cn\tau_{l-1}<\infty) = \Pr(N_{l} = k), \]
\[ \Pr(N_{l} < \infty) = \Pr(\tau_{l} < \infty) = 1, \]
and hence all $N_{l}$, $1\leq l\leq m$ are independent. 
\end{IEEEproof}

With the definitions of $\tau_{l}$ and $N_{l}$ we can write
\[ \Pr(z^{m}\sqsubset Y^{n}(2)) = \Pr(\tau_{m}\leq n) =  \Pr\bigg(\sum_{l=1}^{m}N_{l}\leq n\bigg). \]
Using Chernoff's bound and by the independence of all $N_{l}$,
\begin{align*}
\Pr\bigg(\sum_{l=1}^{m}N_{l}\leq n\bigg) &\leq \infg e^{\gamma n}\cdot\Ex{e^{-\gamma\sum_{l=1}^{m}N_{l}}} \\
& = \infg e^{\gamma n}\cdot\prod_{l=1}^{m}\Ex{e^{-\gamma N_{l}}} \\
& = \infg e^{\gamma n}\cdot\prod_{l=1}^{m}\Phi_{l}(\gamma),
\end{align*}
where $\Phi_{l}(\gamma)$ is the moment generating function of $N_{l}$.

Recall that in Section~\ref{sec:ach-rate}, for $a,b\in\Y$ we define $\tP_{b}$ to be the submatrix of $P$ obtained from removing the $b$-th row and $b$-th column, $\tq_{b}$ to be the column vector obtained from removing the $b$-th entry of the $b$-th column of $P$, and $\tp_{ab}$ to be the row vector obtained from removing the $b$-th entry of the $a$-th row of $P$. In addition, we use $\tpi_{a}$ to denote the row vector obtained from removing the $a$-th entry (i.e., $\pi_{a}$) of $\pi$. With these notations we can express
\[ \Pr(N_{1} = k) = 
\begin{cases}
\pi_{z_{1}} & \text{ if } k = 1 \\
\tpi_{z_{1}}\tP_{z_{1}}^{k-2}\tq_{z_{1}} & \text{ if } k \geq 2
\end{cases}, \]
and for $l>1$
\[ \Pr(N_{l} = k) = \begin{cases}
P_{z_{l-1}z_{l}} & \text{ if } k = 1 \\
\tp_{z_{l-1}z_{l}}\tP_{z_{l}}^{k-2}\tq_{z_{l}} & \text{ if } k \geq 2
\end{cases}. \]
Observe that since $P$ is a stochastic matrix, for any $z\in\Y$ the spectral radius $\rho(\tP_{z}) \leq \| \tP_{z} \|_{\infty} \leq 1$ as the maximum absolute row sum of $\tP_{z}$, i.e., $\| \tP_{z} \|_{\infty}$, is at most 1. Hence for $\gamma > 0$ the series below is (absolutely) convergent, and we have
\[ Q_{z} \triangleq \sum_{k=0}^{\infty}\tP_{z}^{k}e^{-k\gamma} = I + \tP_{z}e^{-\gamma}Q_{z}, \]
\[ Q_{z} = (I-\tP_{z}e^{-\gamma})^{-1} = e^{\gamma}(e^{\gamma}I- \tP_{z})^{-1}. \]
Now applying this result to the moment generating functions for $N_{l}$, we have:
\begin{align*}
\Phi_{1}(\gamma) &= 
\pi_{z_{1}}e^{-\gamma} + \tpi_{z_{1}}\lt(\sum_{k=2}^{\infty}\tP_{z_{1}}^{k-2}e^{-k\gamma}\rt)\tq_{z_{1}} \\
&= \pi_{z_{1}}e^{-\gamma} + e^{-2\gamma}\tpi_{z_{1}}Q_{z_{1}}\tq_{z_{1}} \\
&= e^{-\gamma}\lt[\pi_{z_{1}} + \tpi_{z_{1}}(e^{\gamma}I- \tP_{z_{1}})^{-1}\tq_{z_{1}}\rt],
\end{align*}
\[ \Phi_{l}(\gamma) = e^{-\gamma}\lt[P_{z_{l-1}z_{l}} + \tp_{z_{l-1}z_{l}}(e^{\gamma}I- \tP_{z_{l}})^{-1}\tq_{z_{l}}\rt] \]
for $l>1$. Note that by irreducibility of $P$, these functions are all positive for $\gamma > 0$. Hence we can take the logarithm (base $e$) of them and furthermore, define the following functions for all $a,b\in\Y$:
\[ E_{a}(\gamma) = \ln\lt[\tpi_{a} + \tpi_{a}(e^{\gamma}I- \tP_{a})^{-1}\tq_{a}\rt], \]
\[ E_{ab}(\gamma) = \ln\lt[P_{ab} + \tp_{ab}(e^{\gamma}I- \tP_{b})^{-1}\tq_{b}\rt]. \]
With these definitions we have
\[ \Phi_{1}(\gamma) = \exp(-\gamma + E_{z_{1}}(\gamma)), \]
\[ \Phi_{l}(\gamma) = \exp(-\gamma + E_{z_{l-1}z_{l}}(\gamma)), \]
and so
\begin{align*}
\Pr&\bigg(\sum_{l=1}^{m}N_{l}\leq n\bigg) \\
& \leq \infg e^{\gamma n}\cdot \exp\lt[-m\gamma + E_{z_{1}}(\gamma) + \sum_{l=2}^{m}E_{z_{l-1}z_{l}}(\gamma)\rt] \\
&= \infg \exp\lt[(n-m)\gamma + E_{z_{1}}(\gamma) + \sum_{l=2}^{m}E_{z_{l-1}z_{l}}(\gamma)\rt]. \\
&\leq \infg \exp\lt[(n-m)\gamma + E_{1}(\gamma) + \sum_{l=2}^{m}E_{z_{l-1}z_{l}}(\gamma)\rt],
\end{align*}
where we define $E_{1}(\gamma) = \max_{a\in\Y}E_{a}(\gamma)$.

For $a,b\in\Y$ Let $M_{ab}(z^{m})$ denote the number of transitions from $a$ to $b$, i.e.,
\[ M_{ab}(z^{m}) = \sum_{l=2}^{m}\1{(z_{l-1},z_{l}) = (a,b)}. \]
Then
\[ \sum_{l=2}^{m}E_{z_{l-1}z_{l}}(\gamma) = \sum_{a,b\in\Y}M_{ab}(z^{m})E_{ab}(\gamma). \]
Since $z^{m}$ is doubly-strongly typical, we have
\[ \lt| \frac{1}{m} M_{ab}(z^{m}) - \pi_{a}\bP_{ab} \rt| \leq \epsilon,\quad \forall a,b\in\Y. \]
Hence
\[ \frac{1}{m}M_{ab}(z^{m})E_{ab}(\gamma) \leq \pi_{a}\bP_{ab}E_{ab}(\gamma) + \epsilon|E_{ab}(\gamma)|. \]
Therefore, we have
\[ \sum_{l=2}^{m}E_{z_{l-1}z_{l}}(\gamma) \leq m[E(\gamma) + \epsilon E_{0}(\gamma)], \]
where
\[ E(\gamma) \triangleq \sum_{a,b\in\Y}\pi_{a}\bP_{ab}E_{ab}(\gamma),\quad E_{0}(\gamma) \triangleq \sum_{a,b\in\Y}|E_{ab}(\gamma)|. \]

With all the derivations above, we have
\begin{align*}
\Pr(z^{m}&\sqsubset Y^{n}(2)) =  \Pr\bigg(\sum_{l=1}^{m}N_{l}\leq n\bigg) \\
&\leq \infg \exp\lt[(n-m)\gamma + E_{1}(\gamma) + \sum_{l=2}^{m}E_{z_{l-1}z_{l}}(\gamma)\rt] \\
&\leq \beta_{m},
\end{align*}
where we define
\[ \beta_{m} = \infg \exp\lt[(n-m)\gamma + E_{1}(\gamma) +  mE(\gamma) + m\epsilon E_{0}(\gamma)\rt]. \]

\ifCLASSOPTIONcaptionsoff
  \newpage
\fi



\bibliographystyle{IEEEtran}
\bibliography{IEEEabrv,NanoporeIT}
\end{document}